\documentclass[prx,twocolumn,amsmath,amssymb]{revtex4}

\usepackage{color}
\usepackage{bm}
\usepackage{dcolumn}
\usepackage[pdftex]{graphicx}

\usepackage{slashed}
\usepackage{amsmath}\usepackage{accents}
\usepackage{ulem}
\usepackage{mathdots}
\usepackage{multirow}
\newcommand{\cin}[1]{{#1}}

\newcommand{\1}{\mbox{1}\hspace{-0.25em}\mbox{l}}

\begin{document}

\title{
Theory of edge states based on the Hermiticity of tight-binding Hamiltonian operators
}

\author{Takahiro  Fukui}
\affiliation{Department of Physics, Ibaraki University, Mito 310-8512, Japan}

\date{\today}

\begin{abstract}
We develop a theory of edge states based on the Hermiticity of Hamiltonian operators for tight-binding models defined on lattices with boundaries. We describe  Hamiltonians using shift operators which serve as differential operators in continuum theories. It turns out that such Hamiltonian operators are not necessarily Hermitian on lattices with boundaries, which is due to the boundary terms associated with the summation by parts.
The Hermiticity of Hamiltonian operators leads to natural boundary conditions, and for models with nearest-neighbor (NN) hoppings only, there are reference states that satisfy the Hermiticity and boundary conditions simultaneously. Based on such reference states, we develop a Bloch-type theory for edge states of NN models on a half-plane. This enables us to extract Hamiltonians describing edge-states at one end, which are separated from the bulk contributions. It follows that  we can describe edge states at the left and right ends separately by distinct Hamiltonians for systems of cylindrical geometry.  We show various examples of such edge state Hamiltonians (ESHs), including Hofstadter model, graphene model, and higher-order topological insulators
(HOTIs), etc.
\end{abstract}

\pacs{
}

\maketitle

\section{Introduction}

In the studies of topological properties of matter 
\cite{Thouless:1982uq,Kane:2005aa,Qi:2008aa,Hasan:2010fk,Qi:2011kx}, 
the bulk-edge correspondence \cite{Hatsugai:1993fk,Hatsugai:1993aa} plays a central role.
Although topological invariants are well-defined for the Bloch wave functions,
such bulk topological invariants are not directly related with the 
physical observables, except for the integer quantum Hall effect \cite{Thouless:1982uq,kohmoto:85}. 
Therefore, one usually judges topological phases of matter 
from observation of gapless edge states embedded in the bulk insulating states \cite{Hasan:2010fk,Qi:2011kx,Konig:1977fk}.

Usually, edge states mean  localized states on $d-1$ dimensional boundaries for $d$ dimensional bulk systems.
However, recent discovery of higher-order topological insulators (HOTIs)
\cite{Slager:2015aa,Benalcazar:2017aa,Benalcazar:2017ab,Schindler:2018ab,Hayashi:2018aa,Hashimoto:2016aa,Hashimoto:2017aa}
have drawn our attention to unconventional 
$d-2,\cdots$ dimensional higher-order  edge states
such as corner states, hinge states, etc.
Therefore, the concept of edge states,  including higher-order edge states, would become more and more important
when one investigates topological phases for various kinds of systems.

Bulk states are described by the Bloch states. 
Let us consider a tight-binding model defined on a lattice which is 
characterized by a unit cell with $n$ species as well as $d$ dimensional translation vectors. 
Then, each unit cell is assigned by set of integers $j=(j_1,j_2,\cdots)$ as its position.  
After the $d$-dimensional Fourier transformation for $j$, the Hamiltonian
becomes a $n\times n$  matrix with wave numbers $k=(k_1,k_2,\cdots)$
which gives $n$ Bloch bands describing the bulk system.
Their wave functions 
would define bulk topological invariants.

On the other hand, when one introduces boundaries to this model, one usually 
calculates the edge states and bulk states together.
Let us consider $d-1$ dimensional boundaries perpendicular to 1-direction, $(1,0,\cdots,0)$, e.g., 
at $j_1=1$ and $N$.
Fourier-transforming all $j$ except for $j_1$, 
one obtains a one-dimensional Hamiltonian with $N$ unit cells along 1-direction specified by $j_1$.
The resultant Hamiltonian is a $nN\times nN$ matrix, which  
includes not only the edge states but also the bulk states. 
In order to obtain the well-defined edge states, one should choose $N$ as $N\gg n$, so that 
among $nN$ states of the Hamiltonian, the number of edge states is only of order $n$, and
most of them are bulk states.
Although edge states and bulk states are coupled together for finite $N$ system,
it is very desirable to separate the edge states from the bulk states
and to derive an effective theory of the edge states only.

Contrary to such tight-binding Hamiltonians,
the continuum Dirac models allow us to calculate edge states analytically without considering bulk states
\cite{witten16,Hashimoto:2016aa,Hashimoto:2017aa,Asaga:2020aa}. 
In particular, as pointed out by Witten \cite{witten16} as well as Hashimoto {\it et al}.  
\cite{Hashimoto:2016aa,Hashimoto:2017aa},
boundary conditions have intimate relationship with the Hermiticity of continuum Hamiltonians.
Along this line,  higher-order topological insulators based on continuum models have been 
developed, introducing the idea of ``edge of edge" states \cite{Hashimoto:2017aa}.
In Ref. \cite{Asaga:2020aa}, we investigated the edge states of the continuum Dirac model
which is derived as a continuum limit of the tight-binding model defined on the square lattice.
Then, it was pointed out that the boundary conditions of the lattice model also serve as those of 
the continuum model. This suggests an intimate relationship between the continuum models and tight-binding models.

The role of differential operators in continuum models 
is played by difference (or shift) operators in lattice models.
It is then expected that  Hamiltonians expressed by such difference (or shift) operators on lattices
are not necessarily Hermitian, as in the case of the continuum Dirac Hamiltonians, if systems have boundaries.
Then, even for lattice models, 
the Hermiticity of Hamiltonians would enable us to calculate edge states without considering the bulk states.
This implies that the Hermiticity gives us edge state Hamiltonians (ESHs) describing edge states only.

In this paper, 
we derive first-quantized Hamiltonian operators  on lattices including shift operators.
When we solve the Schr\"odinger equations using the Hamiltonian operators thus defined,
it turns out that Hamiltonians are not necessarily Hermitian when systems have boundaries.
We show that like the continuum Dirac theories, 
the Hermiticity of Hamiltonians would basically determine edge states.
To be concrete, when we can choose boundary wave functions that satisfy the Hermiticity of Hamiltonians,
we can develop a Bloch-type theory for edge states based on the boundary wave functions, which enables us to extract ESHs
separated from bulk states.
We show that this can be carried out for models with nearest-neighbor (NN) hoppings only. 
The ESHs thus obtained are defined for a single end on a half line or a half-plane.
Therefore, for systems of cylindrical geometry with two ends, the edge states at the left end and right ends
can be described separately by two distinct ESHs.
Even if models include next-nearest-neighbor (NNN) hoppings, ESHs can be formally defined.
However,  in these models, simple Bloch-type wave functions
ensuring the Hermiticity of Hamiltonians are not enough: Their linear combinations are needed to 
ensure the boundary conditions,
implying that for an edge state to exist,
the ESH must allow degenerate two or more Bloch-type wave functions.

There are several recent papers which develop theory for edge states \cite{Dwivedi:2016aa,Duncan:2018aa,Alase:2017aa,Cobanera:2018aa,Kunst:2017aa,Kunst:2018aa,Kunst:2019aa,Kunst:2019ab,Pletyukhov:2020aa,Pletyukhov:2020ab}.
In particular, Dwivedi and Chua generalized the transfer matrix method 
so that it can be applied to generic noninteracting tight-binding models \cite{Dwivedi:2016aa},
in which many examples treated in the present paper have been  studied.
Kunst {\it et. al.} \cite{Kunst:2017aa,Kunst:2018aa,Kunst:2019aa,Kunst:2019ab}
proposed systematic approach toward exact solutions of the edge states including 
higher-order edge states defined on semi-infinite lattice spaces.
Pletyukhov {\it et. al.} \cite{Pletyukhov:2020aa,Pletyukhov:2020ab} 
developed topological arguments based on exact edge states of a variant of 
the Hofstadter model solved also on semi-infinite lattice spaces.
The method proposed in this paper is a simple and unified theory based on the {\it Hermiticity} of  
{\it first-quantized Hamiltonian operators}. In our approach, all information is
included in the first-quantized Hamiltonian operators in any dimensions, 
and moreover, their relationship with continuum theories such as Dirac fermions is clear.

This paper is organized as follows.
In Sec. \ref{s:1D}, we formulate our method using a generalized version of the Su-Schrieffer-Heeger (SSH) model 
\cin{\cite{Su:1979aa,Duncan:2018aa,Alase:2017aa,Kunst:2019aa}}
as an example.
Almost all concepts can be developed by this simple one-dimensional (1D) model, 
including the Hermiticity of Hamiltonian,  its relationship with the boundary condition, 
a Bloch-type edge state \cite{Kunst:2019aa}, 
and classes of models categorized according to how the present formulation  can be applied, etc.

We then proceed to two-dimensional (2D) models 
\cin{\cite{Dwivedi:2016aa,Duncan:2018aa,Alase:2017aa,Cobanera:2018aa,Kunst:2017aa,Kunst:2018aa,Kunst:2019aa,Kunst:2019ab,Pletyukhov:2020aa,Pletyukhov:2020ab}}.
We first discuss models with NN hoppings only.
In Sec. \ref{s:hofstadter}, we apply our method to the Hofstadter model 
\cin{\cite{Hatsugai:1993fk,Hatsugai:1993aa,Dwivedi:2016aa,Duncan:2018aa,Hofstadter:1976aa}}
and give explicit ESHs for the left as well as the right ends \cite{Pletyukhov:2020aa,Pletyukhov:2020ab}. 
Sec. \ref{s:graphene} is devoted to edge states of the graphene 
in the absence/presence of a magnetic field 
\cin{\cite{Fujita:1996aa,Hatsugai:2006aa,Cobanera:2018aa}}.
As is well-known, the graphene allows various edge states depending on the shapes of edges.
We present the ESHs describing zigzag and bearded edges. 
For an armchair edge in the presence of a magnetic field, it is too difficult to obtain the edge state by use of our method,
since this is a NNN model.
Instead,  we present our attempt to obtain these edge states, by introducing some defects into the models.

Next, we apply out method to NNN models.
In Sec. \ref{s:wd}, we consider the Wilson-Dirac model \cin{\cite{Dwivedi:2016aa,Cobanera:2018aa}}, 
as an example of simpler models including NNN hoppings.
In Sec. \ref{s:haldane}, we consider the Haldane model \cite{Haldane:1988aa}
as a most difficult case for our formulation.
From a practical point of view, our method is not useful for this class of models.
Nevertheless, we would like to point out that the Hermiticity of Hamiltonians basically determines edge states 
even in this class.

Finally, we switch to 2D HOTI models 
\cin{\cite{Kunst:2018aa,Kunst:2019aa}}. 
In Sec. \ref{s:hoti}, we study corner states as well as edge states for typical HOTI models defined on the square lattice 
and on the breathing kagome lattice
\cite{Benalcazar:2017aa,Benalcazar:2017ab,Liu:2017aa,Ezawa:2018aa}.
Although edge states are governed by the same SSH Hamiltonian, different Hermiticity conditions lead to different edge states.
In Sec. \ref{s:con}, we give summary  and discussion.

\section{Hamiltonian operators for lattice models}\label{s:1D}

For continuum models with boundaries, the Hermiticity of their Hamiltonians is nontrivial, since momentum operators
yield boundary terms associated with the integration by parts \cite{witten16,Hashimoto:2017aa}. 
In tight-binding models defined on lattices, the differential operators are replaced by difference (or shift) operators,
implying that the summation by parts plays a similar role.
Therefore, even for lattice models, the Hermiticity of Hamiltonians is expected also nontrivial 
when they are in the lattice space representation.

In \ref{s:summation}, we introduce the shift operators and related  summation by parts, 
and in Sec. \ref{s:ssh}, we demonstrate, using the generalized SSH model,  
that with a boundary the Hamiltonian becomes Hermitian 
only if a suitable condition is imposed.

\subsection{Summation by parts}\label{s:summation}

In order to study conventional tight-binding models in condensed matter physics, it may be convenient to
introduce the shift operators, instead of the difference operators.
Let $f_j$ be a function of discrete sequence of integers denoted by $j$. 
Then, the forward and backward shift operators are defined by
\begin{alignat}1
\delta f_j
=f_{j+1},
\quad\delta^* f_j
=f_{j-1}.
\end{alignat}
One can show the following summation by parts
\begin{alignat}1
&\sum_{j=-\infty}^\infty f_j\delta g_j=\sum_{j=-\infty}^\infty ({\delta^*}f_j) g_j, 
\nonumber\\
&\sum_{j=-\infty}^\infty ({\delta}f_j) g_j=\sum_{j=-\infty}^\infty f_j\delta^* g_j.
\end{alignat}
Thus, we have $\delta^\dagger =\delta^*$ for a bulk system, where bulk means 
a system defined over $j=-\infty$ to $+\infty$.
If the system has a boundary, summation by parts yields a boundary term, 
\begin{alignat}1
&\sum_{j=1}^\infty f_j\delta g_j=\sum_{j=2}^\infty ({\delta^*}f_j) g_j
=\sum_{j=1}^\infty (\delta^*f_j) g_j-f_{0}g_1, 
\nonumber\\
&\sum_{j=1}^\infty (\delta f_j)g_j=\sum_{j=2}^\infty f_j\delta^* g_j
=\sum_{j=1}^\infty f_j\delta^* g_j-f_1g_{0}.
\label{SumByPar}
\end{alignat}
These properties can be regarded as a lattice version of the integration by parts.

\subsection{Typical example: 1D SSH model}\label{s:ssh}
\cin{This model may be one of the simplest but most typical examples of models with edge states
to consider first 
\cite{Dwivedi:2016aa,Duncan:2018aa,Cobanera:2018aa,Kunst:2018aa,Pletyukhov:2020aa}.}

\subsubsection{First-quantized Hamiltonian operator}
\begin{figure}[htb]
\begin{center}
\begin{tabular}{c}
\includegraphics[width=.7\linewidth]{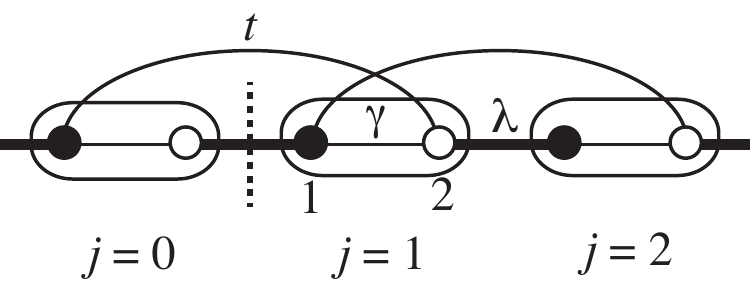}
\end{tabular}
\caption{
Generalized SSH model including $t$-hopping.  The condition \cin{$(\psi_{j=0})_2=0$}
cuts the chain between $j=0$ and $j=1$.
}
\label{f:ssh}
\end{center}
\end{figure}

Let $c_j\equiv(c_{1j},c_{2j})$ be the annihilation operators of the fermions on the $j$-th unit cell,
as shown in Fig. \ref{f:ssh}.
Then, the second-quantized Hamiltonian is given by
\begin{alignat}1
H=&
\sum_{j}\Big[
c_j^\dagger \left(\begin{array}{cc}&\gamma\\\gamma&\end{array}\right)c_j
+c_j^\dagger\left(\begin{array}{cc}&t\\\lambda&\end{array}\right)c_{j+1}
\nonumber\\&\quad
+c_{j+1}^\dagger\left(\begin{array}{cc}&\lambda\\t&\end{array}\right)c_{j}
\Big]
\nonumber\\
=&
\sum_{j} c_j^\dagger
\left(\begin{array}{cc}&\gamma+t\delta+\lambda\overleftarrow\delta\\
\gamma+\lambda\delta+t\overleftarrow\delta&\end{array}\right)c_j
\nonumber\\
\equiv&
\sum_{j} c_j^\dagger
\overleftrightarrow{{\cal H}}
c_j,
\label{SshHamDef}
\end{alignat}
where $\overleftarrow\delta$ acts on the left. The operator $\overleftrightarrow{\cal H}$ defined on the lattice
will be  referred to as  the (first-quantized) Hamiltonian (operator).

For the bulk system, 
the Hamiltonian operator (\ref{SshHamDef}) can be rewritten as
\begin{alignat}1
&H=\sum_{j=-\infty}^\infty c_j^\dagger\overleftrightarrow{\cal H}c_j
=\sum_{j=-\infty}^\infty c_j^\dagger \hat{\cal H}c_j, 
\end{alignat}
where $\hat{\cal H}$ defined by
\begin{alignat}1
\hat{\cal H}&=\left(\begin{array}{cc}&\gamma+\lambda\delta^*+t\delta\\
\gamma+\lambda\delta+t\delta^*&\end{array}\right)
\nonumber\\
&\equiv {\cal K}\delta^*+{\cal K}^\dagger \delta+{\cal V},
\label{HamSSH_Bul}
\end{alignat}
operates on the right only, and hence, it is appropriate for the Schr\"odinger eigenvalue equation. 
The symbol hat on ${\cal H}$ emphasizes that ${\cal H}$ is not a simple matrix but a matrix-valued operator.
$\hat{\cal H}$ will be also referred to as the (first-quantized) Hamiltonian (operator).

On the other hand, for the system with a boundary, there appears a boundary term.
Let us introduce a boundary between $j=0$ and $j=1$ as in Fig. \ref{f:ssh}, and consider the system on 
the semi-infinite line, $j_1\ge1$.
Then, 
\begin{alignat}1
H&=\sum_{j=1}^\infty c_j^\dagger\overleftrightarrow{\cal H}c_j
=\sum_{j=1}^\infty c_j^\dagger\hat{\cal H}c_j 
-c_1^\dagger\left(\begin{array}{cc}&\lambda\\t&\end{array}\right)c_0
\nonumber\\
&\equiv\sum_{j=1}^\infty c_j^\dagger\hat{\cal H}c_j 
-c_1{\cal K}c_0,
\label{HamSSH_Bou}
\end{alignat}
where the operator $c_0$ out of the boundary has been taken into account, and 
the subtracted term in the right-hand-side is due to the summation by parts in Eq. (\ref{SumByPar}).
Thus, for a system with a boundary, $\overleftrightarrow{\cal H}$ and $\hat {\cal H}$ differ in
the boundary term.
Generically, the boundary term can be written by non-Hermitian matrix $\cal K$
associated with the operator $\overleftarrow\delta$ in $\overleftrightarrow{\cal H}$ or $\delta^*$
in $\hat{\cal H}$.

In what follows, we will discuss various aspect of the  Schr\"odinger equation for $\hat{\cal H}$,
\begin{alignat}1
\hat{\cal H}\psi_{jn}=\varepsilon_n\psi_{jn}, 
\label{SshEigEqu}
\end{alignat}
where $\hat{\cal H}$ is defined by Eq. (\ref{HamSSH_Bul}), $\delta$ and $\delta^*$ operate on $j$ of $\psi_{jn}$, and
$n$ denotes the energy quantum number.

\subsubsection{Hermiticity of $\hat{\cal H}$}

First of all, let us discuss the Hermiticity of this operator.
For any wave functions $\psi_j$ and $\phi_j$ including those at $j=0$, $\psi_0$ and $\phi_0$,  we have
\begin{alignat}1
\langle\phi| \hat{\cal H}\psi\rangle
&\equiv\sum_{j=1}^\infty\phi^\dagger_j \hat{\cal H}\psi_j
=\sum_{j=1}^\infty\phi^\dagger_j ({\cal K}\delta^*+{\cal  K}^\dagger\delta+{\cal V})\psi_j
\nonumber\\
&
=\sum_{j=1}^\infty\phi^\dagger_j ({\cal K}\overleftarrow\delta+{\cal  K}^\dagger\overleftarrow\delta^*+{\cal V})\psi_j
-\phi_0^\dagger {\cal K}^\dagger\psi_{1}+\phi_1^\dagger {\cal K}\psi_{0},
\nonumber\\
&=\langle
\hat{\cal H}\phi|\psi\rangle
-\phi_0^\dagger {\cal K}^\dagger\psi_{1}+\phi_1^\dagger {\cal K}\psi_{0}.
\end{alignat}
Therefore, only by imposing the condition 
\begin{alignat}1
\phi_0^\dagger {\cal K}^\dagger\psi_{1}=0,\quad
\phi_1^\dagger {\cal K}\psi_{0}=0, \quad {\cal K}=\left(\begin{array}{cc}&\lambda\\t&\end{array}\right),
\label{HerCon}
\end{alignat}
will the Hamiltonian become Hermitian.

Now, assume that we have a set of eigenfunctions of $\hat{\cal H}$, $\psi_{jn}$, with the Hermiticity 
(\ref{HerCon}) imposed.
\begin{alignat}1
\psi^\dagger_{m1}{\cal K}\psi_{0n}=0,
\label{HerConWav}
\end{alignat}
for all $m,n$. 
Then, they form a complete orthonormal set of functions $\psi_j$ on the semi-infinite line $j\geq1$. 
Hence, we have
\begin{alignat}1
\sum_{j=1}^\infty\psi^\dagger_{mj}\psi_{jn}=\delta_{mn},\quad
\sum_n\psi_{in}\psi^\dagger_{nj}=\delta_{ij}.
\end{alignat}
Define new operators $d_n$ and $d_n^\dagger$ such that
\begin{alignat}1
c_j\equiv \sum_n\psi_{jn}d_n,\quad c_j^\dagger\equiv\sum_nd_n^\dagger \psi^\dagger_{nj},
\end{alignat}
and substitute these into the second quantized Hamiltonian (\ref{HamSSH_Bou}),
\begin{alignat}1
H&=\sum_{j=1}^\infty 
\sum_{m,n}d_m^\dagger \psi^\dagger_{mj}\hat{\cal H}\psi_{jn}d_n
-\sum_{m,n}d_m^\dagger \psi^\dagger_{m1}{\cal K}
\psi_{0n}d_n
\nonumber\\
&=\sum_n\varepsilon_n d_n^\dagger d_n.
\end{alignat}
Note that the boundary term in the Hamiltonian vanishes due to the Hermiticity Eq. (\ref{HerConWav}) imposed 
on the wave functions, and
we have a desired second-quantized  Hamiltonian.  

\subsubsection{Boundary conditions}

Let us now discuss the eigenvalue equation (\ref{SshEigEqu}).
For the bulk system, the Fourier transformation can be readily carried out just by replacing 
$\delta\rightarrow e^{ik}$ and $\delta^*\rightarrow e^{-ik}$.
Then, $\hat{\cal H}$ becomes a simple $2\times2$ matrix whose eigenstates are nothing but the Bloch states.

Let us next consider the system with a boundary.
Assume that the system is defined on the semi-infinite line, $j\ge1$.
Then, we need to specify the boundary condition at $j=1$.
To this end, let us write  the Schr\"odinger equation (\ref{SshEigEqu}) explicitly,
\begin{alignat}1
\hat{\cal H}\psi_j={\cal K}\psi_{j-1}+{\cal V}\psi_{j}+{\cal K}^\dagger\psi_{j+1}=\varepsilon \psi_j,\quad (j=1,2,\cdots),
\end{alignat}
where the energy quantum number $n$ has been suppressed.
Here, the boundary condition associated with $j=1$ above reads
\begin{alignat}1
{\cal K}\psi_0=0,
\label{BouCon}
\end{alignat}
where we have supplementarily included $\psi_0$ which is out of the boundary.
If one does not take $\psi_0$ into account or set $\psi_0=0$ from the beginning, the boundary condition 
at $j=1$ is given by
\begin{alignat}1
{\cal V}\psi_{1}+{\cal K}^\dagger\psi_{2}=\varepsilon\psi_{1},
\label{BouConNo}
\end{alignat}
with a suitable initial value $\psi_1$.

In this paper, we adopt Eq. (\ref{BouCon}) as a boundary condition. 
This may be much simpler than (\ref{BouConNo}), if we can find $\psi_0$. 
Based on such a $\psi_0$, we can develop Bloch-type techniques,  as we show below.
Even if such a $\psi_0$ cannot exist, we can take the following route: 
Without considering  the boundary condition (\ref{BouCon}),
we solve the Schr\"odinger equation (\ref{SshEigEqu}) only imposing the Hermiticity  (\ref{HerCon}).
Then, using the wave functions thus obtained, we construct
suitable eigenstates satisfying the  condition (\ref{BouCon}) by
taking their linear combination.

In this sense, we regard the Hermiticity of Hamiltonians (\ref{HerConWav}) 
as a guiding principle to choose $\psi_0$. In what follows,  
we develop a Bloch-type theory for the edge states based on $\psi_0$ satisfying the Hermiticity (\ref{HerCon}).
In the case of NN hopping models, such a $\psi_0$ naturally satisfies the boundary condition (\ref{BouCon}), 
whereas for other models including NNN hoppings, we have to take linear combinations of degenerate 
Bloch-type wave functions to construct the wave functions satisfying the boundary condition
(\ref{BouCon}), as will be argued below.

\subsubsection{Edge states}\label{s:ssh_edge}

For the system defined on the semi-infinite line $j_1\ge1$,
let us solve the eigenvalue equation (\ref{SshEigEqu}) assuming wave functions decaying exponentially.
To this end, we introduce a Bloch-type wave function with a complex wave number $K$,
\begin{alignat}1
\psi_{j}=\psi_{0} e^{iKj}, \quad K=k+i\kappa,
\label{Blo}
\end{alignat}
where we have utilized the supplemental wave function $\psi_{0}$ as a reference state, and by definition, 
$\kappa>0$ is required. Since this model allows at the most one edge state, the quantum number $n$ has been  suppressed.
Then, substituting Eq. (\ref{Blo}) into Eq. (\ref{SshEigEqu}), the eigenvalue equation becomes
\begin{alignat}1
&\left(\begin{array}{cc}&\gamma+\lambda e^{-iK}+te^{iK}\\
\gamma+\lambda e^{iK}+te^{-iK}&\end{array}\right)\psi_{0}
=\varepsilon_n\psi_{0}.
\label{EigValEquSSH}
\end{alignat}
Since $\psi_{1}=\psi_{0}e^{iK}$, the Hermiticity condition (\ref{HerConWav})
can be written solely by $\psi_0$,
\begin{alignat}1
\psi_{0}^\dagger {\cal K}\psi_{0}=0,
\label{HerConBlo}
\end{alignat}
This equation requires that
\begin{alignat}1
\psi_0
\propto\mbox{ either }\left(\begin{array}{c}1\\0\end{array}\right)\equiv \psi_{\uparrow}
 \mbox{ or }\left(\begin{array}{c}0\\1\end{array}\right)\equiv \psi_{\downarrow}.
\label{BouWav}
\end{alignat}
The above vectors can also be interpreted as follows: Since $\{{\cal K}, \sigma^3\}=0$, the eigenstates of $\sigma^3$,
$\psi_{\uparrow,\downarrow}$,
satisfy the Hermiticity condition, as discussed in Refs. \cite{witten16,Hashimoto:2017aa}.

\noindent
$\bullet$ NN model ($t=0$)

Let us first consider the case with nearest neighbor hopping only, setting $t=0$.
Here, consider the boundary in Fig. \ref{f:ssh}. The system defined on the infinite line can be separated at 
the dotted-line if the second component of the wave function at $j=0$ is set 0.
Thus, from the point of view of such a lattice termination, the appropriate wave function at $j=0$ is 
the former $\psi_\uparrow$
in Eq. (\ref{BouWav}).  Namely, 
among the functions satisfying the Hermiticity (\ref{HerConBlo}), we have to choose the reference function 
$\psi_0=\psi_\uparrow$. 
What is important here is that
\begin{alignat}1
{\cal K}\psi_\uparrow=0,
\label{EigSSH}
\end{alignat}
holds. This guarantees that the boundary condition (\ref{BouCon}) is automatically satisfied.
For NN models, the Hermiticity, the lattice termination, and the boundary condition are consistently 
satisfied by a suitable reference state $\psi_0$, as will be seen in various examples.
Substituting $\psi_0=\psi_\uparrow$ into the eigenvalue equation (\ref{EigValEquSSH}), we have
\begin{alignat}1
\varepsilon=0, \quad \gamma+\lambda e^{iK}=0.
\label{SSHt0}
\end{alignat}
The latter equation gives
\begin{alignat}1
k=0,\pi, \quad e^{-\kappa}=\mp\frac{\gamma}{\lambda}.
\end{alignat} 
Thus, an edge state exists when $0<e^{-\kappa}<1$. Explicitly, it is given by
\begin{alignat}1
\varepsilon=0,\quad \psi_j=\left(\begin{array}{c}1\\0\end{array}\right)(-\mbox{sgn } \lambda\gamma)^j
\left|\frac{\gamma}{\lambda}\right|^j ,
\end{alignat}
valid for $|\gamma/\lambda|<1$.

\noindent
$\bullet$ NNN model ($t\ne0$).

In this case, ${\cal K}\psi_{\uparrow,\downarrow}\ne0$, implying that the Hermiticity does not ensure the 
boundary condition. 
Even in this case, 
if Eq. (\ref{EigValEquSSH}) has degenerate two solutions $K_1\neq K_2$, we have a wave function  of the form
\begin{alignat}1
\tilde\psi_j=\psi_{0}(e^{iK_1j}-e^{iK_2j}),
\end{alignat}
which becomes $\tilde\psi_0=0$, and therefore, the boundary condition (\ref{BouCon}) holds.

\begin{figure}[htb]
\begin{center}
\begin{tabular}{c}
\includegraphics[width=.5\linewidth]{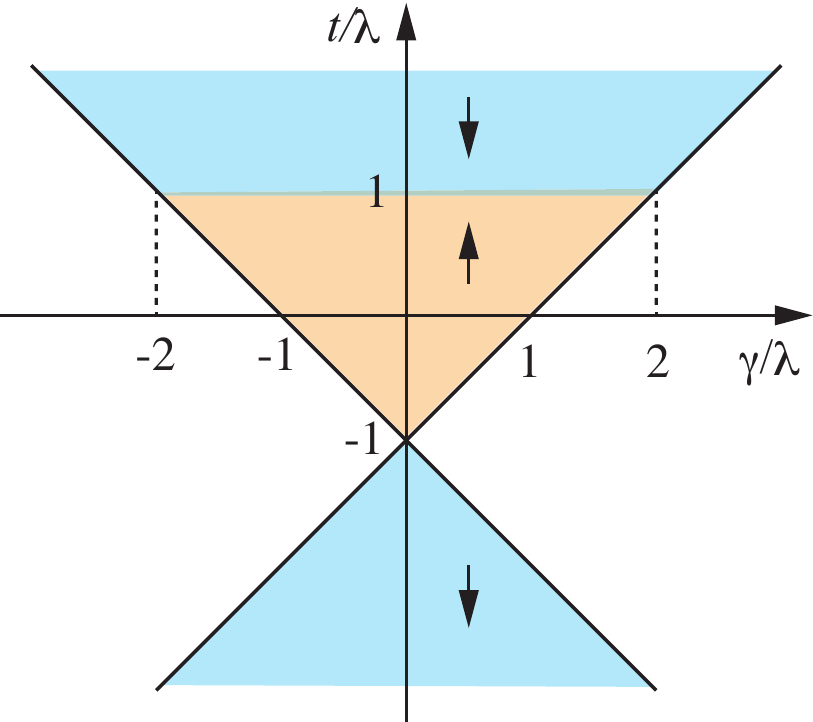}
\end{tabular}
\caption{
Phase diagram of the generalized SSH model.  Colored regions denote the topological phase with an edge state.
The arrows means the reference wave functions $\psi_{\uparrow,\downarrow}$.
}
\label{f:ssh_pd}
\end{center}
\end{figure}

Let us choose $\psi_0=\psi_\uparrow$. Then, similarly to Eq. (\ref{SSHt0}), we have
\begin{alignat}1
\varepsilon=0, \quad \gamma+\lambda e^{iK}+te^{-iK}=0.
\end{alignat}
If we choose $\psi_0=\psi_\downarrow$, we have
\begin{alignat}1
\varepsilon=0, \quad \gamma+\lambda e^{-iK}+te^{iK}=0.
\end{alignat}
When these equations have two solutions $|e^{iK}|=e^{-\kappa}<1$, the model has an edge state.
We show in Fig. \ref{f:ssh_pd} the phase diagram of the generalized SSH model.

\subsubsection{Bulk states}
\cin{The bulk states for finite systems 
have been investigated in Refs. \cite{Duncan:2018aa,Kunst:2019ab}}.
For the bulk state, we know that 
the conditions associated with the boundaries are not very important, although they should be satisfied.
Therefore, we firstly solve the eigenvalue equation for the Bloch states with a real wave vector $k$,
\begin{alignat}1
\psi_j=\psi_k e^{ikj},
\end{alignat}
where at the moment, we do not consider the Hermiticity  and the boundary condition.
Then, from the Schr\"odinger equation, we have 
\begin{alignat}1
\varepsilon_\pm(k)=\pm\sqrt{\gamma^2+\lambda^2+2\gamma\lambda\cos k},
\end{alignat}
Note that $\varepsilon_\pm(k)=\varepsilon_\pm(-k)$.
Therefore, for each state with $\varepsilon_\pm(k)$ we have left-going and right-going waves, and 
their superposition allows the 
wave function satisfying the boundary condition (\ref{BouCon}):
\begin{alignat}1
\psi_{j,n}=\alpha_{n}\psi_{k,n} e^{ikj}+\beta_{n}\psi_{-k,n}e^{-ikj},
\end{alignat}
where $\alpha_{n}\psi_{k,n}+\beta_{n}\psi_{-k,n}\propto \psi_\uparrow \mbox{ or }\psi_{\downarrow}$,
depending on the phases in Fig. \ref{f:ssh_pd}.
This is the bulk state satisfying the Hermiticity.
In what follows, we will not consider the bulk states.

\subsection{Classes of models}\label{s:class}

In the following sections, we will apply the method developed above to various 2D models. 
For this purpose, it is useful to classify models into following three types.
 
A:  Reference state $\psi_0$ which satisfies ${\cal K}\psi_0=0$ can be chosen. 
Models with a boundary whose Hamiltonian can be represented only by NN hoppings belong to this class. 
Here, by NN hopping, we mean that if one of the bonds connected by finite hoppings between unit cells is cut, 
the system is divided into two pieces. The generalized SSH model with $t=0$ is one of  the examples, and various other models such as
the Hofstadter model in Sec. \ref{s:hofstadter}, the graphene with the zigzag or bearded edges
in the absence/presence of a magnetic field in Sec. \ref{s:graphene},
HOTIs in Sec. \ref{s:hoti} belong to A.
In this class, it is very easy to extract Bloch-type ESHs free from the bulk contributions.

B: Models including NNN hoppings, but there exist a matrix $\cal A$ with ${\cal A}^2=1$ which 
anti-commutes with  Hermiticity matrix $\cal K$, $\{{\cal A},{\cal K}\}=0$.
The generalized SSH model with $t\ne 0$ is one of the typical examples, 
and the Wilson-Dirac model in Sec. \ref{s:wd} is another example in this paper.
In this class,  we choose the reference state $\psi_0$ as either eigenstates of ${\cal A}$, 
${\cal A}\psi_0=\pm\psi_0$, which guarantees the Hermiticity of the Hamlitonians.
Then, we can develop a Bloch-type theory based on $\psi_0$.
However, we need two independent Bloch-type states to yield edge states satisfying the boundary condition.

C:  If one cannot find any matrix ${\cal A}$ that anti-commute with $\cal K$, one has to solve the 
Bloch-type eigenvalue equation based on the reference state $\psi_0$ which satisfies the Hermiticity (\ref{HerCon}).
Like the class B, two independent solutions are needed.
As an example of this class, we will discuss the Haldane model in Sec. \ref{s:haldane}. 
From a practical point of view, our method is not useful for these models. Nevertheless, the idea that the edge states are determined by the Hermiticity seems to be important.

\section{Application to 2D models: Hofstadter model}

In 2D systems, second-quantized Hamiltonians are generically given by
\begin{alignat}1
H&=\sum_jc_j^\dagger \overleftrightarrow{\cal H}(\delta_1,\overleftarrow\delta_1,\delta_2,\overleftarrow\delta_2)c_j
\nonumber\\
&=\sum_jc_j^\dagger \hat{\cal H}(\delta_1,\delta_1^*,\delta_2,\delta_2^*)c_j,
\label{2DSecHam}
\end{alignat} 
where $j=(j_1,j_2)$ denotes 2D lattice points, and $\delta_\mu$ and $\delta_\mu^*$ ($\mu=1,2$) are,
respectively,  the forward and backward shift operators to the $\mu$-direction,
$\delta_\mu f_j=f_{j+\hat\mu}$  and $\delta_\mu^* f_j=f_{j-\hat\mu}$ 
with $\hat\mu$ being the unit vector toward the $\mu$-direction.

Without loss of generality, we consider the system defined on the half-plane $j_1\geq1$, and discuss the edge states
along the boundary $j_1=1$.
Then, we can make the Fourier transformation in the $2$-direction, 
\begin{alignat}1
H&=\sum_{k_2}H_{k_2},
\nonumber\\
H_{k_2}&=
\sum_{j_1=1}^\infty c_{j_1k_2}^\dagger \hat{\cal H}(\delta_1,\delta_1^*,e^{ik_2},e^{-ik_2})c_{j_1k_2}
\nonumber\\&
\equiv\sum_{j=1}^\infty c_{j}^\dagger \hat{\cal H}(\delta)c_{j}.
\label{2DHam}
\end{alignat} 
The last equality is the abbreviation for notational simplicity: 
The operators $\delta\equiv\delta_1$ and $\delta^*\equiv\delta_1^*$  act on $j\equiv j_1$,
and $k_2$ dependence is suppressed. 
Note that $H_{k_2}$ is nothing but a 1D Hamiltonian, 
so that we can obtain the edge states using the technique in Sec. \ref{s:ssh}.

In the following sections, we will show that the method in Sec. \ref{s:ssh} can apply various 2D models.
In particular, for NN models (class A in Sec. \ref{s:class})
ESHs can be explicitly obtained, 
which is exemplified in detail by the Hofstadter model in the following Sec. \ref{s:hofstadter}.

\begin{figure}[htb]
\begin{center}
\begin{tabular}{c}
\includegraphics[width=.7\linewidth]{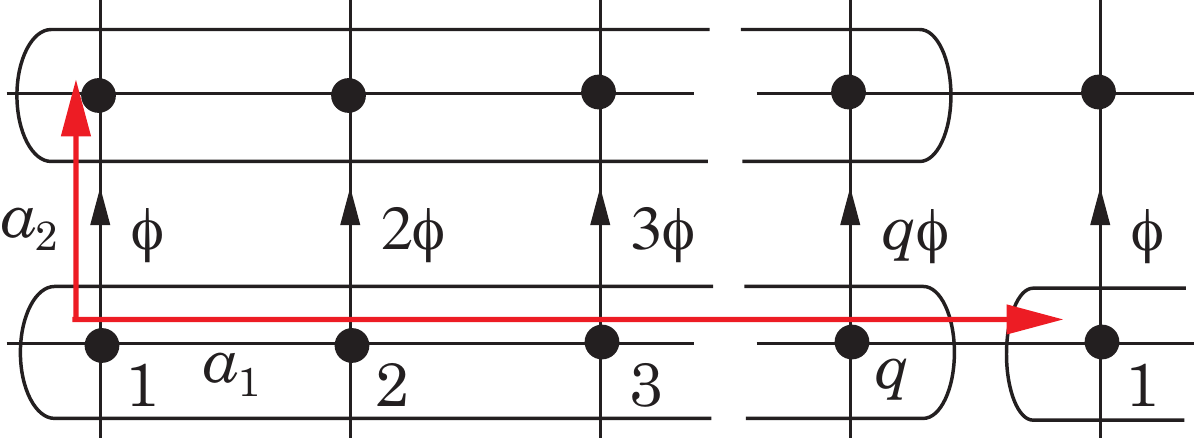}
\end{tabular}
\caption{
The Hofstadter model on the square lattice with a uniform magnetic flux $\phi=2\pi p/q$ per plaquette.
The arrow with $n\phi$ denotes the link with the phase $e^{in\phi}$.
Large boxes denote a magnetic unit cell.
}
\label{f:hof}
\end{center}
\end{figure}

\subsection{Hofstadter model}\label{s:hofstadter}


As a typical example of 2D models, 
let us consider the Hofstadter model  
\cin{\cite{Hofstadter:1976aa,Hatsugai:1993aa,Dwivedi:2016aa,Duncan:2018aa,Cobanera:2018aa,Pletyukhov:2020aa}.}
Despite its simplicity, the model has revealed various aspects and properties of the topological phase,
including TKNN integers and associated Diophantine equation \cite{Thouless:1982uq,kohmoto:85},
the bulk-edge correspondence \cite{Hatsugai:1993fk,Hatsugai:1993aa}, 
the Streda formula \cite{Streda:1982aa,Koshino:2006aa}, etc.
In particular, using the transfer matrix method, Hatsugai showed that edge states give winding numbers
around a complex energy surface \cite{Hatsugai:1993fk,Hatsugai:1993aa}.
On the other hand, since this model is a NN model, 
the present formulation enables us to treat the edge states at the left and right ends separately
and to  extract the ESH at each end, as will be shown momentarily.
Based on these, we can discuss the bulk-edge correspondence,
which will be published elsewhere \cite{inpreparation}.

We consider the square lattice with $\phi/(2\pi)= p/q$ flux per plaquette, where 
$p$ and $q$ are coprime integers.
For the Landau gauge illustrated in Fig. \ref{f:hof},
the first-quantized Hamiltonian operator $\hat{\cal H}(\delta_1,\delta_2)$ in Eq. (\ref{2DSecHam}) is given by
\begin{widetext}
\begin{alignat}1
\hat{\cal H}(\delta_1,\delta_2)=
t\left(
\begin{array}{cccccc}
e^{i\phi}\delta_2^*+e^{-i\phi}\delta_2&1&&&&\delta_1^*\\
1&e^{2i\phi}\delta_2^*+e^{-2i\phi}\delta_2&1&&&\\
&1&\ddots&&&\\
&&&&e^{i(q-1)\phi}\delta_2^*+e^{-i(q-1)\phi}\delta_2&1\\
\delta_1&&&&1&\delta_2^*+\delta_2
\end{array}
\right).
\end{alignat}
After the Fourier transformation with respect to the 2-direction as in Eq. (\ref{2DHam}),  the Hamiltonian operator for the 
1D chain toward the 1-direction is given by
\begin{alignat}1
\hat{\cal H}=
t\left(
\begin{array}{cccccc}
2\cos(k_2-\phi)&1&&&&\delta^*\\
1&2\cos(k_2-2\phi)&1&&&\\
&1&\ddots&&&\\
&&&&2\cos(k_2-(q-1)\phi)&1\\
\delta&&&&1&2\cos k_2
\end{array}
\right),
\label{HamOpeHof}
\end{alignat}
\end{widetext}
where the operators $\delta\equiv\delta_1$ and $\delta^*\equiv\delta_1^*$  act on $j\equiv j_1$,
as was already mentioned. 

\subsubsection{Edge states at the left end}

Paying attention to the backward operator $\delta^*$ of  the above Hamiltonian,  
we find the Hermiticity matrix $\cal K$,
\begin{alignat}1
\newfont{\bg}{cmr10 scaled\magstep4}
\newcommand{\bigzero}{\smash{\hbox{\bg 0}}}
{\cal K}=
t\left(
\begin{array}{ccccc}
&&&&1\\
&&&0&\\
&&\iddots&&\\
&&&&\\
0&&&&
\end{array}
\right).
\label{HofK}
\end{alignat}
The Hermiticity condition Eqs. (\ref{HerConWav}) or (\ref{HerConBlo}), i.e., $ \psi^\dagger_{m0}{\cal K}\psi_{0n}=0$, leads to
\begin{alignat}1
\psi_{0n}\propto\left(\begin{array}{c}\chi_{n}\\0\end{array}\right) 
\mbox{ or }\left(\begin{array}{c}0\\\chi_{n}\end{array}\right),
\label{HofWavFun}
\end{alignat}
where $\chi_{n}$ is a vector with $q-1$ components.
From the point of view of the lattice termination, we see that the former wave function matches the boundary 
condition for the system defined on the half-plane $j_1\geq1$.
To see this, it may be more convenient not to use the magnetic unit cell, but to label each site as $j$.
Then, the Hamiltonian can be expressed by
$H=t\sum_j(c_{j+1}^\dagger c_{j}+c_j^\dagger c_{j+1}+v_jc_j^\dagger c_j)$, 
where $c_{j=1},c_{j=2},\cdots$ corresponds to
$c_{1,j_1=1},c_{2,j_1=1},\cdots$ and 
$v_j=2\cos(k_2-\phi j)$.
Using such a new labeling of sites and assuming the wave function
$|\psi\rangle=\sum_jc_j^\dagger \psi_j|0\rangle$,
the eigenvalue equation, $H|\psi\rangle=\varepsilon|\psi\rangle$ 
is explicitly given by $t\psi_{j-1}+v_j\psi_j+t\psi_{j+1}=\varepsilon\psi_j$ $(j=1,2,\cdots)$.
Now, let us consider the system defined on the half-plane $j\ge1$. 
Since the above eigenvalue equation at $j=1$ becomes
$t\psi_0+v_1\psi_1+t\psi_2=\varepsilon\psi_1$,
it is natural to require $\psi_0=0$ as a boundary condition, where 
$\psi_{0}$ in the new notation corresponds to $\psi_{q,j_1=0}$.
In passing, we would like to mention that the eigenvalue equation is basically given by
recurrence relation of order 2. Therefore, it is natural to use a $2\times2$ transfer matrix, as 
was used in Refs. \cite{Hatsugai:1993fk,Hatsugai:1993aa}.

On the other hand, in the present formulation, we utilize the magnetic unit-cell representation, and 
develop a Bloch-type theory. 
Here, it should be noted here that wave functions in (\ref{HofWavFun}) satisfy 
the boundary condition (\ref{BouCon}),
\begin{alignat}1
{\cal K}\psi_{0n}=0,
\end{alignat}
implying that a single Bloch-type state $\psi_{jn}=\psi_{0n}e^{iK_nj}$ can be an eigenstate of the Hamiltonian,
where $K_{n}=k_n+i\kappa_n$ with $\kappa_n>0$.
Substituting this
into the eigenvalue equation, we have
\begin{widetext}
\begin{alignat}1
t\left(
\begin{array}{ccccc|c}
2\cos(k_2-\phi)&1&&&&e^{-iK_n}\\
1&2\cos(k_2-2\phi)&1&&&\\
&1&\ddots&&&\\
&&&&2\cos(k_2-(q-1)\phi)&1\\
\hline
e^{iK_n}&&&&1&2\cos k_2
\end{array}
\right)
\left(\begin{array}{c}\begin{array}{c}\\ \\ \chi_{n}\\ \\ \\ \end{array}\\\hline0\end{array}\right) 
=\varepsilon_{n}
\left(\begin{array}{c}\begin{array}{c}\\ \\ \chi_{n}\\ \\ \\ \end{array}\\\hline0\end{array}\right).
\label{HofEqu}
\end{alignat}
This equation tells that the upper diagonal $(q-1)\times(q-1)$ part of the above equation determines the eigenvalues $\varepsilon_n(k_2)$ 
and eigenstates of the edge states,
and the last equation yields the constraint 
$e^{iK_n}\chi_{1,n}+\chi_{q-1,n}=0$ which determines whether the edge states obtained above are localized 
near $j_1=1$. 
Namely, the edge states are eigenstates of the following ESH,
\begin{alignat}1
{\cal H}_{\rm e}=t\left(
\begin{array}{ccccc}
2\cos(k_2-\phi)&1&&&\\
1&2\cos(k_2-2\phi)&1&&\\
&1&\ddots&&\\
&&&&2\cos(k_2-(q-1)\phi)
\end{array}
\right),
\label{HofRedHam}
\end{alignat}
\end{widetext}
Thus, 
we have been able to define the ESH for the Hofstadter model, ${\cal H}_{\rm e}$, 
which determines the eigenvalues and eigenstates of the edge states only,
\begin{alignat}1
{\cal H}_{\rm e}\chi_{n}=\varepsilon_n\chi_{n} .
\label{EdgEigEquHof}
\end{alignat} 
As mentioned, although ${\cal H}_{\rm e}$ looks independent of $K_n$, the last equation of Eq. (\ref{HofEqu})
yields the following constraint ensuring the localization of the edge states, $\kappa_n>0$, 
\begin{alignat}1
|e^{iK_n}|=e^{-\kappa_n}=\left|\frac{\chi_{q-1,n}}{\chi_{1,n}}\right|<1.
\label{HofCon}
\end{alignat}
As we have suppressed the $k_2$-dependence in the above discussion, 
$\kappa_n(k_2)$ depends on $k_2$  through $\chi_n(k_2)$.
Therefore, 
although the energies $\varepsilon_n(k_2)$ and wave functions $\chi_n(k_2)$
in Eq. (\ref{EdgEigEquHof}) form $q-1$ continuous functions of $k_2$
in the Brillouin zone, they do not necessarily describe the edge states:
It is only those satisfying the condition Eq. (\ref{HofCon}) that  can be the edge states
localized near the $j_1=1$ end.

\begin{figure}[htb]
\begin{center}
\begin{tabular}{ccc}
\includegraphics[width=.5\linewidth]{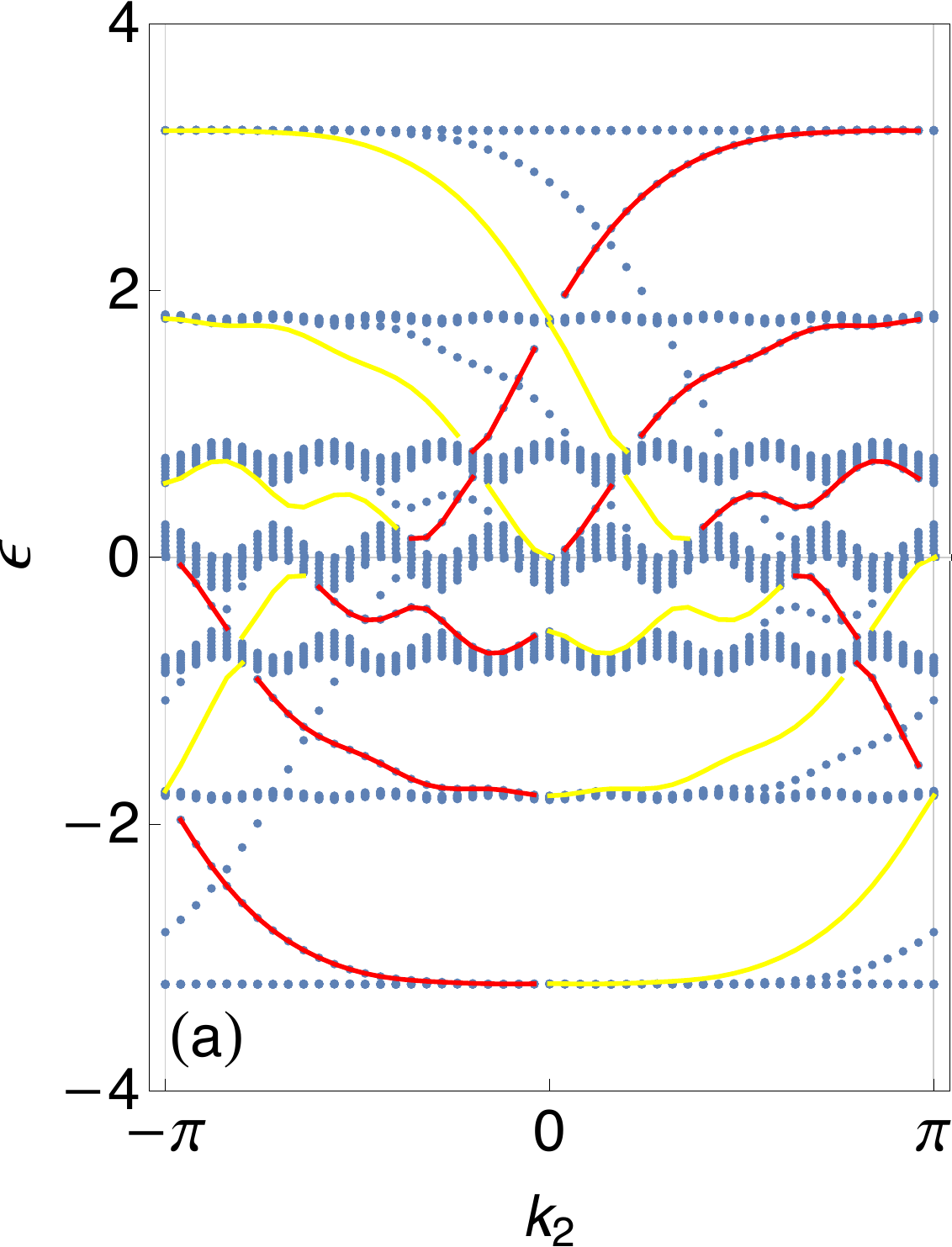}
&
\includegraphics[width=.5\linewidth]{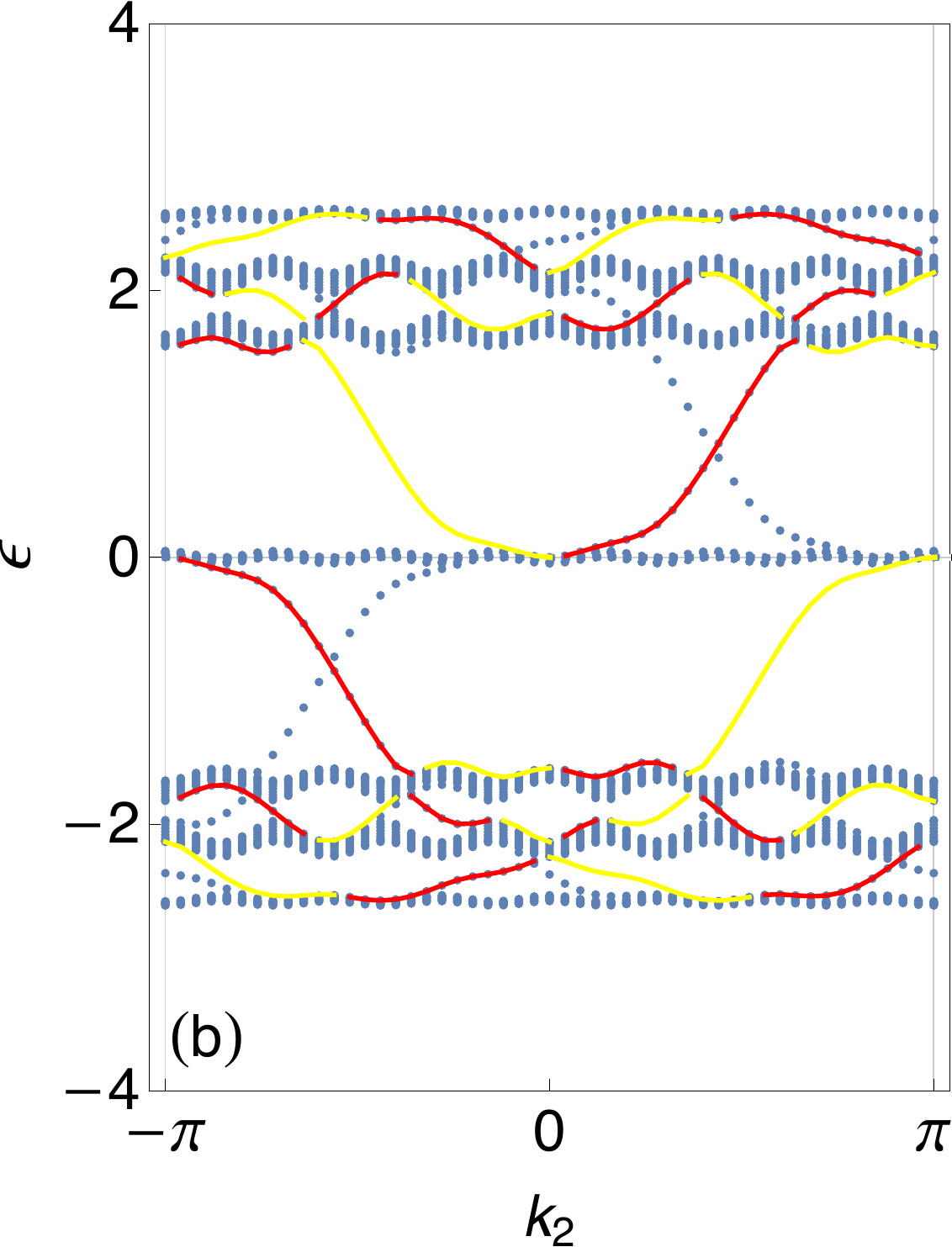}
\end{tabular}
\caption{
Spectra of the Hofstadter model with flux (a) $\phi/(2\pi)=\frac{1}{7}$ and (b) $\frac{3}{7}$.
The red and yellow curves show $\varepsilon_n(k_2)$ in Eq. (\ref{EdgEigEquHof})  
with $\kappa_n(k_2)<0$ and $\kappa_n(k_2)>0$, respectively. 
The gray dots are total spectrum for the system with open boundaries (i.e.,  with the left and right ends) 
in the 1-direction.
}
\label{f:hof_1}
\end{center}
\end{figure}

In passing, we have to mention that the normalization of the wave functions can be determined by
\begin{alignat}1
\sum_{j=1}^\infty|\psi_{j,n}|^2=|\psi_{0,n}|^2
\sum_{j=1}^\infty e^{-2\kappa_nj}
=\frac{|\chi_{0,n}|^2}{e^{2\kappa_n}-1},
\end{alignat}
where  the sum over $j$ converges due to  the condition $e^{-\kappa_n}<1$.  
With thin in mind, we will not pay attention to the normalization of the wave functions below.

In Fig. \ref{f:hof_1}, the spectra of the edge states for the system defined on the half-plane
$1\le j_1\le+\infty$ are shown on the background of the total spectra for the system of
cylindrical geometry, $1\le j_1\le L$.
The red and yellow curves are the eigenvalues of the ESH Eq. (\ref{HofRedHam}) with 
$e^{-\kappa_n(k_2)}<1$ and $e^{-\kappa_n(k_2)}>1$, respectively.
One can see that red curves describe the spectra of the edge states localized at the left boundary, whereas 
there are no states corresponding to the yellow curves, since they cannot be physical states.
Thus, for the theory of the edge states, not only the Hamiltonian (\ref{HofRedHam}) but also
the condition (\ref{HofCon}) are needed to describe the edge states at one end.
In passing, we add a comment that if we consider a cylindrical system without the rightmost $q$th site 
in the rightmost unit cell $j_1=L$,  the yellow curves are edge states localized at the right end.
However, we will consider the edge states at the right end in the way given in the next Sec. \ref{s:right_edge}.

\subsubsection{Edge states at the right end}\label{s:right_edge}

\begin{figure}[htb]
\begin{center}
\begin{tabular}{ccc}
\includegraphics[width=.5\linewidth]{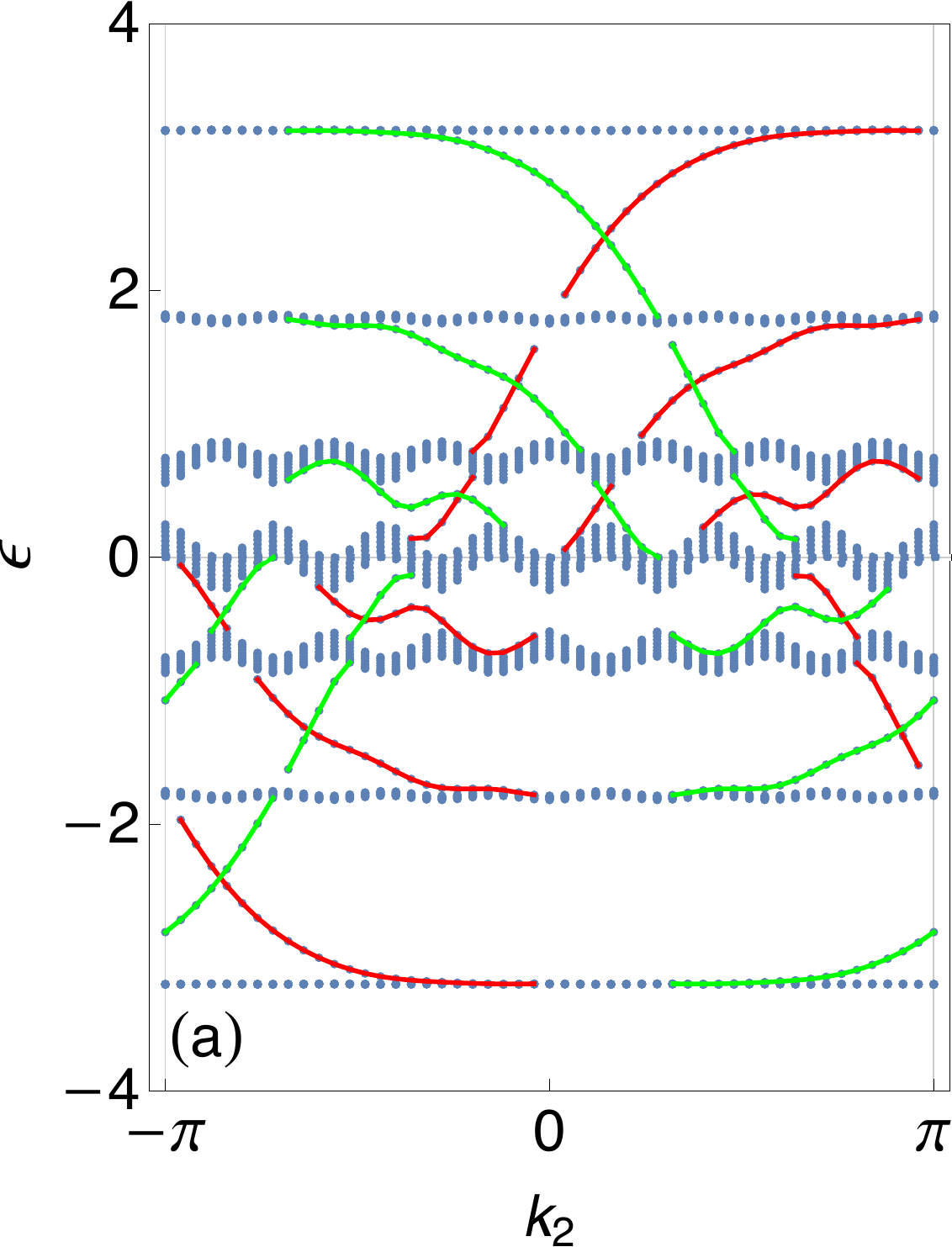}
&
\includegraphics[width=.5\linewidth]{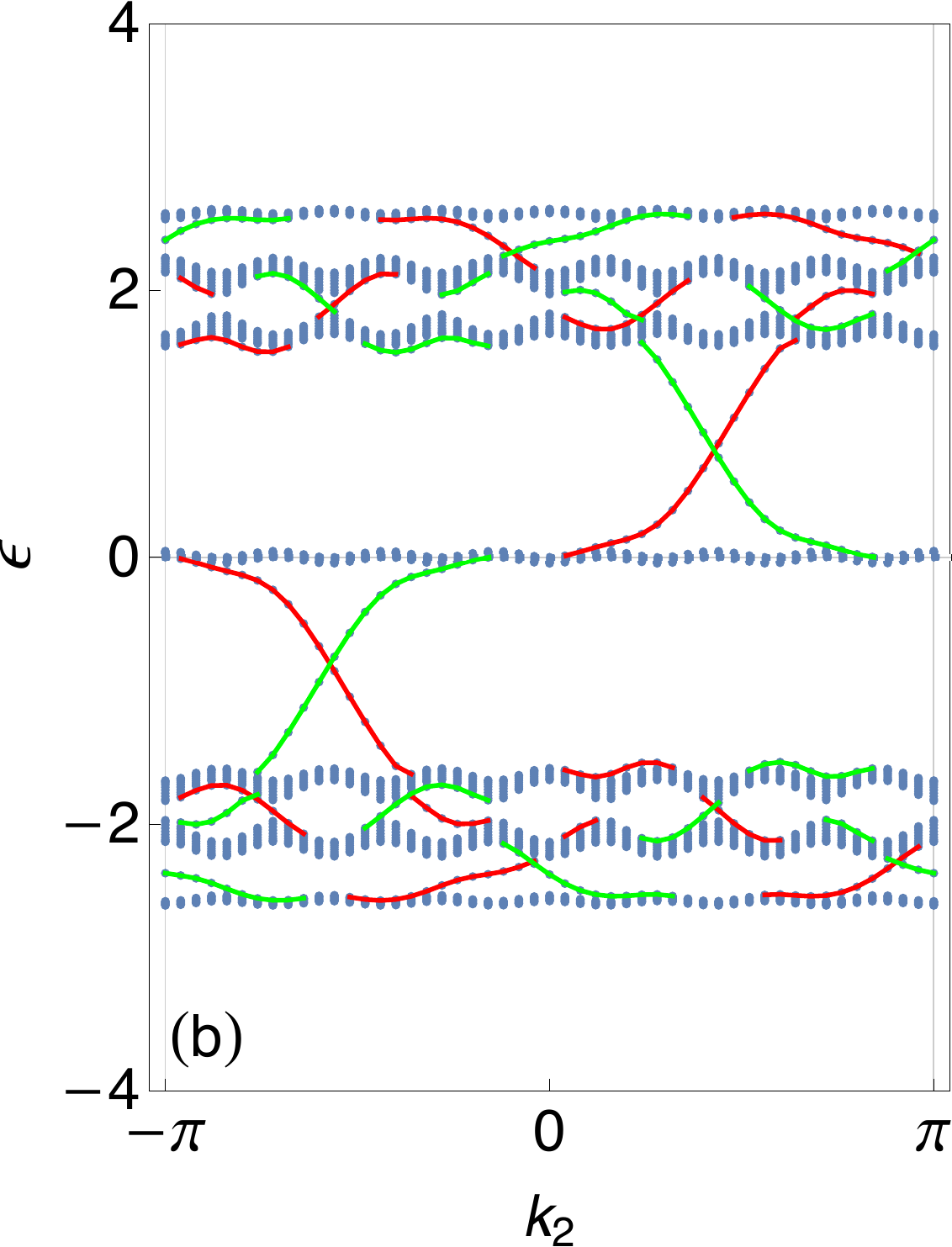}
\end{tabular}
\caption{
Spectra of the Hofstadter model with flux (a) $\phi/(2\pi)=\frac{1}{7}$ and (b) $\frac{3}{7}$.
The red and green curves denote, respectively, the spectrum of the left ESH (\ref{HofRedHam})
satisfying the condition (\ref{HofCon}) and the right ESH (\ref{HofGreHam}) satisfying the condition (\ref{HofConGre}).
%
}
\label{f:hof_2}
\end{center}
\end{figure}

As demonstrated above, when we numerically calculate systems of cylindrical geometry, 
$1\le j_1\le L$, it is inevitable to have both edge states at the left and right ends. 
The Hamiltonian (\ref{HofRedHam}) is only for the edge states at the left boundary, $j_1=1$.
This implies that there exists another Hamiltonian which would describe the edge states at the right edge, $j_1\sim L$.
Below, we derive the ESH at the right end.
It is indeed one of the merit of our method to be able to treat the left end and the right end separately. 

So far we have studied 
the system defined on a half-plane, $j_1\ge1$.
In order to find the right ESH,
let us consider the same system defined on  the opposite half-plane, $j_1\le-1$,  for the edge states 
at the right end, $j_1=-1$. 
Such edge states correspond to those localized at the right end, $j_1=L$, mentioned above.
In this case, the boundary condition should be
\begin{alignat}1
{\cal K}^\dagger\psi_0=0,
\label{BouConR}
\end{alignat}
instead of Eq. (\ref{BouCon}).
Let  $\psi_{jn}=\psi_{0n}e^{iK_nj}$ with $K_n=k_n+i\kappa_n$ be the eigenstate near $j_1=-1$. 
Then, $\kappa_n<0$ is required, and 
from the point of view of the lattice termination, the latter type of $\psi_{0n}$ in Eq. (\ref{HofWavFun}) is suitable in this case, which naturally satisfied the boundary condition (\ref{BouConR}).
Thus, the eigenvalue equation is decomposed into
\begin{widetext}
\begin{alignat}1
t\left(
\begin{array}{c|ccccc}
2\cos(k_2-\phi)&1&&&&e^{-iK_n}\\
\hline
1&2\cos(k_2-2\phi)&1&&&\\
&1&\ddots&&&\\
&&&&2\cos(k_2-(q-1)\phi)&1\\
e^{iK_n}&&&&1&2\cos k_2
\end{array}
\right)
\left(\begin{array}{c}0\\\hline \multirow{4}{*}{$\chi_{n}$}\\ \\ \\ \\ \\\end{array}\right) 
=\varepsilon_{n}
\left(\begin{array}{c}0\\\hline \multirow{4}{*}{$\chi_{n}$}\\ \\ \\ \\ \\\end{array}\right).
\label{HofEquR}
\end{alignat}
\end{widetext}
This equation tells that the lower $(q-1)\times(q-1)$ part of the same Hamiltonian (\ref{HamOpeHof}) can be the
Hamiltonian of the right edge states as long as the condition,
\begin{alignat}1
|e^{-iK_n}|=e^{\kappa_n}=\left|\frac{\chi_{1,n}}{\chi_{q-1,n}}\right|<1,
\label{HofConGre}
\end{alignat}
holds, which is obtained in the first line of the above equation.
Namely, the edge states at the right edge are determined by the following Hamiltonian,
\begin{alignat}1
{\cal H}_{\rm e}=t\left(
\begin{array}{ccccc}
2\cos(k_2-2\phi)&1&&&\\
1&2\cos(k_2-3\phi)&1&&\\
&1&\ddots&&\\
&&&&2\cos k_2
\end{array}
\right).
\label{HofGreHam}
\end{alignat}
with the condition Eq. (\ref{HofConGre}).
This is the right ESH,  actually different from the left ESH (\ref{HofRedHam}).
In Fig. \ref{f:hof_2}, we show, by the green curves, the eigenvalues of the Hamiltonian (\ref{HofGreHam}) with the 
condition (\ref{HofConGre}) satisfied.
One can find that the red curves and green curves
fully describe the edge states for the system of 
cylindrical geometry.
In passing, we mention that the Hamiltonian (\ref{HofGreHam}) differs from (\ref{HofRedHam}) 
in the constant sift of $k_2\rightarrow k_2+\phi$.
Therefore, the yellow curves in (\ref{HofRedHam}) can be the edge states if one chooses a suitable 
boundary condition.

\subsubsection{$1/3$-flux}
In the case of $\phi/(2\pi)=\frac{1}{3}$, we can obtain  analytic solutions of the edge states.
It follows from Eq. (\ref{HofRedHam}) that the left ESH is given by the following $2\times2$ matrix,
\begin{alignat}1
\hat{\cal H}_{\rm e}=t\left(
\begin{array}{cc}
2\cos(k_2-\phi)&1\\
1&2\cos(k_2+\phi)\\
\end{array}
\right),
\label{HofRedHam3}
\end{alignat}
which gives the eigenvalues
\begin{alignat}1
\varepsilon_\pm=-\cos k_2\pm\sqrt{3\sin^2 k_2+1}.
\label{EdgSpe13}
\end{alignat}
The parameter $\kappa_\pm$ is found to be
\begin{alignat}1
e^{-\kappa_\pm}=|\varepsilon_\pm-2\cos (k_2-\phi)|.
\label{EdgCon13}
\end{alignat}
\begin{figure}[htb]
\begin{center}
\begin{tabular}{cc}
\includegraphics[width=.5\linewidth]{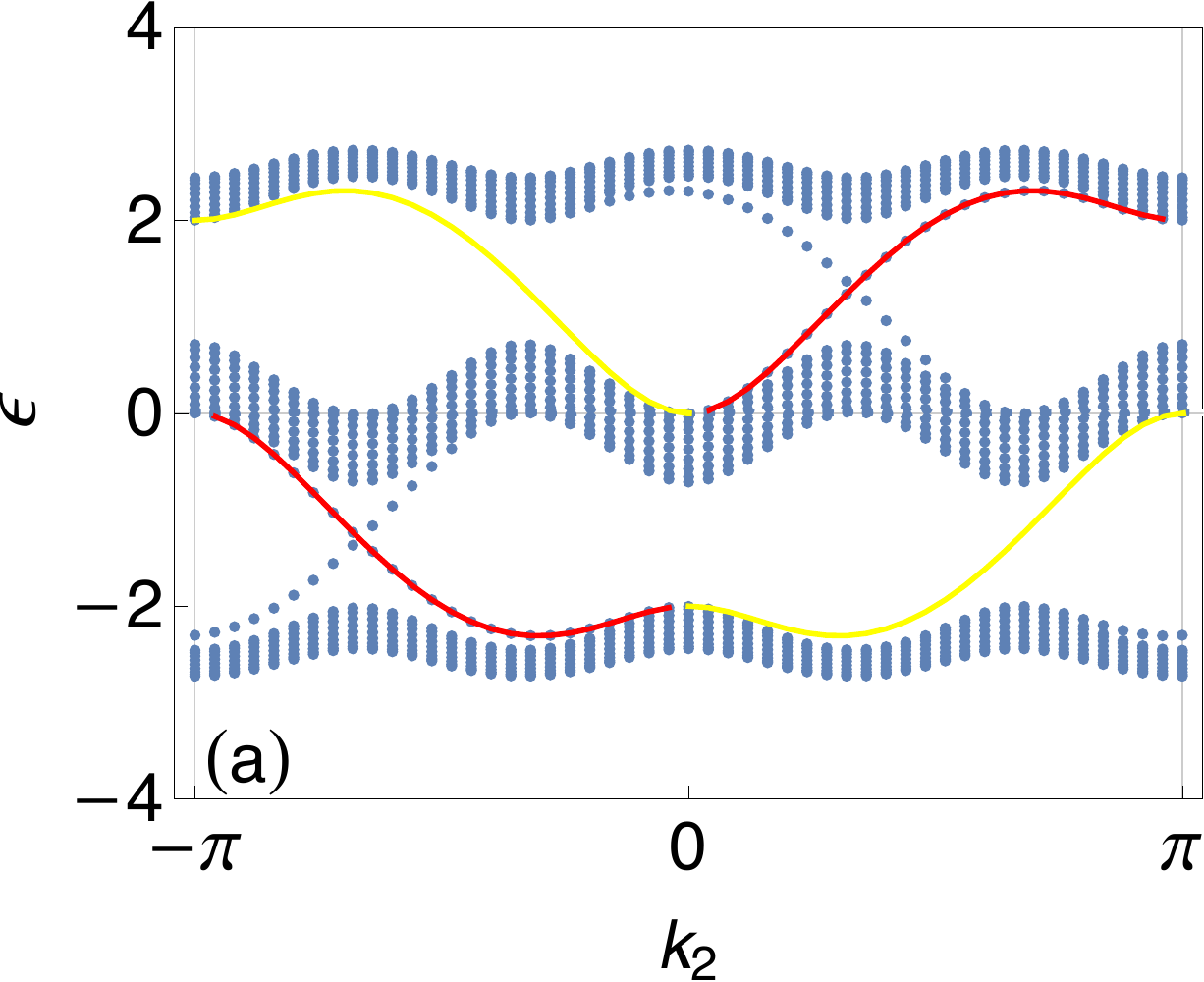}
&
\includegraphics[width=.5\linewidth]{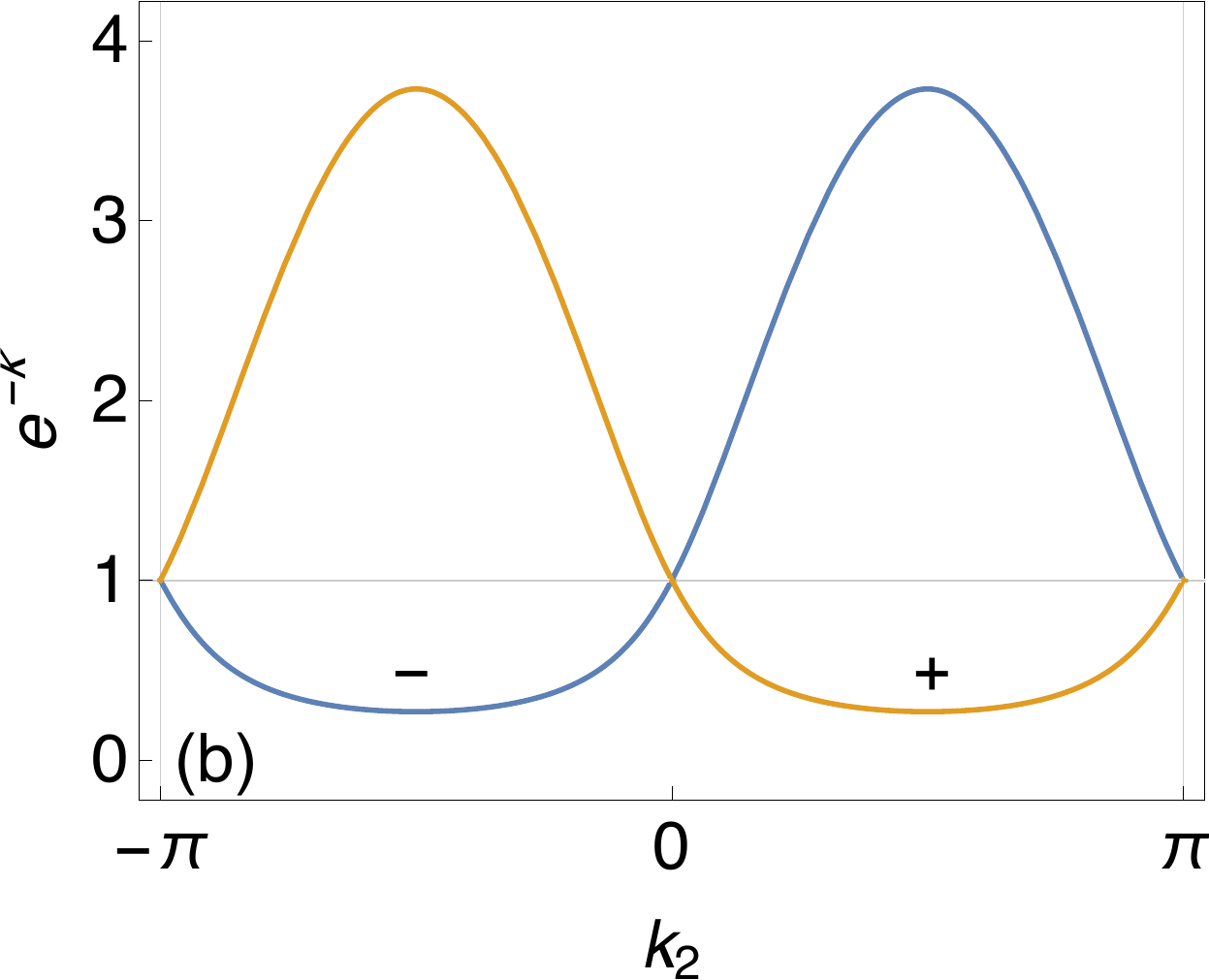}
\end{tabular}
\caption{
(a) The same figure as  Fig. \ref{f:hof} for the model with flux $\phi/(2\pi)=\frac{1}{3}$.
The red and yellow curves denote the spectra $\varepsilon_{\pm}$ in Eq. (\ref{EdgSpe13})
with the constraint $e^{-\kappa_\pm}<1$ and $e^{-\kappa_\pm}>1$, respectively.
(b) Plot of $e^{-\kappa_\pm(k_2)}$ as functions of $k_2$.
}
\label{f:hof_3}
\end{center}
\end{figure}
We show in Fig. \ref{f:hof_3}, the edge spectrum obtained in Eq. (\ref{EdgSpe13})
as well as the plot of $e^{-\kappa_\pm}$ as functions of $k_2$. 
One sees that the edge state in the lower (higher) gap is the physical state in $-\pi<k_2<0~(0<k_2<\pi)$.

\section{Graphene}\label{s:graphene}
Another example in 2D systems treated in this section is  (spinless) graphene. 
In Sec. \ref{s:graphene_nomag}, we consider the model in the absence of a magnetic field
and derive the famous edge states along the zigzag edge, etc \cite{Fujita:1996aa},
based on our Hamiltonian operator formalism.
The zero energy edge states are protected by reflection symmetry characterized 
by quantized Berry phase \cite{Ryu:2002fk}.
In Sec. \ref{s:graphene_mag}, we discuss the edge states of graphene 
in the presence of a magnetic field \cite{Hatsugai:2006aa}. 
In the case of the zigzag edge or bearded edge, we derive the ESHs, 
as discussed in Sec. \ref{s:graphene_mag_zig}.
Since our method treats edge states at the left end and right end separately, we can reproduce 
the edge states of the cylindrical system with the zigzag edge at the left end and bearded edge at the right end by use of
left and right ESHs.
In spite of the same graphene,  the system with the armchair edge belongs to class B or C in Sec. \ref{s:class},
implying that simple Hamiltonian formalism is impossible. 
Nevertheless, if one deforms a model with the armchair edge kept unchanged,
one can derive the ESH which reproduces well the edge states of the undeformed model,
as demonstrated in Sec. \ref{s:graphene_mag_arm}.
This is due to localization property of edge states along the boundary.

\begin{figure}[htb]
\begin{center}
\begin{tabular}{c}
\includegraphics[width=.82\linewidth]{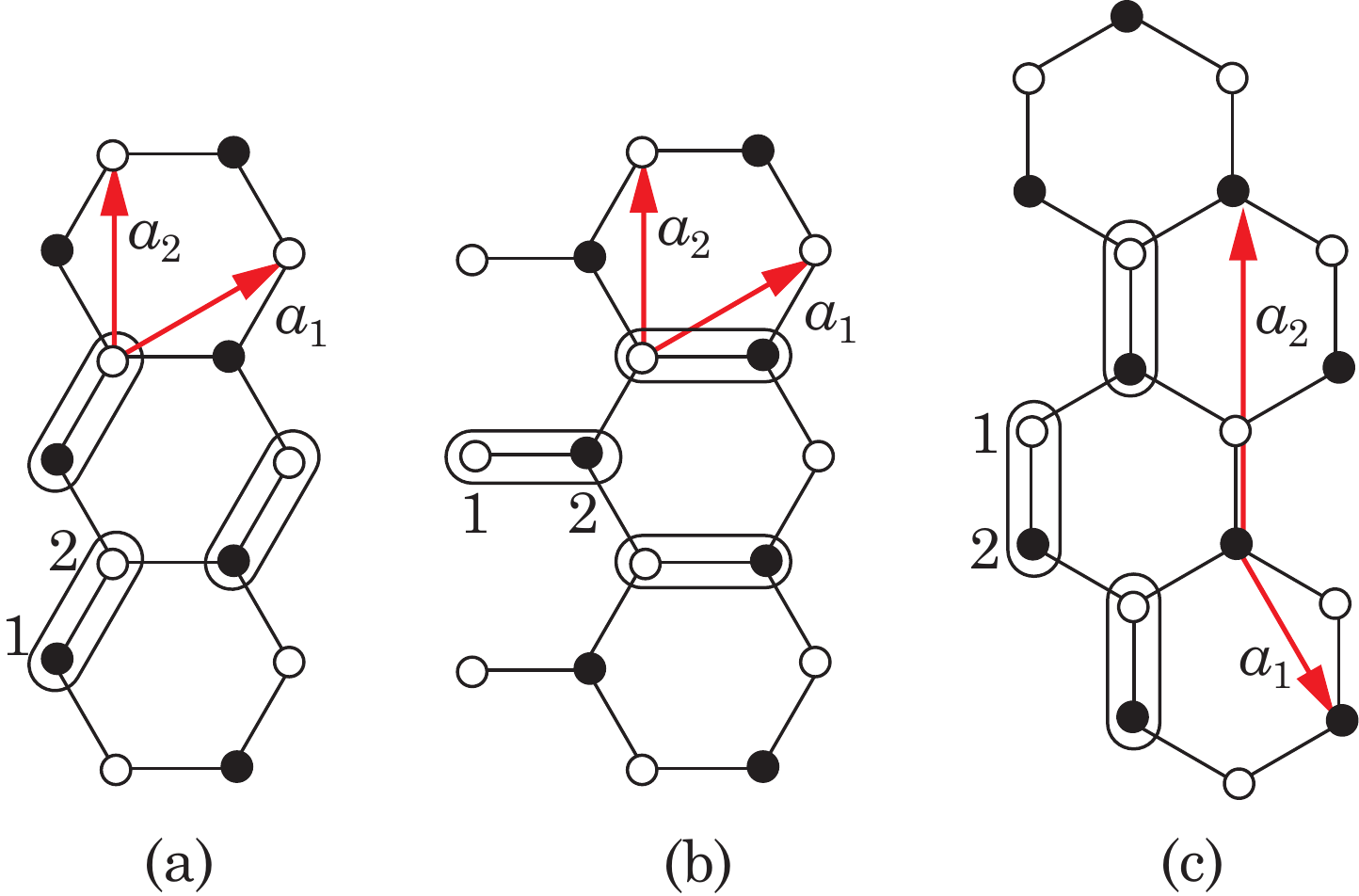}
\end{tabular}
\caption{
Boundaries of graphene, (a) zigzag edge, (b) bearded edge , and (c) armchair edge.
}
\label{f:graphene_lat}
\end{center}
\end{figure}

\subsection{In the absence of a magnetic field}\label{s:graphene_nomag}

\subsubsection{Zigzag edge}\label{s:zigzag}

Let us start with the Hamiltonian in the lattice representation.
It follows from Fig. \ref{f:graphene_lat} (a) that the Hamiltonian operator appropriate for the zigzag edge reads
\begin{alignat}1
H=t\sum_j c_j^\dagger 
\left(
\begin{array}{cc}
&1+\delta_1^*+\delta_2^*\\
1+\delta_1+\delta_2&
\end{array}
\right)c_j.
\label{GraZigHam}
\end{alignat}
Then, this operator becomes 1D Hamiltonian operator 
after the Fourier transformion in the 2-direction by replacing $\delta_2\rightarrow e^{ik_2}$:
\begin{alignat}1
\hat{\cal H}&=t
\left(
\begin{array}{cc}
&1+\delta_1^*+\delta_2^*\\
1+\delta_1+\delta_2&
\end{array}
\right)
\nonumber\\
&\rightarrow t\left(
\begin{array}{cc}
&1+e^{-ik_2}+\delta^*\\
1+e^{ik_2}+\delta&
\end{array}
\right),
\label{GraZigHam2}
\end{alignat}
where $\delta\equiv\delta_1$. Form this Hamiltonian operator, the Hermiticity matrix $\cal K$ reads 
\begin{alignat}1
{\cal K}=t
\left(
\begin{array}{cc}
&1\\
0&
\end{array}
\right) ,
\label{GraZigHer}
\end{alignat}
which is the same matrix as the NN SSH model ($t=0$), implying that this model belongs to class A in 
Sec. \ref{s:class}.
Thus, as in the case of the SSH model in Sec. \ref{s:ssh}, we assume a Bloch-type wave function (\ref{Blo}) based on
$\psi_0=\psi_\uparrow$ in Eq. (\ref{BouWav}) which satisfies (\ref{EigSSH}).
Then, we have
\begin{alignat}1
t
\left(
\begin{array}{cc}
&1+e^{-ik_2}+e^{-iK}\\
1+e^{ik_2}+e^{iK}&
\end{array}
\right)
\left(\begin{array}{c}1\\0\end{array}\right)=\varepsilon\left(\begin{array}{c}1\\0\end{array}\right).
\end{alignat}
This equation separates into two components such that
\begin{alignat}1
\varepsilon=0, \quad 1+e^{ik_2}+ e^{iK}=0.
\end{alignat}
The last equation leads to
\begin{alignat}1
\left|e^{iK}\right|=e^{-\kappa}=|1+e^{ik_2}|<1\quad \rightarrow \frac{2\pi}{3}<|k_2|<\pi.
\label{ZigRan}
\end{alignat}

\begin{figure}[htb]
\begin{center}
\begin{tabular}{ccc}
\includegraphics[width=.33\linewidth]{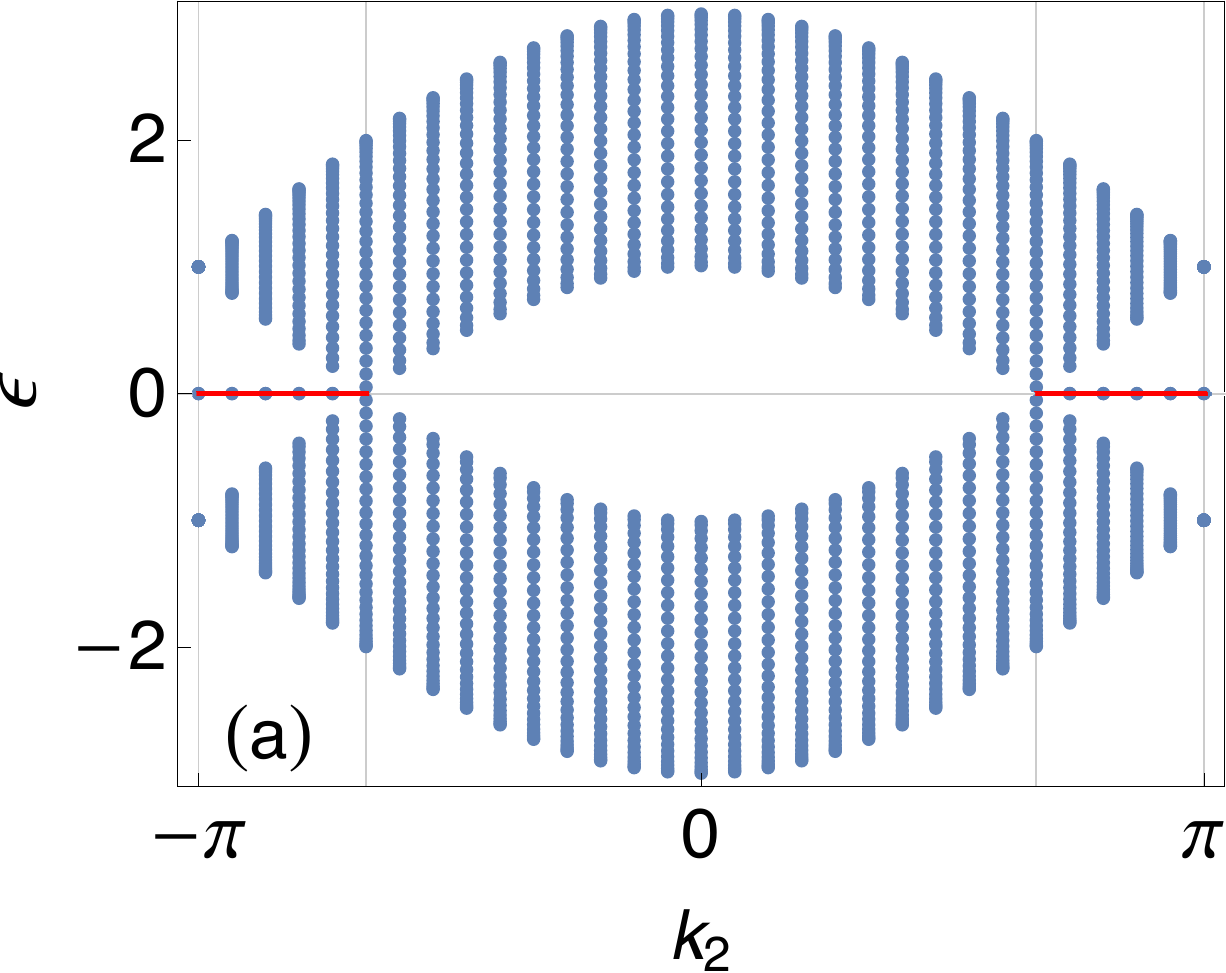}
&
\includegraphics[width=.33\linewidth]{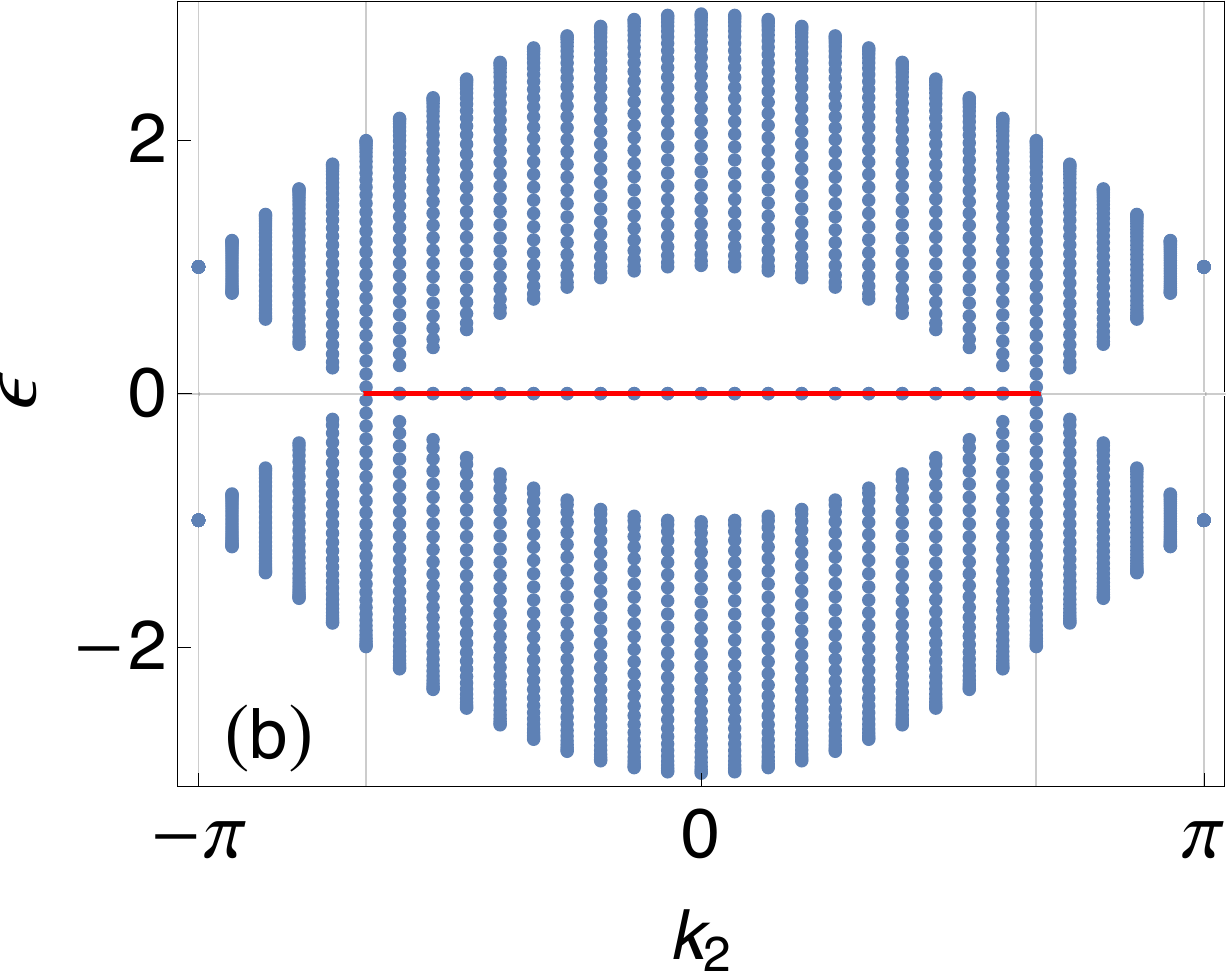}
&
\includegraphics[width=.33\linewidth]{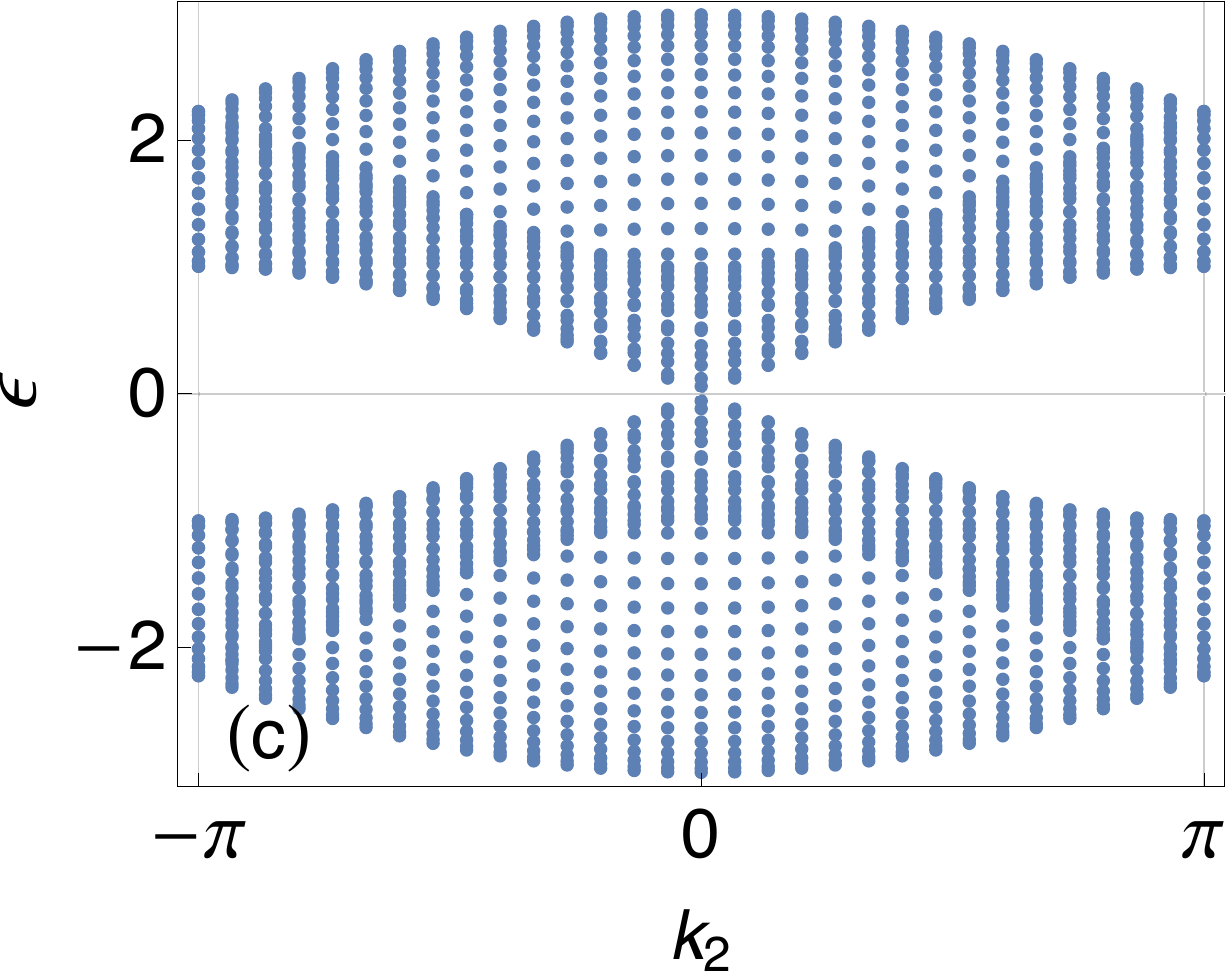}
\end{tabular}
\caption{
The spectra of graphene model with (a) zigzag, (b) bearded, and (c) armchair edges.
The red lines show the zero energy edge states in the range of Eqs. (\ref{ZigRan}) and  (\ref{BeaRan})
embedded in the full spectra of the system
with cylindrical geometry.
The grid lines in (a) and (b) indicate $k_2=\pm 2\pi/3$.
}
\label{f:graphene}
\end{center}
\end{figure}

In Fig. \ref{f:graphene} (a), we show the edge states at zero energy by red lines 
which exists in the range Eq. (\ref{ZigRan}).
This reproduces the results in Ref. \cite{Fujita:1996aa}.

\subsubsection{Bearded edge}

Similarly to the zigzag edge, it follows from Fig. \ref{f:graphene_lat} (b) that
the 2D Hamiltonian operator and Fourier-transformed 1D Hamiltonian operator
appropriate for the bearded edge is given by
\begin{alignat}1
\hat{\cal H}
&=t
\left(
\begin{array}{cc}
&1+\delta_1^*+\delta_1^*\delta_2\\
1+\delta_1+\delta_1\delta_2^*&
\end{array}
\right),
\nonumber\\
&\rightarrow t
\left(
\begin{array}{cc}
&1+(1+e^{ik_2})\delta^*\\
1+(1+e^{-ik_2})\delta&
\end{array}
\right).
\label{GraBeaHam2}
\end{alignat}
Then, the Hermiticity matrix $\cal K$ reads 
\begin{alignat}1
{\cal K}=t\left(
\begin{array}{cc}
&1+e^{ik_2}\\
0&
\end{array}
\right).
\label{GraBeaK}
\end{alignat}
Although ${\cal K}$ depends on $k_2$,
this is basically the same as (\ref{GraZigHer}), and  this model is also a NN model in class A. 
Therefore, we can choose $\psi_0=\psi_\uparrow$ satisfying Eq. (\ref{EigSSH}). 
Form the eigenvalue equation for the same Bloch-type wave function as in Sec. \ref{s:zigzag}, we have
\begin{alignat}1
\varepsilon=0, \quad 1+(1+e^{ik_2}) e^{iK}=0.
\end{alignat}
Hence, it follows from 
\begin{alignat}1
\left|e^{iK}\right|=e^{-\kappa}=\frac{1}{|1+e^{ik_2}|}<1,
\label{BeaRan}
\end{alignat}
that the range of $k_2$ for the bearded edge is the complement of that for the zigzag edge (\ref{ZigRan}).

\subsubsection{Armchair edge}\label{s:graphene_arm}

\begin{figure}[htb]
\begin{center}
\begin{tabular}{c}
\includegraphics[width=.5\linewidth]{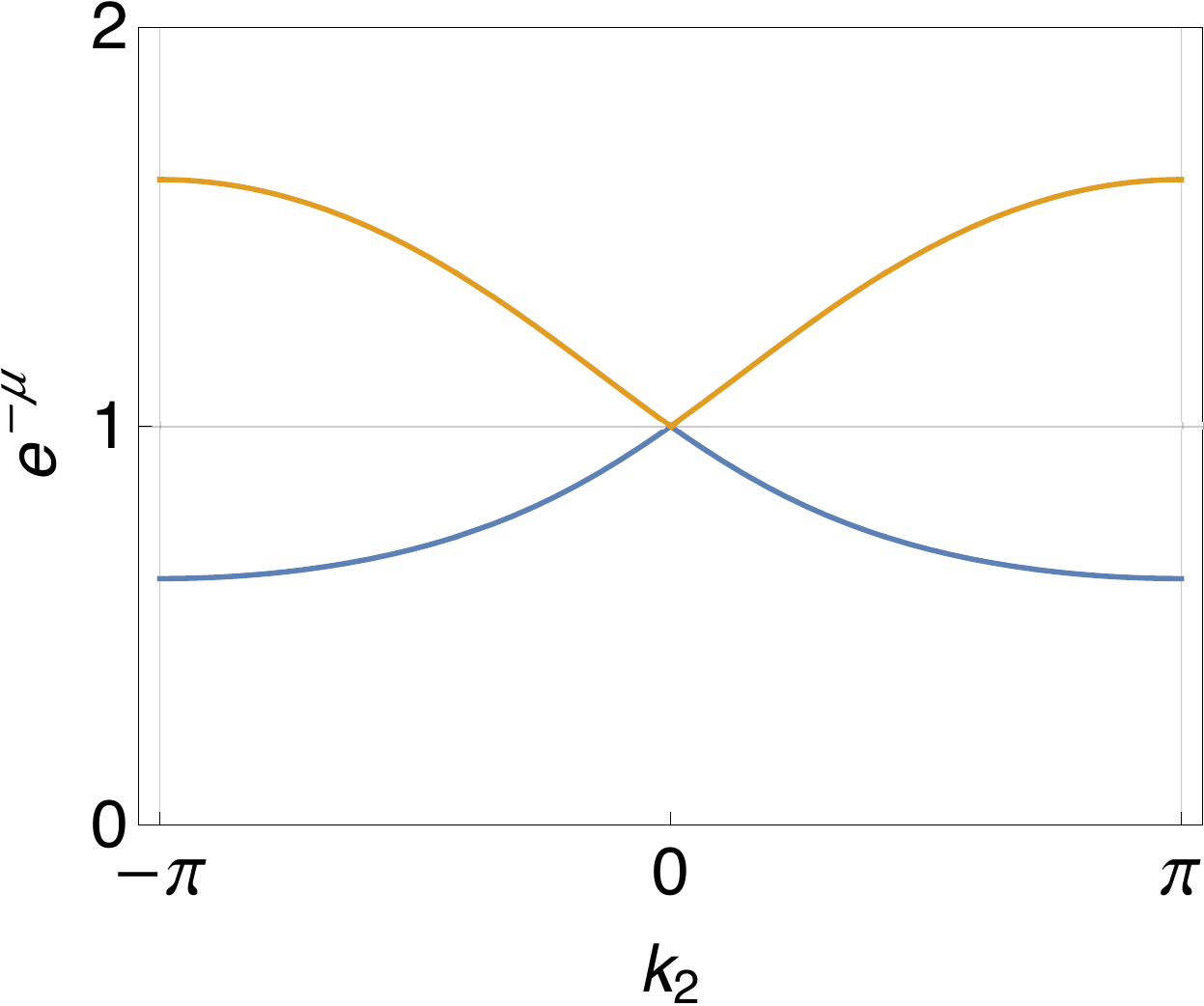}
\end{tabular}
\caption{
The two solutions for Eqs. (\ref{SolArm}) and (\ref{SolArm2}). 
There is only one solution satisfying $e^{-\kappa}<1$ at any $k_2$.
}
\label{f:armsol}
\end{center}
\end{figure}

Lastly, in the case of the armchair edge, Fig.\ref{f:graphene_lat} (c) leads us to
\begin{alignat}1
\hat{\cal H}&=t
\left(
\begin{array}{cc}
&1+\delta_1^*+\delta_1\delta_2\\
1+\delta_1+\delta_1^*\delta_2^*&
\end{array}
\right)
\nonumber\\
&\rightarrow
t
\left(
\begin{array}{cc}
&1+\delta^*+e^{ik_2}\delta\\
1+\delta+e^{-ik_2}\delta^*&
\end{array}
\right).
\label{GraArmHam2}
\end{alignat}
Then, the Hermiticity matrix $\cal K$ reads
\begin{alignat}1
{\cal K}=t\left(
\begin{array}{cc}
&1\\
e^{-ik_2}&
\end{array}
\right),
\end{alignat}
which is the same as that for the generalized SSH model with $t\ne0$ in Sec. \ref{s:ssh_edge}, 
so that this case belongs to the class B in Sec. \ref{s:class}. 
Therefore, it is not enough to choose $\psi_0=\psi_\uparrow$ or $\psi_\downarrow$ ensuring the Hermiticity of the 
Hamiltonian: We need to find two independent solutions to obtain the wave function satisfying the boundary condition
Eq. (\ref{BouCon}).
When one chooses $\psi_0=\psi_\uparrow$, the eigenvalue equation leads to
 \begin{alignat}1
\varepsilon=0, \quad 1+e^{iK}+e^{-ik_2} e^{-iK}=0.
\label{SolArm}
\end{alignat}
and when one chooses $\psi_0=\psi_\downarrow$,
 \begin{alignat}1
\varepsilon=0, \quad 1+e^{-iK}+e^{ik_2} e^{iK}=0.
\label{SolArm2}
\end{alignat}
These equations for $\psi_0=\psi_{\uparrow,\downarrow}$ give the same solutions for real part, $|e^{iK}|=e^{-\kappa}$.
As shown in Fig. \ref{f:armsol}, we find one solution in $e^{-\kappa}<1$ and the other in $e^{-\kappa}>1$ at any $k_2$,
implying that it is impossible to get the wave function satisfying the boundary condition (\ref{BouCon}).
Therefore, we conclude that the armchair edge allows no edge states.
This also reproduces the results in Ref. \cite{Fujita:1996aa}.
It should be stressed that all the differences of the edge structure are incorporated 
in the differences  of the Hamiltonian operators
in Eqs. (\ref{GraZigHam2}), (\ref{GraBeaHam2}), and (\ref{GraArmHam2}).

\subsection{In the presence of a uniform magnetic field}\label{s:graphene_mag}

The edge states and the bulk-edge correspondence of graphene in a uniform magnetic field
were studied in detail in Ref. \cite{Hatsugai:2006aa}, in which
a model with  hybrid edges, i.e., the zigzag edge at one end and the bearded edge at the 
other end was examined.
The merit of our method developed in this paper is that we can examine these edge states separately.
Let us first consider the zigzag edges.

\begin{figure}[htb]
\begin{center}
\begin{tabular}{c}
\includegraphics[width=.7\linewidth]{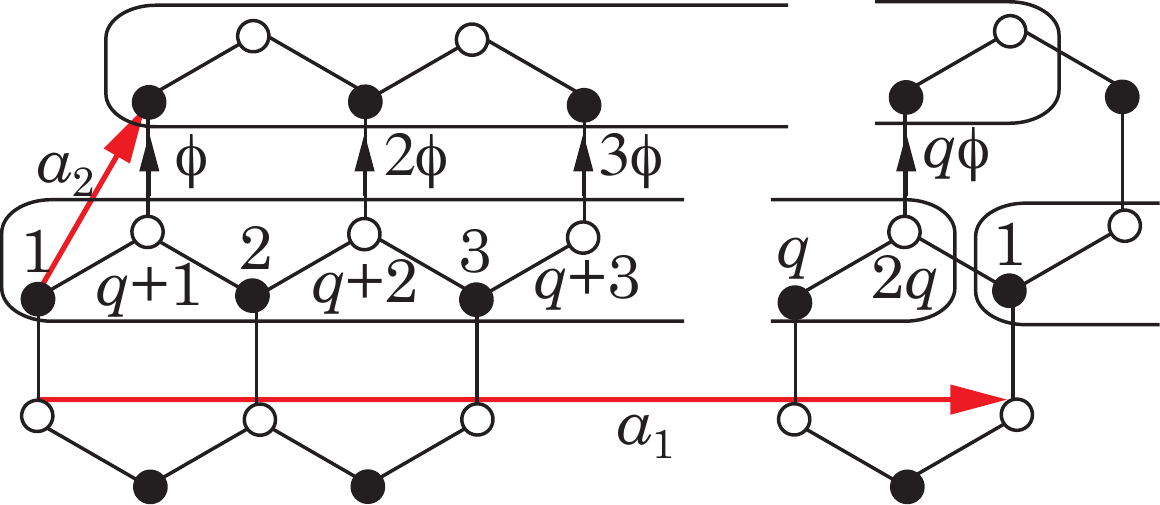}
\end{tabular}
\caption{
Magnetic unit cell for graphene under a uniform magnetic flux $\phi/(2\pi)=p/q$ per hexagon 
suitable for the zigzag edge.
}
\label{f:graphene_qhe_lat}
\end{center}
\end{figure}

\subsubsection{Zigzag edge}\label{s:graphene_mag_zig}

For the choice of the magnetic unit cell and the labeling of sites in it in Fig. \ref{f:graphene_qhe_lat},  
the Hamiltonian operator with the zigzag edge at the left end
under uniform magnetic flux $\phi/(2\pi)=p/q$ per hexagon is given by
\begin{alignat}1
\hat{\cal H}&=\left(\begin{array}{cc} & \hat{\cal D}\\ \hat{\cal D}^\dagger&\end{array}\right),
\label{GraHamOpe}
\end{alignat}
where 
\begin{alignat}1
\hat{\cal D}&=
t\left(
\begin{array}{cccccc}
1+\hat h_{\phi}&&&&&\delta_1^*\\
1&1+\hat h_{2\phi}&&&&\\
&1&&&&\\
&&\ddots&&&\\
&&&1&1+\hat h_{(q-1)\phi}&\\
&&&&1&1+\hat h_{q\phi}
\end{array}
\right).
\label{ZigDOpe}
\end{alignat}
Here, $\hat h_{\phi}=e^{i\phi}\delta_2^*$ denotes the hopping towards the 2-direction, 
and the Hermiticity matrix $\cal K$, which is given by the coefficient of $\delta^*$ in $\hat{\cal H}$, is
the same as that of the Hofstadter model in Eq. (\ref{HofK}):
\begin{alignat}1
{\cal K}=t\left(
\begin{array}{ccc|ccc}
&&&&&1\\
&&&&0& \vspace*{-2.5mm}\\
&&&\iddots&& \\
\hline
&&0&&&\\
&0&&&& \vspace*{-2.5mm} \\
\iddots&&&&&
\end{array}
\right).
\label{KforZig}
\end{alignat}
Therefore, the reference wave function $\psi_0$ is given by Eq. (\ref{HofWavFun}),
and assuming a Bloch-type edge state based on the reference state $\psi_{jn}=\psi_{0n}e^{iK_nj}$
with $K_n=k_n+i\kappa_n$,
the eigenvalue equation becomes
\begin{widetext}
\begin{alignat}1
t\left(
\begin{array}{cccccc|ccccc|c}
&&&&&&g_{\phi-k_2}&&&&&e^{-iK_n}\\
&&&&&&1&g_{2\phi-k_2}&&&\\
&&&&&&&1&&&\\
&&&&&&&&\ddots&&\\
&&&&&&&&&1&g_{(q-1)\phi-k_2}&\\
&&&&&&&&&&1&g_{-k_2}\\
\hline
g_{\phi-k_2}^*&1&&&&&&&&&&\\
&g_{2\phi-k_2}^*&1&&&&&&&&&\\
&&&\ddots&&&&&&&&\\
&&&&1&&&&&&&\\
&&&&g_{(q-1)\phi-k_2}^*&1&&&&&&\\
\hline
e^{iK_n}&&&&&g_{-k_2}^*&&&&&&\\
\end{array}
\right)
\left(\begin{array}{c}\multirow{11}{*}{$\chi_{n}$}\\ \\ \\ \\ \\ \\ \\ \\ \\  \\  \\  \\ \hline0\end{array}\right)
=\varepsilon_n
\left(\begin{array}{c}\multirow{11}{*}{$\chi_{n}$}\\ \\ \\ \\ \\ \\ \\ \\  \\ \\ \\ \\ \hline0\end{array}\right),
\label{GraQheEeq}
\end{alignat}
\end{widetext}
where $g_{\theta}=1+e^{i\theta}$.
Similarly to the Hofstadter model, the upper diagonal $2q-1$ part of this $2q$ equation defines 
the ESH ${\cal H}_{\rm e}$ for the zigzag edge  such that 
\begin{alignat}1
{\cal H}_{\rm e}&=
\left(\begin{array}{cc} & {\cal D}_{\rm e}\\{\cal D}_{\rm e}^\dagger&\end{array}\right),
\label{GraQheZigHam}
\end{alignat}
where ${\cal D}_{\rm e}$ is $q\times(q-1)$ matrix defined by
\begin{alignat}1
{\cal D}_{\rm e}&=
t\left(
\begin{array}{ccccc}
g_{\phi-k_2}&&&&\\
1&g_{2\phi-k_2}&&\\
&1&&\\
&&\ddots&\\
&&&1&g_{(q-1)\phi-k_2}\\
&&&&1
\end{array}
\right).
\label{GraQheZigHam2}
\end{alignat}
The equation of the last component in Eq. (\ref{GraQheEeq}) yields the constraint 
$e^{iK_n}\chi_{1,n}+g_{-k_2}^*\chi_{q,n}=0$,
from which it follows that the localization condition is given by
\begin{alignat}1
|e^{iK_n}|=e^{-\kappa_n}=\left|\frac{g_{-k_2}^*\chi_{q,n}}{\chi_{1,n}}\right|<1.
\label{GraQheMu}
\end{alignat}
Thus, we have successfully derived  the ESH
describing the edge states along the 
zigzag edge at the left end, which are completely separated from the bulk states.
It should be stressed again that 
not only the Hamiltonian given by Eqs. (\ref{GraQheZigHam}) and (\ref{GraQheZigHam2}) 
but also the localization condition (\ref{GraQheMu})
describe the edge states.

\begin{figure}[htb]
\begin{center}
\begin{tabular}{cc}
\includegraphics[width=.5\linewidth]{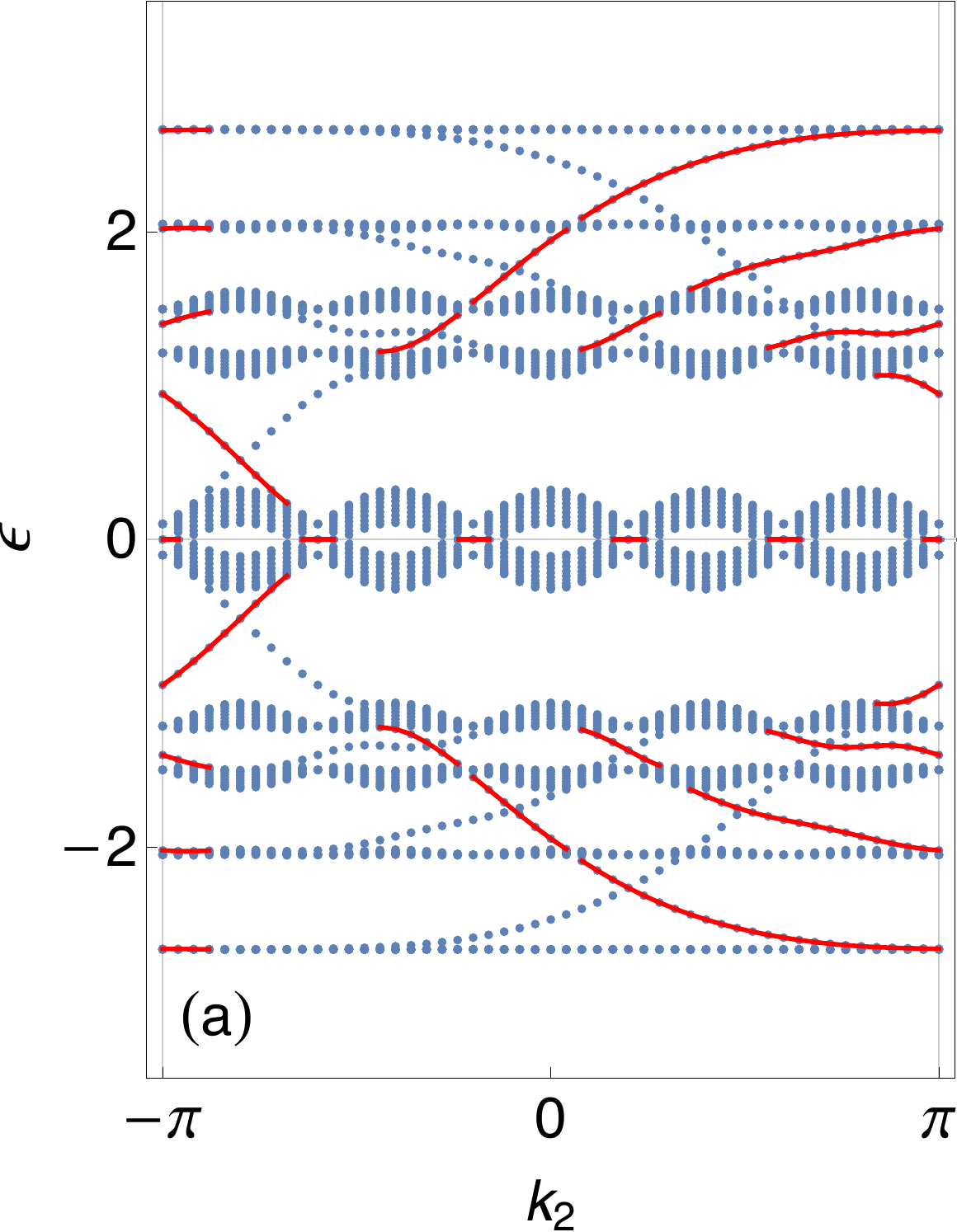}
&
\includegraphics[width=.5\linewidth]{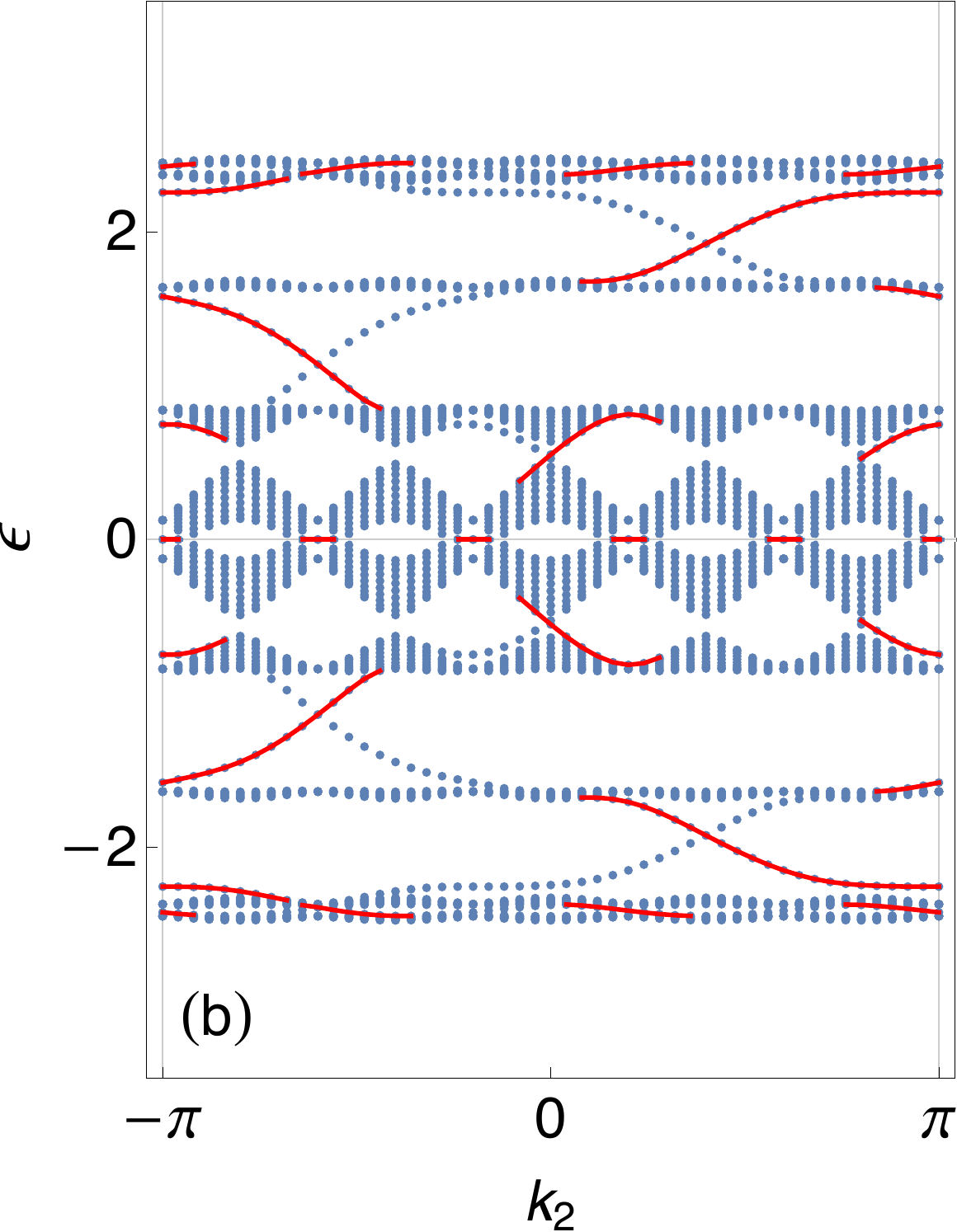}
\end{tabular}
\caption{
The spectra of the graphene model with the zigzag edge
under magnetic flux (a) $\phi/(2\pi)=1/5$ and (b) $2/5$.
The red curves show the eigenvalues of the Hamiltonian (\ref{GraQheZigHam}) satisfying (\ref{GraQheMu}).
}
\label{f:graphene_QHE}
\end{center}
\end{figure}

We show in Fig. \ref{f:graphene_QHE}, two examples of the QHE of graphene.
We plot, by red curves, the eigenvalues of the Hamiltonian (\ref{GraQheZigHam}) satisfying (\ref{GraQheMu})
on the background of the total spectra for the cylindrical systems with the zigzag edges at both ends.
Half of the edge state spectra are well reproduced, from which we find that they are edge states
localized at the left end.
One of characteristic properties of the edge states of graphene is the existence of the zero energy flat bands \cite{Hatsugai:2006aa}.
Actually, in Fig. \ref{f:graphene_QHE}, at several intervals of $k_2$, we find such bands.

The zero-energy edge states can be readily understood from the ESH in Eq. (\ref{GraQheZigHam}):
Because of the reference state $\psi_0$ satisfying the hemiticity, 
the $q\times q$ square matrix $\hat{\cal D}$ in Eq. (\ref{ZigDOpe})
reduces to $q\times(q-1)$ {\it imbalanced} matrix ${\cal D}_{\rm e}$ in Eq. (\ref{GraQheZigHam2}).
This reveals the existence of 
one zero-energy edge state at each $k_2$, as long as the condition Eq. (\ref{GraQheMu}) is satisfied.
To be more specific, let us write  the eigenvalue equation for the ESH, 
\begin{alignat}1
\left(\begin{array}{cc} & {\cal D}_{\rm e}\\{\cal D}_{\rm e}^\dagger&\end{array}\right)
\left(\begin{array}{c}\zeta_n\\\eta_n\end{array}\right)
=\varepsilon_n\left(\begin{array}{c}\zeta_n\\\eta_n\end{array}\right)
\begin{array}{ll} \}\, \scriptstyle q\\\}\, \scriptstyle q-1\end{array}, 
\label{GraQheChiEq}
\end{alignat}
where $2q-1$ component vector $\chi_n$ in Eq. (\ref{GraQheEeq})
is divided into $q$ component vector $\zeta_n$ and $q-1$ component vector $\eta_n$.
Owing to chiral symmetry, it may be convenient to solve the eigenvalue equation by considering the square of 
the Hamiltonian,
\begin{alignat}1
{\cal H}_{\rm e}^2\chi_n=\left(\begin{array}{cc} 
{\cal D}_{\rm e}{\cal D}_{\rm e}^\dagger&\\
&{\cal D}_{\rm e}^\dagger{\cal D}_{\rm e}\end{array}\right)
\left(\begin{array}{c}\zeta_n\\\eta_n\end{array}\right)=
\varepsilon_n^2\left(\begin{array}{c}\zeta_n\\\eta_n\end{array}\right).
\end{alignat}
Note that ${\cal D}_{\rm e}{\cal D}_{\rm e}^\dagger$ and ${\cal D}_{\rm e}^\dagger{\cal D}_{\rm e}$
are $q\times q$ and $(q-1)\times(q-1)$ matrices and they have the same energy eigenvalues except for 
zero energy, from which it follows that 
the zero-energy edge states are included in the upper part, ${\cal D}_{\rm e}{\cal D}_{\rm e}^\dagger$.
Other nonzero energy edge states are obtained by diagonalizing 
the lower part, ${\cal D}_{\rm e}^\dagger{\cal D}_{\rm e}\eta_n=\varepsilon_n^2\eta_n$.
Using Eq. (\ref{GraQheChiEq}), 
we find that the eigenfunction $\chi_n$ for nonzero energy states is given by
\begin{alignat}1
\varepsilon_{\pm n}=\pm\varepsilon_n,\quad 
\chi_{\pm n}=
\left(\begin{array}{c}\displaystyle \pm\frac{{\cal D}_{\rm e}\eta_n}{\varepsilon_n}\\ \eta_n\end{array}\right).
\end{alignat} 
It then follows that the localization condition Eq. (\ref{GraQheMu}) can be rewritten as
\begin{alignat}1
e^{-\kappa_n}=\left|\frac{g^*_{-k_2}\eta_{q-1}}{g_{\phi-k_2}\eta_1}\right|<1.
\label{GraQheMuNonZ}
\end{alignat}
Therefore, if ${\cal D}_{\rm e}^\dagger{\cal D}_{\rm e}$ has no zero energy eigenvalues,  
${\cal D}_{\rm e}{\cal D}_{\rm e}^\dagger$ has one zero energy state.
Such a zero energy state should obey ${\cal D}_{\rm e}^\dagger\zeta_0=0$ and $\eta_0=0$, and 
the localization condition is given by Eq. (\ref{GraQheMu}), i.e.,
\begin{alignat}1
|e^{iK_0}|=\left|\frac{g^*_{-k_2}\zeta_{q,0}}{\zeta_{1,0}}\right|<1.
\label{LocCon0}
\end{alignat}

For small-$q$ systems, we can obtain  exact spectra as well as exact wave functions of the edge states.
These are so didactic that we describe them separately below.

\begin{figure}[htb]
\begin{center}
\begin{tabular}{cc}
\includegraphics[width=.5\linewidth]{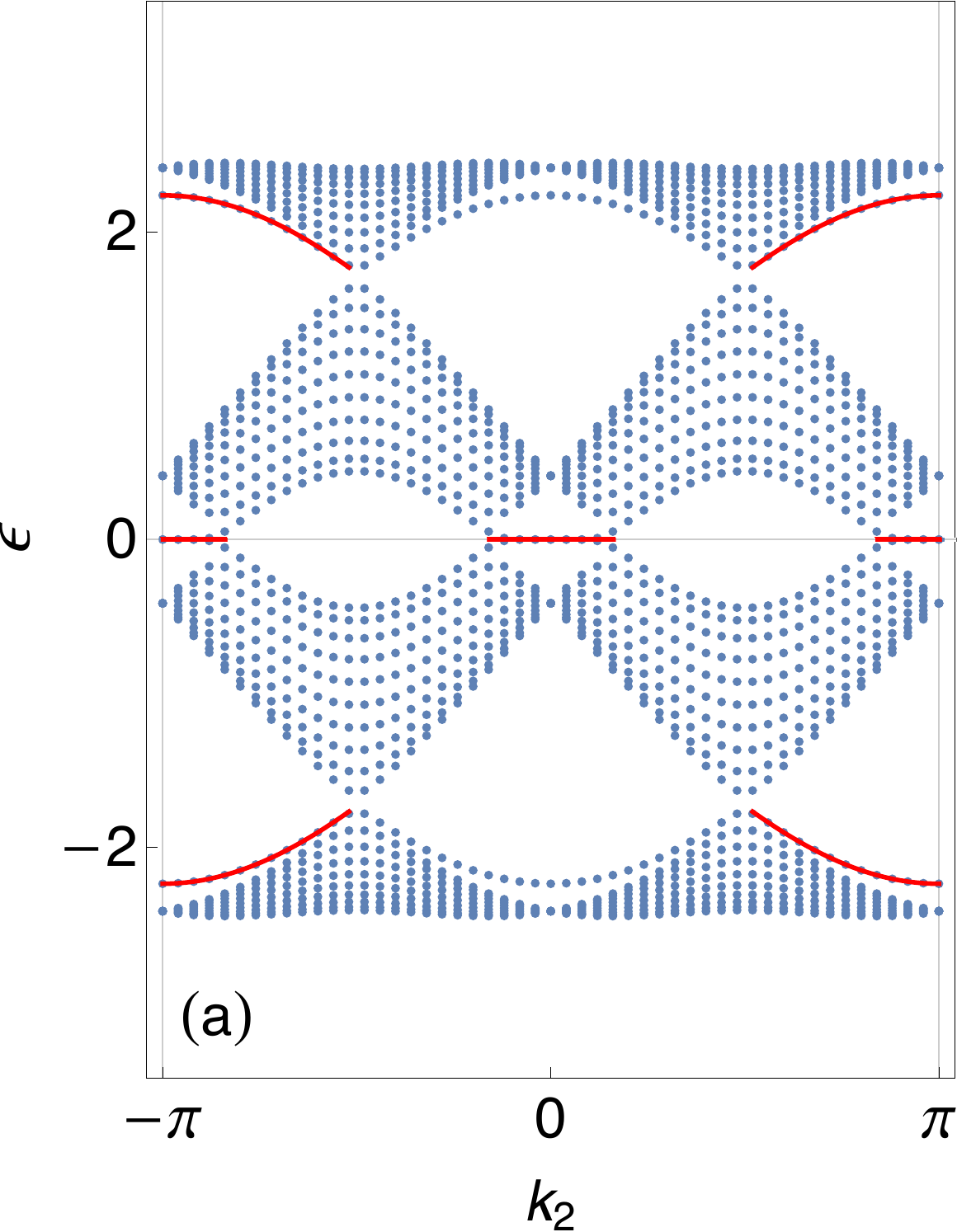}
&
\includegraphics[width=.5\linewidth]{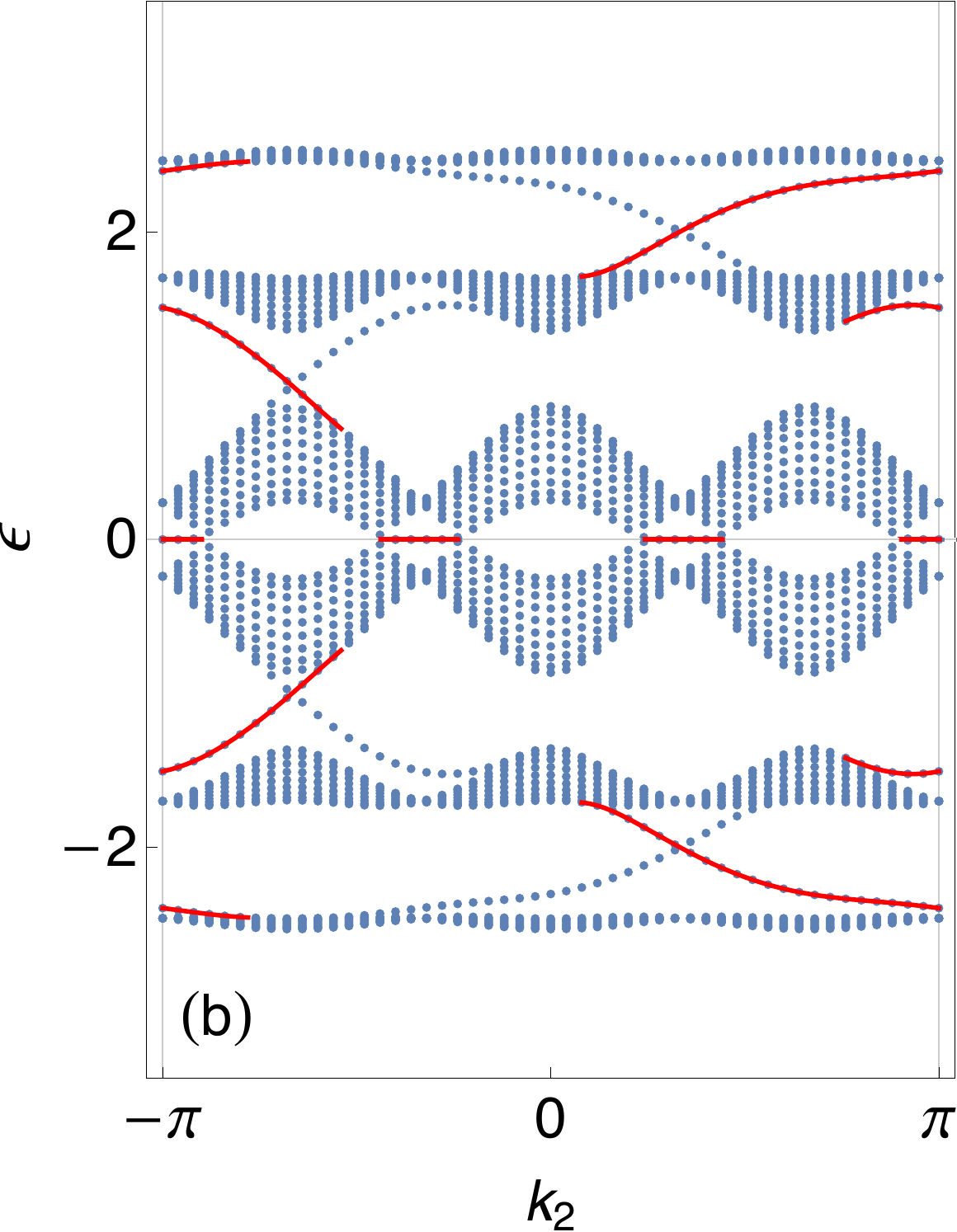}
\end{tabular}
\caption{
Spectra of graphene with the zigzag edge
under magnetic flux (a) $\phi/(2\pi)=1/2$ and (b) $1/3$.
The red curves show the eigenvalues  of the ESH. 
}
\label{f:graphene_QHE_2}
\end{center}
\end{figure}

\noindent $\bullet$ {$\pi$-flux}

Let us first consider case with $\pi$ flux per unit hexagon ($p/q=1/2$). 
Although we have already derived the ESH for generic $q$, it may still be convenient to write  the
total Schr\"odinger equation (\ref{GraQheEeq}), since this includes the condition to determine $e^{iK}$:
\begin{alignat}1
\left(
\begin{array}{cc|c|c} 
&&g_{\pi-k_2}&e^{-iK}\\
&&1&g_{-k_2}\\ 
\hline
g_{\pi+k_2}&1&&\\
\hline e^{iK}&g_{k_2}&&
\end{array}
\right)
\left(\begin{array}{c}\zeta_1\\ \zeta_2\\ \eta\\ \hline0\end{array}\right)
=\varepsilon\left(\begin{array}{c}\zeta_1\\ \zeta_2\\ \eta\\ \hline0\end{array}\right),
\label{Gra1o2}
\end{alignat}
where $g_{\pi\pm k_2}=1-e^{\pm ik_2}$ and $g_{\pm k_2}=1+e^{\pm ik_2}$, and
the energy quantum number has been suppressed.
The upper $3\times3$ matrix determines the spectrum of the edge states, from which we find
$2\times1$ ${\cal D}_{\rm e}$ 
\begin{alignat}1
{\cal D}_{\rm e}=
\begin{pmatrix} g_{\pi-k_2}\\1\end{pmatrix}.
\end{alignat}
The nonzero-energy states are readily obtained such that
\begin{alignat}1
{\cal D}_{\rm e}^\dagger{\cal D}_{\rm e}=|1-e^{-ik_2}|^2+1=3-2\cos k_2
\equiv \varepsilon^2(k_2).
\end{alignat}
whose eigenstate can be set $\eta=1$. 
Thus, nonzero-energy eigenvalues are $\varepsilon_\pm(k_2)=\pm\varepsilon(k_2)$, which should satisfy 
the localization condition (\ref{GraQheMuNonZ}),
\begin{alignat}1
|e^{iK_\pm}|=\left|\frac{(1+e^{ik_2})\zeta_2}{\zeta_1}\right|=\left|\frac{1+e^{ik_2}}{1-e^{-ik_2}}\right|<1.
\label{Gra1o2K}
\end{alignat}
Namely, both of the nonzero edge states in the same range of $k_2$ such that
\begin{alignat}1
\varepsilon_\pm=\pm\sqrt{3-2\cos k_2},\quad (\pi/2<|k_2|<\pi).
\label{GraQheZigNonz}
\end{alignat}

The zero energy edge state lives in $\zeta$ sector satisfying ${\cal D}_{\rm e}^\dagger\zeta_0=0$, 
so that we set $\eta_0=0$. Then, we have the following edge state at zero energy,
\begin{alignat}1
\varepsilon=0, \quad\zeta_0=\left(\begin{array}{c}-1\\ 1-e^{ik_2}\end{array}\right),
\label{Zig0}
\end{alignat}
and $e^{iK_0}$ in this case in Eq. (\ref{LocCon0}) or directly from the last component in 
Eq. (\ref{Gra1o2}) becomes
\begin{alignat}1
|e^{iK_0}|=|(1+e^{ik_2})(1-e^{ik_2})|<1,
\label{GraQheZig0}
\end{alignat}
which yields $0<|k_2|<\pi/6$, $5\pi/6<|k_2|<\pi$. 

In Fig. \ref{f:graphene_QHE_2} (a), the spectra Eqs. (\ref{GraQheZigNonz}) and (\ref{Zig0}) with (\ref{GraQheZig0})
obtained from the ESH are shown as red curves on the background of the total 
spectrum of the cylindrical system with the zigzag edges at both ends.
This figure suggests that the left and right ends yield the (non)zero energy edge states at the same (different) range of $k_2$.

\noindent$\bullet${$2\pi/3$-flux}

In the case of  $\phi/(2\pi)=1/3$, the eigenvalue equation for the edge states is 
\begin{widetext}
\begin{alignat}1
\newfont{\bg}{cmr10 scaled\magstep3}
\newcommand{\bigzero}{\smash{\hbox{\bg 0}}}
\left(
\begin{array}{ccc|cc|c}
&&&g_{\phi-k_2}&&e^{-iK_n}\\
&&&1&g_{-\phi-k_2}&\\
&&&&1&g_{-k_2}\\
\hline
 g_{\phi-k_2}^*&1&&&&\\
& g_{-\phi-k_2}^*&1&&&\\
\hline 
e^{iK_n}&&g_{-k_2}^*&&&
\end{array}
\right)
\left(\begin{array}{c}\multirow{3}{*}{$\zeta_n$}\\ \\  \\ \hline \multirow{2}{*}{$\eta_n$}\\ \\ \hline0\end{array}\right)
=\varepsilon_n
\left(\begin{array}{c}\multirow{3}{*}{$\zeta_n$}\\ \\  \\ \hline \multirow{2}{*}{$\eta_n$}\\ \\ \hline0\end{array}\right),
\label{Gra1o3}
\end{alignat}
\end{widetext}
where $\xi_n$ and $\eta_n$ denote three- and two-component vectors, respectively.
It follows that ${\cal D}_{\rm e}$ matrix reads
\begin{alignat}1
{\cal D}_{\rm e}=
\begin{pmatrix} g_{\phi-k_2}&0\\
1&g_{-\phi-k_2}\\
0&1
\end{pmatrix}.
\end{alignat}
The nonzero energies of the edge states are thus determined by 
\begin{alignat}1
{\cal D}_{\rm e}^\dagger{\cal D}_{\rm e}=
\begin{pmatrix} |1+e^{-i(k_2-\phi)}|^2&1+e^{-i(k_2+\phi)}\\
1+e^{i(k_2+\phi)}&|1+e^{i(k_2+\phi)}|^2
\end{pmatrix},
\end{alignat}
which gives two eigenenergies
\begin{alignat}1
\varepsilon_\pm^2=3-\cos k_2\pm\sqrt{3\sin^2k_2+2[1+\cos(k_2+\phi)]}.
\label{GraQheZig1o3}
\end{alignat}
On the other hand, the zero energy state is determined by ${\cal D}_{\rm e}^\dagger\zeta_0=0$, from which
one finds $\zeta_0\propto(1,-g_{\phi-k_2}^*,g_{\phi-k_2}^*g_{-\phi-k_2}^*)^T$. 
Therefore, the localization condition (\ref{LocCon0}) is 
\begin{alignat}1
|e^{iK_0}|=|(1+e^{-ik_2})(1+e^{i(\phi-k_2)})(1+e^{-i(\phi+k_2)})|<1.
\end{alignat}
It follows that around three points $k_2\sim\pi,\pm \pi/3$, the zero energy state appears.
We show in Fig. \ref{f:graphene_QHE_2} (b) the energies (\ref{GraQheZig1o3}) and zero energies by red curves,
whose localization conditions are numerically computed.
One sees that half of the edge states are well reproduced, which are localized at the left end.

\subsubsection{Bearded edge}\label{s:graphene_mag_beard}

\begin{figure}[htb]
\begin{center}
\begin{tabular}{c}
\includegraphics[width=.7\linewidth]{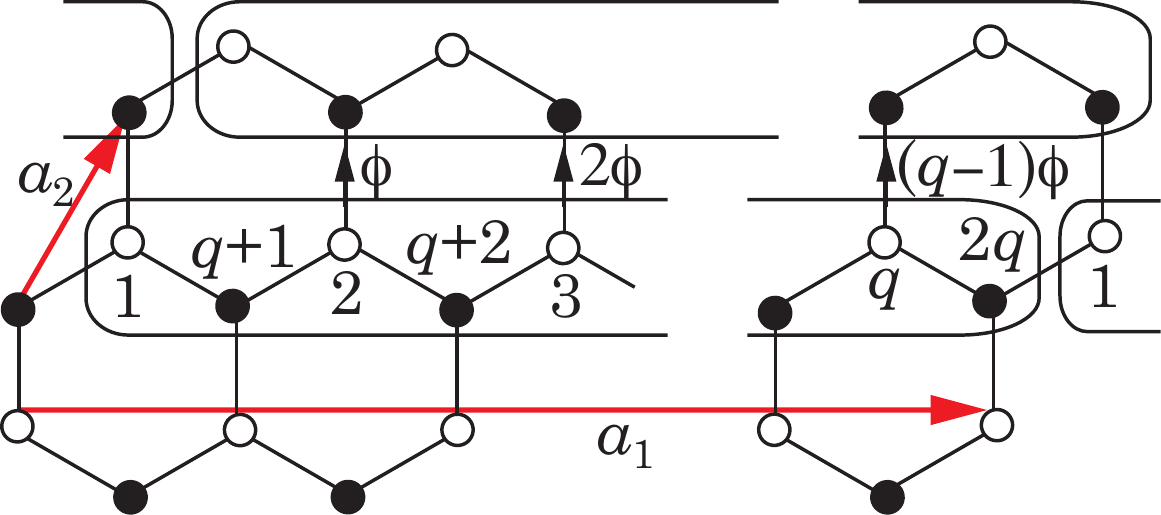}
\end{tabular}
\caption{
Magnetic unit cell suitable 
for graphene with the bearded edge under a uniform magnetic flux $\phi/(2\pi)=p/q$.
}
\label{f:graphene_qhe_beard_lat}
\end{center}
\end{figure}

Let us next consider the bearded edge in the presence of a uniform magnetic flux.
When we choose the magnetic unit cell as illustrated in Fig. \ref{f:graphene_qhe_beard_lat}, we have 
the Hamiltonian Eq. (\ref{GraHamOpe}) by use of the following new operator $\hat{\cal D}$
\begin{alignat}1
\hat{\cal D}&=
t\left(
\begin{array}{cccccc}
1&&&&&\delta_1^*(1+\delta_2)\\
1+\hat h_{\phi}^\dagger&1&&&&\\
&1+\hat h_{2\phi}^\dagger&1&&\\
&&\ddots&&&\\
&&&&1&\\
&&&&1+\hat h_{(q-1)\phi}^\dagger&1\\
\end{array}
\right).
\end{alignat}
Thus, the $\cal K$ matrix is basically the same as the zigzag edge, although the single nonzero component 
depends on $k_2$.
Therefore, from a similar eigenvalue equation to 
Eq. (\ref{GraQheEeq}), we find the Hamiltonian of the edge states at the left end is basically 
the same structure as Eq. (\ref{GraQheZigHam}),
\begin{alignat}1
\begin{array}{cc}
\vspace*{-1mm}\hspace*{17mm}\overbrace{\scriptscriptstyle}^{ q-1}&\\ 
{\cal H}_{\rm e}^{\rm L}=\Bigg(\begin{array}{cc}
& {\cal D}_{\rm e}^{\rm L}\\
{\cal D}_{\rm e}^{\rm L\dagger}&
\end{array}
\Bigg) &\hspace*{-4mm}
\begin{array}{l}\big\}\scriptstyle\,q\\~\end{array}
\vspace*{5mm}
\end{array} ,
\label{GraQheBeaHamL}
\end{alignat}
where $q\times(q-1)$ matrix $\cal D_{\rm e}^{\rm L}$ is in this case defined by
\begin{alignat}1
{\cal D}_{\rm e}^{\rm L}&=
t\left(
\begin{array}{ccccc}
1&&&&\\
g_{k_2-\phi}&1&&&\\
&g_{k_2-2\phi}&1&&\\
&&&\ddots&\\
&&&&1\\
&&&&g_{k_2-(q-1)\phi}\\
\end{array}
\right).
\end{alignat}
This is the ESH at the left end. Here, in Eq. (\ref{GraQheBeaHamL}), we have 
indicated the numbers of row and columns explicitly to compare the ESH at the right end below.
The localization condition is
\begin{alignat}1
|e^{iK_n}|=\left|\frac{\chi_{q,n}}{g^*_{k_2}\chi_{1,n}}\right|<1.
\end{alignat}

Next, let us consider the right edge states. In the same way as in Sec. \ref{s:right_edge},
we choose the reference states as in Eq. (\ref{HofEquR}).
Assuming a Bloch-type wave function for the edge states $\psi_{jn}=\psi_0e^{iK_nj}$,
we find the following Schr\"odinger equation,
\begin{widetext}
\begin{alignat}1
&t\left(
\begin{array}{c|ccccc|cccccc}
&&&&&&1&&&&&e^{-iK_n}g_{k_2}\\
\hline
&&&&&&g_{k_2-\phi}&1&&&&\\
&&&&&&&g_{2\phi-k_2}&1&&&\\
&&&&&&&&&\ddots&&\\
&&&&&&&&&&1&\\
&&&&&&&&&&g_{k_2-(q-1)\phi}&1\\
\hline
1&g_{k_2-\phi}^*&&&&&&&&&&\\
&1&g_{k_2-2\phi}^*&&&&&&&&&\\
&&1&&&&&&&&&\\
&&&\ddots&&&&&&&&\\
&&&&1&g_{k_2-(q-1)\phi}^*&&&&&&\\
e^{iK_n}g_{k_2}^*&&&&&1&&&&&&\\
\end{array}
\right)
\left(\begin{array}{c}0\\\hline\\ \\ \\ \\ \\ \chi_{n}\\ \\ \\ \\  \\  \\  \\  \end{array}\right)
=\varepsilon_n
\left(\begin{array}{c}0\\\hline\\ \\ \\ \\ \\ \chi_{n} \\ \\ \\ \\  \\  \\  \\  \end{array}\right).
\label{GraQheBeaEeq}
\end{alignat}
\end{widetext}
This equation, from its lower part, leads to the following ESH at the right end,
\begin{alignat}1
\begin{array}{cc}
\vspace*{-1mm}\hspace*{17mm}\overbrace{\scriptscriptstyle}^{ q}&\\ 
{\cal H}_{\rm e}^{\rm R}=\Bigg(\begin{array}{cc}
& {\cal D}_{\rm e}^{\rm R}\\
{\cal D}_{\rm e}^{\rm R\dagger}&
\end{array}
\Bigg) &\hspace*{-4mm}
\begin{array}{l}\big\}\scriptstyle\,q-1\\~\end{array}
\end{array} ,
\label{GraQheBeaHamR}
\end{alignat}
with $(q-1)\times q$ matrix $\cal D_{\rm e}^{\rm R}$ defined by
\begin{alignat}1
{\cal D}_{\rm e}^{\rm R}&=
t\left(
\begin{array}{cccccc}
g_{k_2-\phi}&1&&&&\\
&g_{k_2-2\phi}&1&&&\\
&&&\ddots&&\\
&&&&1&\\
&&&&g_{k_2-(q-1)\phi}&1\\
\end{array}
\right).
\label{GraQheBeaHamR2}
\end{alignat}
The localization condition, which is the equation of the first component in Eq. (\ref{GraQheBeaEeq}),
becomes
\begin{alignat}1
|e^{-iK_n}|=e^{-\kappa_n}=\left|\frac{\chi_{q,n}}{g_{k_2}\chi_{2q-1,n}}\right|<1.
\label{GraQheBeaHamR3}
\end{alignat}

\begin{figure}[htb]
\begin{center}
\begin{tabular}{cc}
\includegraphics[width=.5\linewidth]{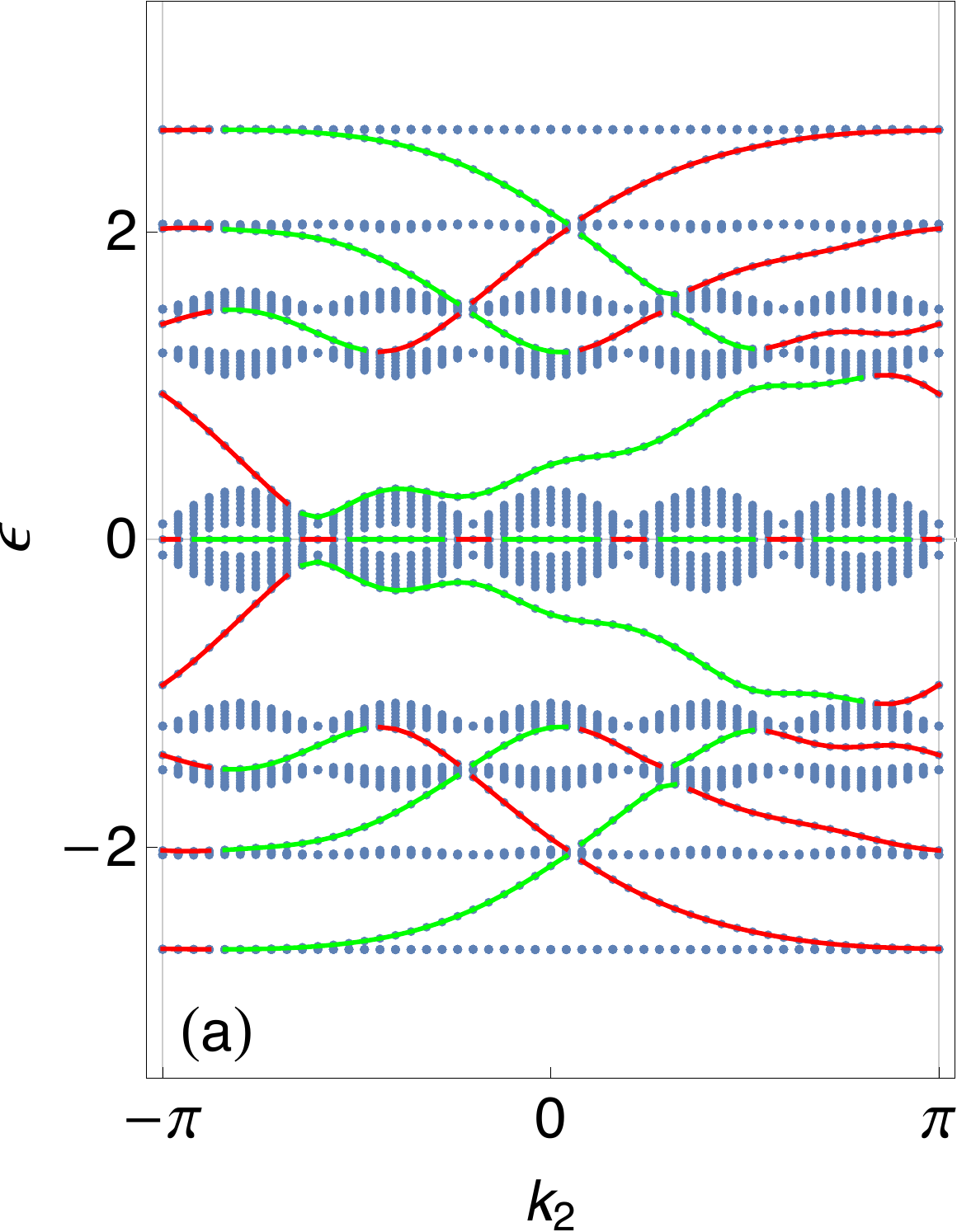}
&
\includegraphics[width=.5\linewidth]{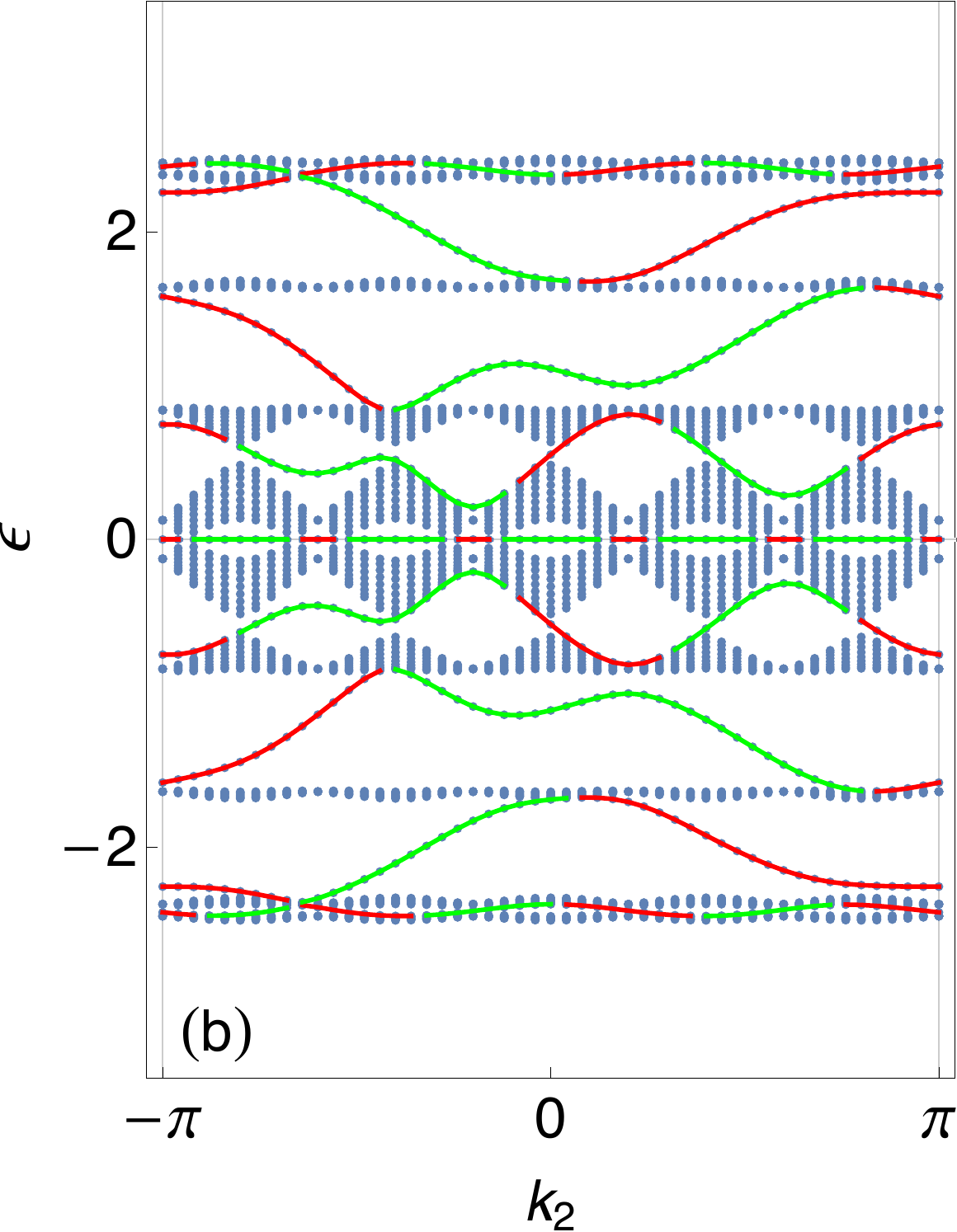}
\end{tabular}
\caption{
Spectra of the graphene model with the zigzag edge at the left end and bearded edge at 
the right end under magnetic flux (a) $\phi/(2\pi)=1/5$ and (b) $2/5$.
The red and green curves show the spectra of the Hamiltonians (\ref{GraQheZigHam}) and (\ref{GraQheBeaHamR}), 
respectively, whereas 
the background gray dots represent the total eigenvalues of the system of 
the cylindrical geometry with the zigzag and bearded edges.
}
\label{f:graphene_QHE_zb}
\end{center}
\end{figure}

So far we have derived the two ESHs with the bearded edges, one is at the left end and the other is at the
right end. In the previous Sec. \ref{s:graphene_mag_zig}, we have also derived two ESHs with the zigzag edges.
These are independent, since we have considered a half-plane which has only one end to derive each ESH.
In order to check this explicitly, we introduce a model with the zigzag edge at the left end and the bearded edge
at the right end, which was studied in Ref. \cite{Hatsugai:2006aa}.
Figure \ref{f:graphene_QHE_zb} shows various spectra of such a model.
The background dots are total energy eigenvalues of the cylindrical system with the zigzag and bearded edges.
The red curves are eigenvalues of the left ESH for the zigzag edge 
defined by  Eqs. (\ref{GraQheZigHam}) and (\ref{GraQheZigHam2}) with the localization condition Eq. (\ref{GraQheMu}),
whereas the green curves are those of the right ESH for the bearded edge
defined by Eqs. (\ref{GraQheBeaHamR}) and (\ref{GraQheBeaHamR2}) with the localization condition (\ref{GraQheBeaHamR3}).
We see that the red and green curves indeed reproduce the edge states of the graphene model with the hybrid edges 
at two ends.

\subsubsection{Armchair edge}\label{s:graphene_mag_arm}

\begin{figure}[htb]
\begin{center}
\begin{tabular}{c}
\includegraphics[width=.9\linewidth]{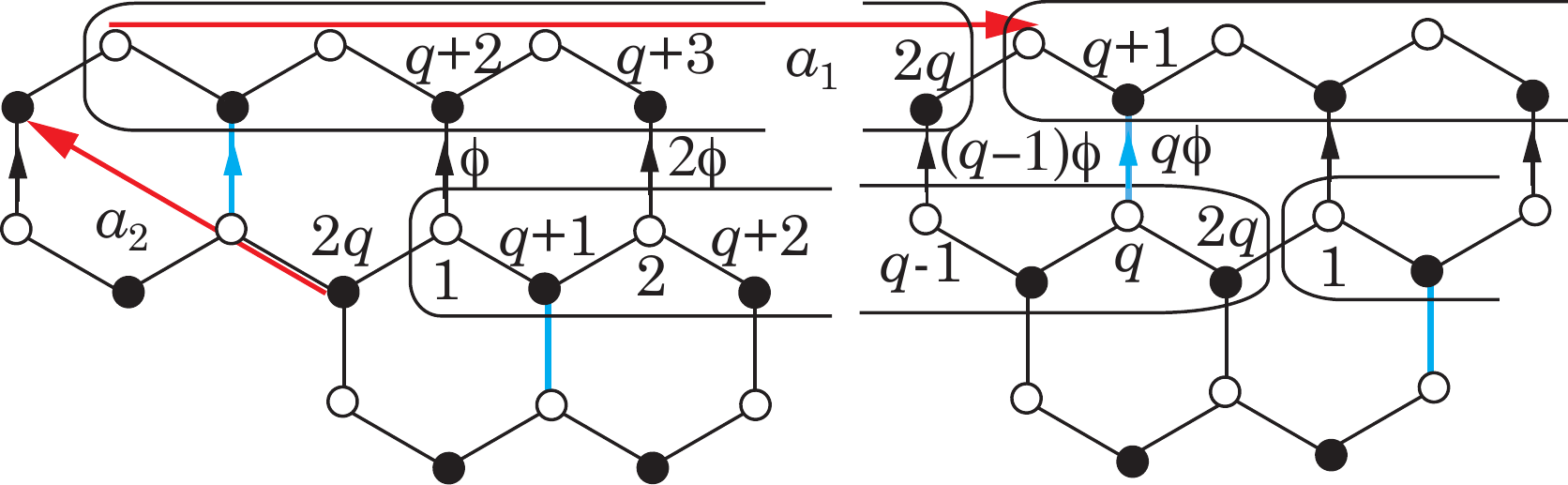}
\end{tabular}
\caption{
Magnetic unit cell suitable for the graphene  with the armchair edge 
under a uniform magnetic flux $\phi/(2\pi)=p/q$.
Blue bonds denote the hoppings with amplitude $rt$. 
When $r=0$, the model belongs to A.
}
\label{f:graphene_arm_qhe_lat}
\end{center}
\end{figure}

Irrespective of the presence or absence of a magnetic field,
the choices of the unit cell for the zigzag and bearded edges lead to the Hamiltonian operators with NN hoppings only.
This enables us to extract the exact ESHs at the left and right ends separately.
In the case of the armchair edge, on the other hand, 
the model belongs to class B or C in Sec. \ref{s:class}, since 
the Hermiticity matrix ${\cal K}$ has two nonzero matrix elements.
In the absence of a magnetic field, the model is in class B. 
Indeed, we have a unique matrix $\cal A$ that anti-commutes with $\cal K$. 
As a consequence, we have been able to  solve the eigenvalue equation with the Hermiticity condition imposed. 
However, if a magnetic field is introduced, the problem is much more difficult.
One of the reasons is that the matrix $\cal A$  is not unique. Even if we treat the model as the one in class C, 
it is quite difficult to carry out numerical search for solutions in a huge parameter space.

In this section, we provide a different perspective on the edge states at the armchair edge.
Let us try to deform the model so that the Hamiltonian 
belongs to class A without changing  the edge.
If such a deformation is far from the edge, edge states are not affected so much.
Of course, such a deformation would modify the topological properties of the original model. 
Nevertheless, it could be useful to reproduce the  edge state bands located at one end using a simple ESH.
In this section, we propose such an attempt. 

The Hamiltonian operator with the armchair edge at the left end 
is the same chiral structure as Eq.  (\ref{GraHamOpe}), where
$\hat{\cal D}$ operator in the present choice of the magnetic unit cell in Fig. \ref{f:graphene_arm_qhe_lat} is given by
\begin{alignat}1
\hat{\cal D}=t
\left(
\begin{array}{cccccc}
1&\hat h^\dagger_\phi&&&&\delta_1^*\\
1&1&\hat h^\dagger_{2\phi}&&&\\
&1&1&\hat h^\dagger_{3\phi}&&\\
&&&&\ddots&\\
&&&&1&\hat h^\dagger_{(q-1)\phi}\\
r\delta_1\delta_2&&&&1&1
\end{array}
\right),
\end{alignat}
with $r=1$ for the uniform honeycomb lattice. 
As in the case of zero field, this model cannot be a simple NN model 
classified as A, since the ${\cal K}$ matrix is 
\begin{alignat}1
{\cal K}=t\left(
\begin{array}{ccc|ccc}
&&&&&1\\
&&&&0& \vspace*{-2.5mm}\\
&&&\iddots&& \\
\hline
&&r e^{-ik_2}&&&\\
&0&&&& \vspace*{-2.5mm} \\
\iddots&&&&&
\end{array}
\right).
\end{alignat}
It follows that if one sets $r=0$, the above $\cal K$ matrix reduces to the one for the zigzag edge (\ref{KforZig}), and 
the edge states can be obtained in the same way as the zigzag and bearded edges.
In this case, $r=0$, it is easy to see that the ESH becomes Eq. (\ref{GraQheZigHam}) with
\begin{alignat}1
{\cal D}_{\rm e}=t
\left(
\begin{array}{ccccc}
1&e^{-(k_2-\phi)}&&&\\
1&1&e^{i(k_2-2\phi)}&&\\
&1&1&\\
&&&\ddots&\\
&&&&1\\
&&&&1
\end{array}
\right).
\label{ArmEdgD}
\end{alignat}

\begin{figure}[htb]
\begin{center}
\begin{tabular}{cc}
\includegraphics[width=.5\linewidth]{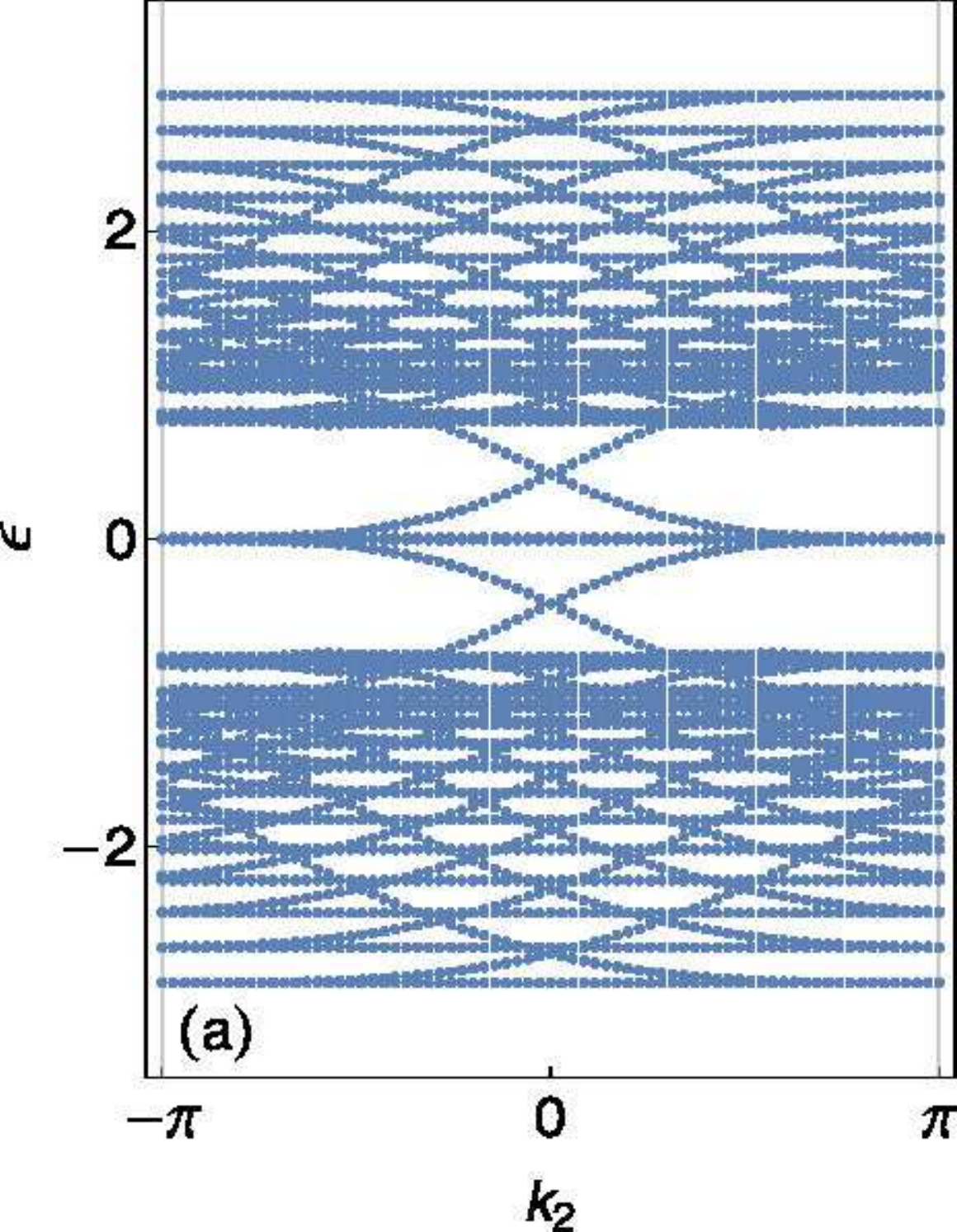}
&
\includegraphics[width=.5\linewidth]{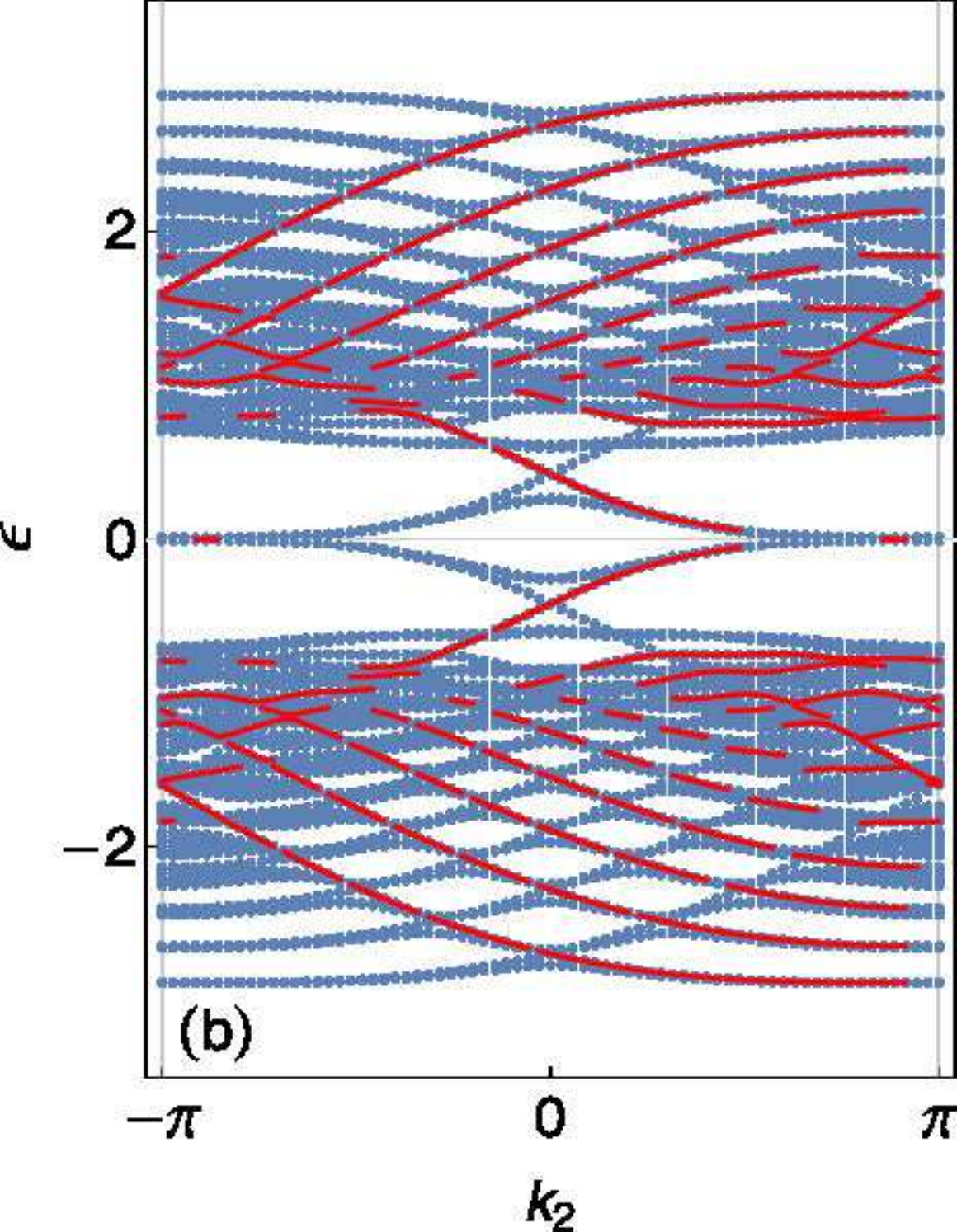}
\\
\includegraphics[width=.5\linewidth]{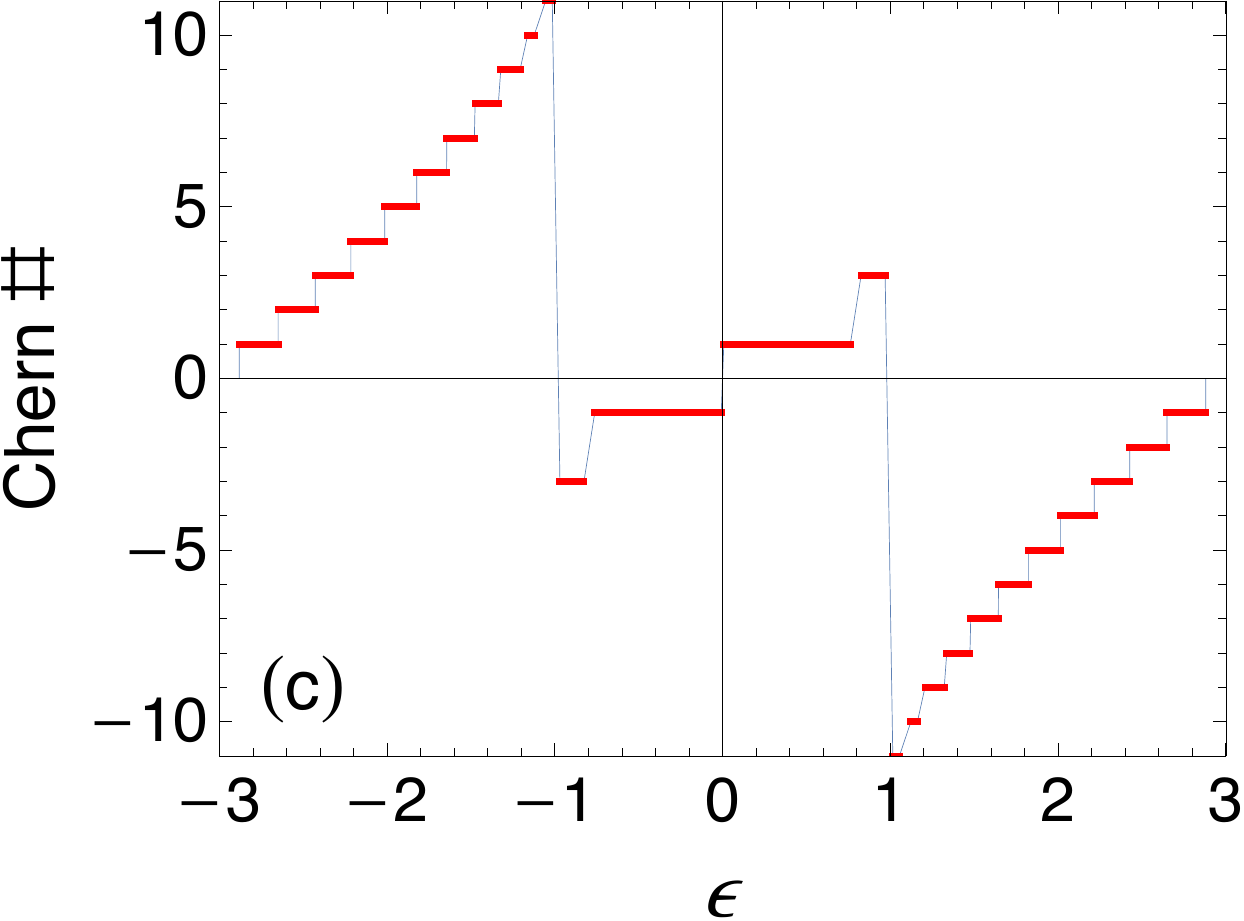}
&
\includegraphics[width=.5\linewidth]{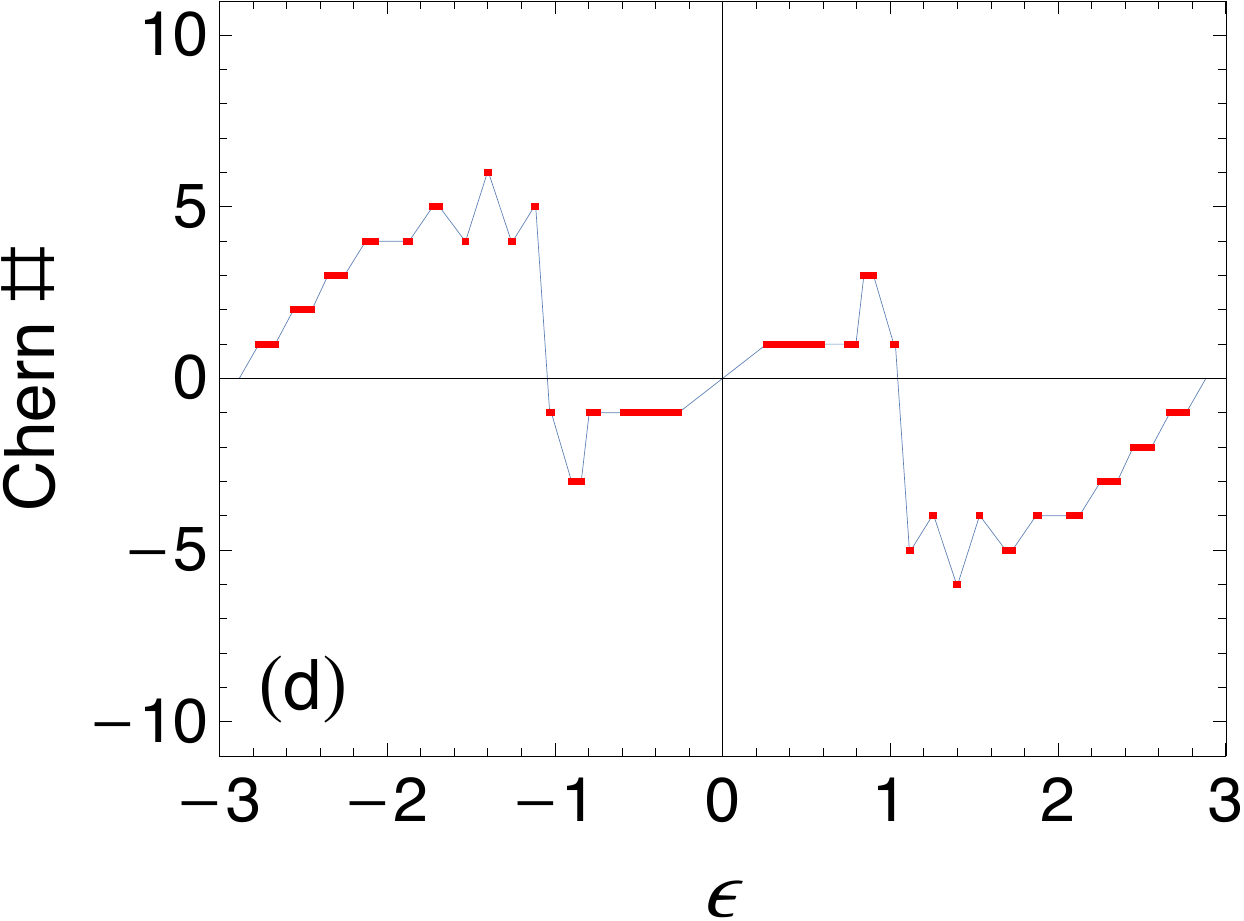}\end{tabular}
\caption{
(a) shows spectrum of the graphene model with the armchair edge
under magnetic flux $\phi/(2\pi)=1/15$. (b) is the same as (a) but with defects ($r=0$) as described in 
the text. The red curves are eigenvalues of the ESH Eq. (\ref{GraQheZigHam}) with (\ref{ArmEdgD}).
(c) and (d) are Chern numbers (Hall conductivities) as a function of the Fermi energy $\varepsilon$.
}
\label{f:graphene_QHE_arm}
\end{center}
\end{figure}

In Fig. \ref{f:graphene_QHE_arm}, we show spectra of graphene with the armchair edges.
Fig. \ref{f:graphene_QHE_arm} 
(a) is the full spectrum of the uniform system with $r=1$ of cylindrical geometry, which is compared with
(b) of the system with $r=0$. These two systems are different topological band structures, as can be seen from 
Figs. \ref{f:graphene_QHE_arm} (c) and (d), which are the Chern numbers 
as functions of the fermi energy \cite{Sheng:2006aa,FHS05} computed according to Ref. \cite{FHS05}.
Nevertheless, the behavior of the edge states is rather similar. This is because for large-$q$ system, the left end and
the defects at $r=0$ bonds are well separated, so that edge states localized at the end and the impurities states
localized at defects are almost decoupled. 
Although these edge states cannot be used for the discussion of the topological arguments such as the bulk-edge 
correspondence, they may be helpful, e.g.,  to determine at which end the edge states of the uniform system are 
localized, etc.

\begin{figure}[htb]
\begin{center}
\begin{tabular}{cc}
\includegraphics[width=.5\linewidth]{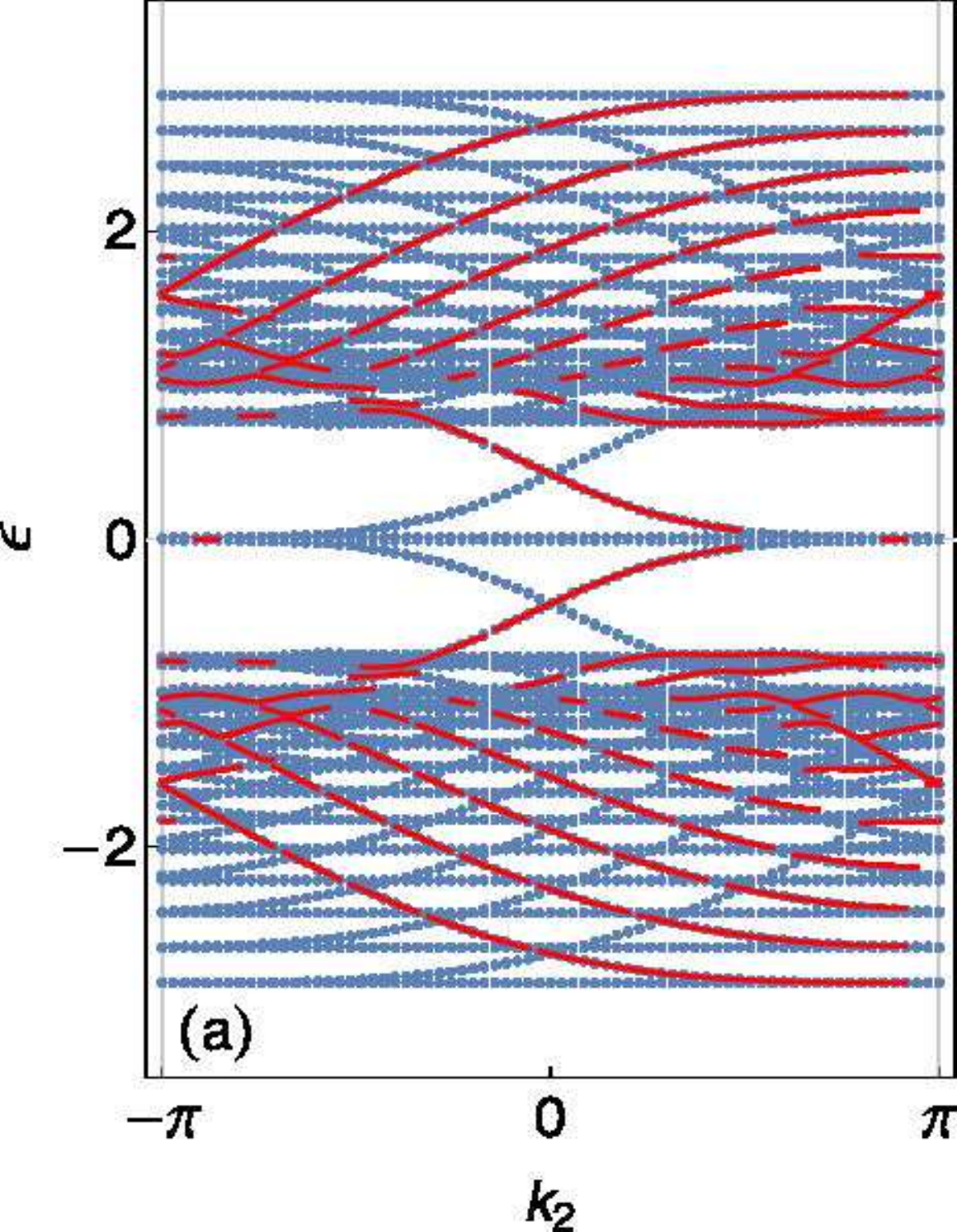}
&
\includegraphics[width=.5\linewidth]{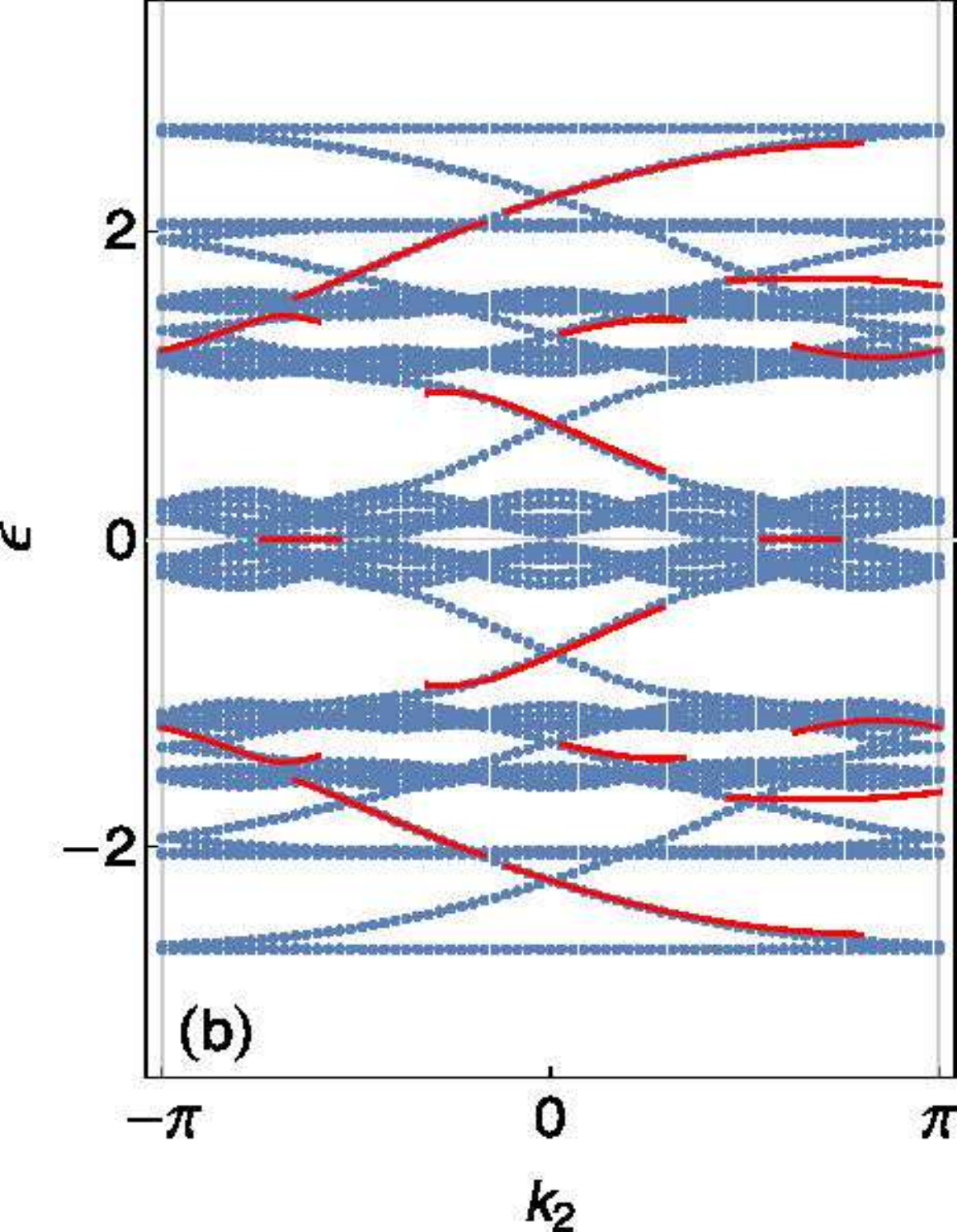}
\end{tabular}
\caption{
(a) Total spectra Fig. \ref{f:graphene_QHE_arm} (a) overwritten by
edge state spectra (red curves) in Fig. \ref{f:graphene_QHE_arm} (b).
(b) The same as (a) in the case of $\phi/(2\pi)=1/5$.
}
\label{f:graphene_QHE_arm_syn}
\end{center}
\end{figure}

In fact, it may be interesting to overlay the edge states (the red curves) of the defect model in (b) and 
the full spectrum of the uniform system in (a), which is carried out in Fig, \ref{f:graphene_QHE_arm_syn} (a).
Of course, the red curves include impurity states, for example, those at zero energy 
near $k_2\sim\pm\pi$ are localized states at the 
defects of the $r=0$ bonds. Nevertheless, the overall spectrum of the edge states is well reproduced.
On the other hand, for small-$q$ systems, one expects that the coupling between the edge states and impurity states  
becomes stronger. In Fig.  \ref{f:graphene_QHE_arm_syn} (b), we show the model with flux $\phi/(2\pi)=1/5$.
One can indeed observe slight deviation of the red curves from the background spectrum, as expected.

\section{Wilson-Dirac model}\label{s:wd}

So far we have investigated 2D models mainly in class A defined in Sec. \ref{s:class}.
In this section, we study a simple example, Wilson-Dirac model,  
in class B which allows some matrix ${\cal A}$ anti-commuting with the 
Hermiticity matrix ${\cal K}$.

The Wilson-Dirac model is a typical model whose edge states have been obtained analytically 
\cin{\cite{Konig:1977fk,Mao:2010aa,Dwivedi:2016aa,Cobanera:2018aa}}.
The Hamiltonian operator for the Wilson-Dirac model is given by
\begin{alignat}1
\hat{\cal H}=&
\sum_{\mu=1,2}\left[
\frac{it}{2}\sigma^\mu(\delta^*_\mu-\delta_\mu)+\frac{b}{2}\sigma^3(\delta_\mu^*+\delta_\mu-2)\right]
\nonumber\\
\rightarrow&\frac{it}{2}\sigma^1(\delta^*-\delta)+\frac{b}{2}\sigma^3(\delta^*+\delta)
\nonumber\\
&+t\sigma^2\sin k_2+\sigma^3\left[m+b(\cos k_2-1)-b\right],
\end{alignat}
where in the second line $j_2$ has been Fourier-transformed by $\delta_2\rightarrow e^{ik_2}$, 
and $\delta\equiv\delta_1$ acts on $j\equiv j_1$.
Without loss of generality, we assume $t>0$.
Then, from the coefficients of $\delta^*$, we find 
\begin{alignat}1
{\cal K}=\frac{it}{2}\sigma^1+\frac{b}{2}\sigma^3,
\label{KMatWD}
\end{alignat}
For this $\cal K$ matrix, one finds $\{{\cal K},\sigma^2\}=0$, so that this model is classified into B.
Actually, the $\cal K$ matrix (\ref{KMatWD}) is the unitary equivalent to that of the SSH model in the case of $t\ne 0$ 
in Eq. (\ref{HerConBlo}). 
Thus, one can choose the reference state as 
$\psi_0=\psi_{\pm}$, where $\psi_{\pm}$ is the eigenstates of $\sigma^2$, $\sigma^2\psi_\pm=\pm\sigma_\pm$.
Using these states, let us introduce a Bloch-type wave function for the edge states, 
$\psi_j=\psi_\pm e^{i Kj}$, where $K=k+i\kappa$.
Here, it should be noted that for a given model parameters, either one of $\psi_\pm$ can be the wave function,
provided that it allows two independent solutions, $K_1$ and $K_2$, for the boundary condition (\ref{BouCon})
to be satisfied.
If both $\psi_\pm$ do not allow two solutions, one concludes that the model has no edge states.
Substituting a Bloch-type wave function into the eigenvalue equation, $\hat{\cal H}\psi_j=\varepsilon\psi_j$, 
one has
\begin{alignat}1
&\left[\frac{\mp t}{2}(e^{-iK}-e^{iK})+\frac{b}{2}(e^{-iK}+e^{iK})+bM(k_2)\right]\psi_\mp
\nonumber\\
&\qquad\pm t\sin k_2\psi_\pm=\varepsilon\psi_\pm,
\end{alignat}
where $M(k_2)=\cos k_2-2+\frac{m}{b}$, and the energy quantum number $n$ has been suppressed.
One sees that $\psi_\pm$ is indeed the eigenstate of $\hat{\cal H}$ with energy eigenvalue 
$\varepsilon=\pm t \sin k_2$, provided that
\begin{alignat}1
\frac{\mp t}{2}(e^{-iK}-e^{iK})+\frac{b}{2}(e^{-iK}+e^{iK})+bM(k_2)=0.
\label{WDFli}
\end{alignat}
The imaginary parts of the above equation yields
\begin{alignat}1
k=0, \pi, e^{-2\kappa}=\frac{b\mp t}{b\pm t} .
\label{WDIma}
\end{alignat}
In what follows, we calculate the case $\psi_0=\psi_+$ (the case of upper sign in the above equations).
Let us substitute the solutions (\ref{WDIma}) into the real part of Eq. (\ref{WDFli}).
Then, one has
\begin{alignat}1
&(b+t)e^{-\kappa}+(b-t)e^{\kappa}\pm bM(k_2)=0,
\nonumber\\
&\left[(b+t)\sqrt{\frac{b- t}{b+ t}}+(b- t)\sqrt{\frac{b+ t}{b- t}}\right]\cos k+2bM(k_2)=0,
\end{alignat}
where $\pm$ in the upper equation corresponds to $k=0,\pi$.
It then turns out that
there exist two solutions when $|M(k_2)|<1$ for $0<b$ and when $|M(k_2)|<\sqrt{1-(t/b)^2}$ for $b<-t$,
and otherwise, two solutions are not allowed.
Likewise, in the case of $\psi_0=\psi_-$, an edge state exist  when $|M(k_2)|<1$ for $b<0$ and 
when $|M(k_2)|<\sqrt{1-(t/b)^2}$ for $t<b$.
As a consequence, edge states exist when $|M(k_2)|<1$, and 
the condition that there exists $k_2$ that satisfies $|M(k_2)|<1$  becomes
\begin{alignat}1
0<\frac{m}{b}<2,\quad 2<\frac{m}{b}<4.
\end{alignat}
This determines the topological phase of the Wilson-Dirac model.

Moreover, from the types of $K$, one can get more information on the wave function of the edge state. 
In addition to $|M(k_2)|<1$, further conditions below classify the types of $K$ as follows: 
For $|b|<t$, the two solutions $K_{1,2}$ are of the types
\begin{alignat}1
&M(k_2)>0,\quad K_{1,2}=i\kappa_{1,2},
\nonumber\\
&M(k_2)<0,\quad K_{1,2}=\pi+i\kappa_{1,2},
\label{WDCla1}
\end{alignat}
and for $|b|>t$, 
\begin{alignat}1
&|M(k_2)|<\sqrt{1-\left(t/b\right)^2},\quad K_{1,2}=k_{1,2}+i\kappa,
\nonumber\\ 
&M(k_2)>\sqrt{1-\left(t/b\right)^2},\quad K_{1,2}=i\kappa_{1,2},
\nonumber\\
&M(k_2)<-\sqrt{1-\left(t/b\right)^2}, \quad K_{1,2}=\pi+i\kappa_{1,2}.
\end{alignat}
\begin{figure}[htb]
\begin{center}
\begin{tabular}{cc}
\includegraphics[width=.5\linewidth]{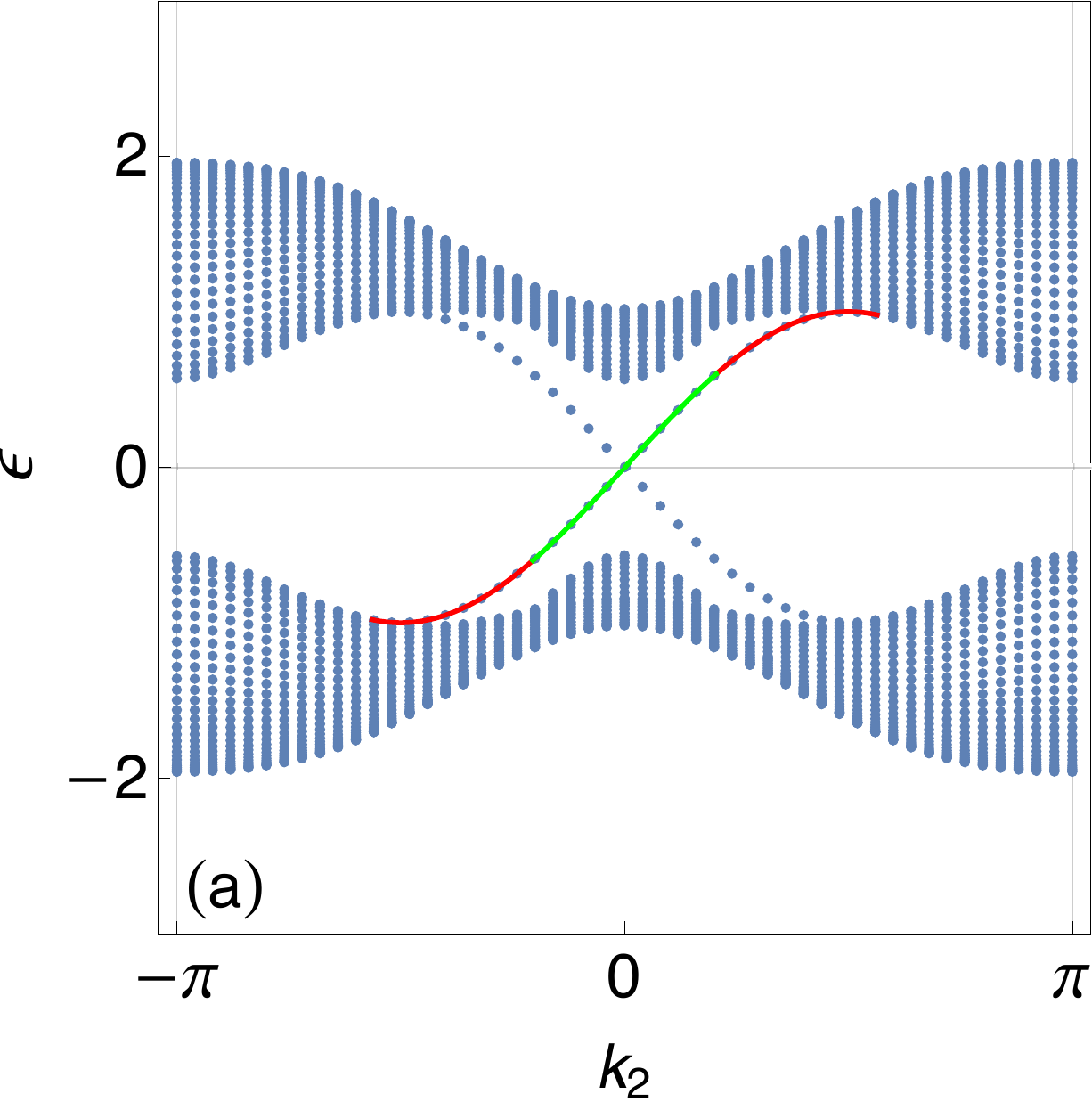}
&
\includegraphics[width=.5\linewidth]{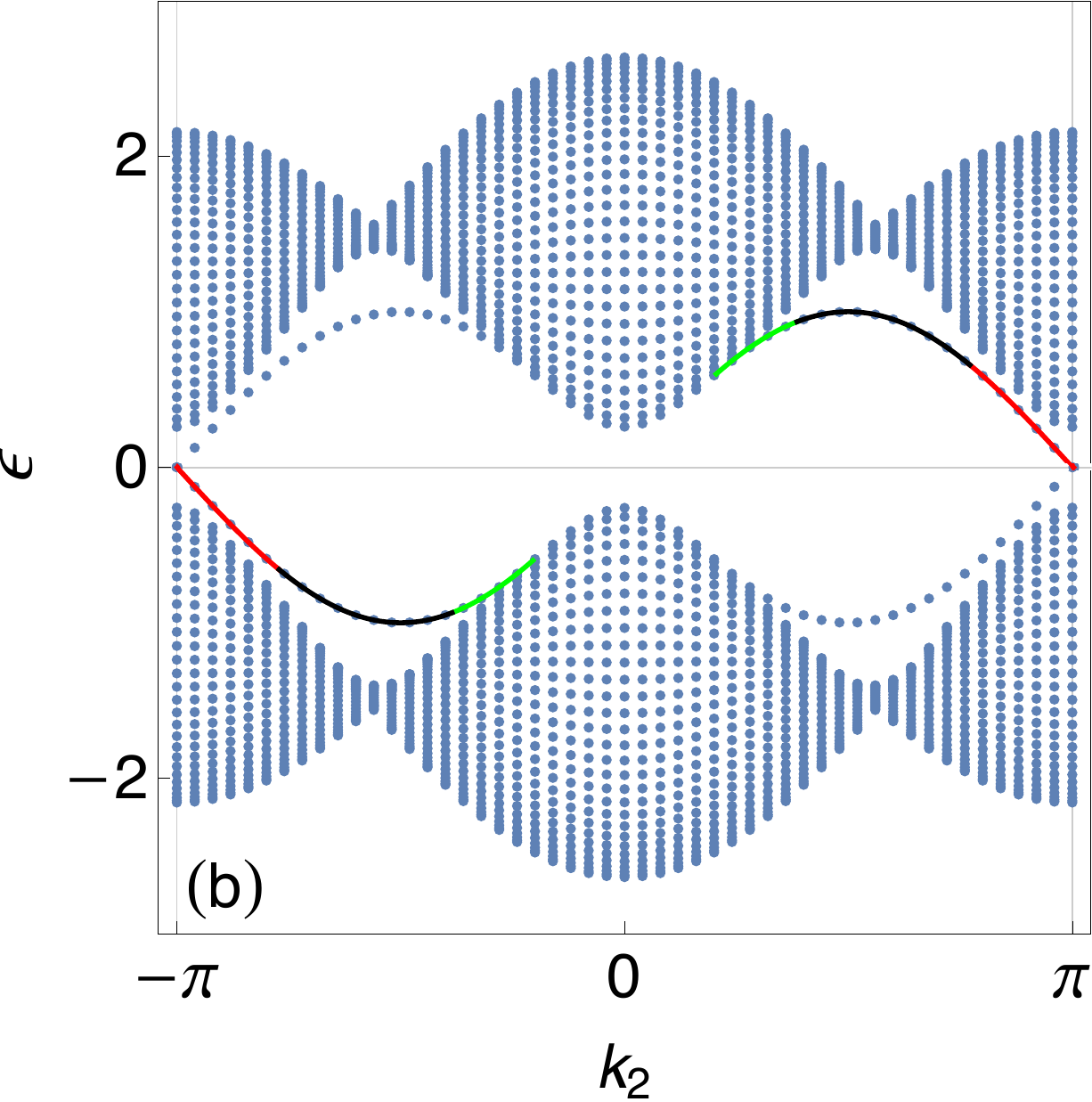}
\end{tabular}
\caption{
Plot of the edge state energies $\varepsilon=t\sin k_2$ satisfying $|M(k_2)|<1$.
Red, green, black curves mean $K=i\kappa_{1,2}$, $K=\pi+i\kappa_{1,2}$, and $K=k_{1,2}+i\kappa$ type
edge states, which are plotted on the total spectra of cylindrical geometry.
(a) $b/t=0.7$ and $m/b=1.2$.
(b) $b/t=1.2$ and $m/b=2.2$.
}
\label{f:wd}
\end{center}
\end{figure}
For example, according to Eq. (\ref{WDCla1}), when $0<b<t$ and for $k_2$ that satisfies $M(k_2)<0$, 
the wave function of the edge state is
\begin{alignat}1
\psi_j=\psi_+(-)^j(e^{-\kappa_1j}-e^{-\kappa_2j}) ,
\end{alignat}
whereas $k_2$ that satisfies $M(k_2)>0$,
\begin{alignat}1
\psi_j=\psi_+(e^{-\kappa_1j}-e^{-\kappa_2j}) .
\end{alignat}
In Fig. \ref{f:wd}, we show some examples that have different types of edge states dependent on $k_2$ 
in one model.


\section{Haldane model}\label{s:haldane}

In this section, we study the Haldane model as an example in class C in Sec.\ref{s:class}.
For practical purpose, the present method may not be very useful in order to obtain the edge states in class C. 
Nevertheless, we would like to claim that the Hermiticity of Hamiltonians would determine the edge states
in principle.

\begin{figure}[htb]
\begin{center}
\begin{tabular}{c}
\includegraphics[width=.35\linewidth]{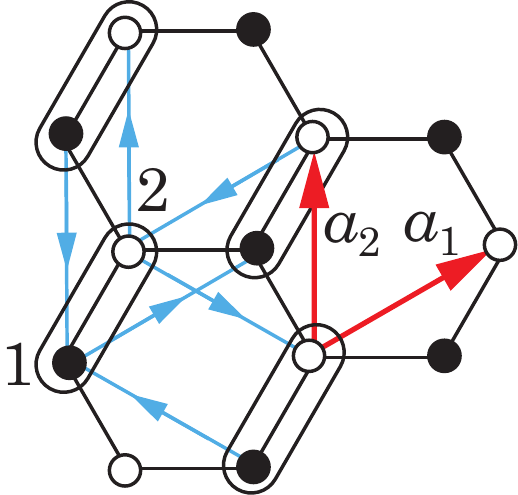}
\end{tabular}
\caption{
Haldane model on the honeycomb lattice. The black and blue bonds show 
NN and NNN hoppings with amplitudes $t$ and $t'$, respectively.
Blue arrows mean the phase factor $e^{i\phi}$ attached to $t'$.
On the sites labeled by 1 and 2, a staggered potential, $\pm m$, is introduced.
}
\label{f:haldane_lat}
\end{center}
\end{figure}

From the unit cell defined in Fig. \ref{f:haldane_lat},   
the Hamiltonian operator reads
\begin{alignat}1
\hat{\cal H}=
\left(
\begin{array}{cc}
\hat{\cal H}_{1}+m& t(1+\delta_1^*+\delta_2^*)\\
t(1+\delta_1+\delta_2)&\hat{\cal H}_{2}-m
\end{array}
\right), 
\label{HalHamOpe}
\end{alignat}
where $\hat{\cal H}_{1,2}$ denote the NNN hopping operators defined by
\begin{alignat}1
&\hat{\cal H}_{1}=t'(e^{-i\phi}\delta_1+e^{i\phi}\delta_2+e^{i\phi}\delta_1\delta_2^*+h.c),
\nonumber\\
&\hat{\cal H}_{2}=t'(e^{i\phi}\delta_1+e^{-i\phi}\delta_2+e^{-i\phi}\delta_1\delta_2^*+h.c).
\end{alignat}
In order to find the edge states localized near the left end $j_1=1$, the 2-direction is 
Fourier-transformed to obtain
\begin{alignat}1
\hat{\cal H}=
\left(
\begin{array}{cc}
m_1+t_1\delta+t_1^*\delta^*& t(1+e^{-ik_2}+\delta^*)\\
t(1+e^{ik_2}+\delta)&m_2+t_2\delta+t_2^*\delta^*
\end{array}
\right), 
\label{HalHamOpe2}
\end{alignat}
where $\delta\equiv\delta_1$ operates to $j\equiv j_1$ and
\begin{alignat}1
&m_1=m+2t'\cos(k_2+\phi),
\nonumber\\
&m_2=-m+2t'\cos(k_2-\phi),
\nonumber\\
&t_1=t'(e^{-i\phi}+e^{-i(k_2-\phi)}),
\nonumber\\
&t_2=t'(e^{i\phi}+e^{-i(k_2+\phi)}).
\end{alignat}
The coefficient of $\delta^*$ leads to the following ${\cal K}$ matrix,
\begin{alignat}1
{\cal K}=\left(
\begin{array}{cc}
t_1^*& t\\
0&t_2^*
\end{array}
\right).
\end{alignat}
This matrix never anti-commutes with any other nontrivial matrix, so that the model 
belongs to class C.
Thus, there is no simple way to choose $\psi_0$; rather, it should be determined using the Hermiticity condition.
To be more specific, we first set the wave function of the edge state $\psi_j=\psi_{0n} e^{iK_nj}$ ($K_n=k_n+i\kappa_n$)
and require the Hermiticity
\begin{alignat}1
\psi_{n0}^\dagger {\cal K}\psi_{0n}=t_1^*+t\chi_n+t_2^*|\chi_n|^2=0,
\label{HalHerCon}
\end{alignat}
where we have set $\psi_{0n}=(1,\chi_n)^T$ for simplicity.
Under this condition, we solve the eigenvalue equation
\begin{alignat}1
&\left(
\begin{array}{cc}
m_1+t_1e^{iK_n}+t_1^*e^{-iK_n}& t(1+e^{-ik_2}+e^{-iK_n})\\
t(1+e^{ik_2}+e^{iK_n})&m_2+t_2\delta_1+t_2^*e^{-iK_n}
\end{array}
\right)
 \left(\begin{array}{c}1\\ \chi_n \end{array}\right)
 \nonumber\\
& \qquad=\varepsilon_n 
 \left(\begin{array}{c}1\\ \chi_n \end{array}\right).
\label{HalEigEqu}
\end{alignat}
Thus, we have three equations (\ref{HalHerCon}) and (\ref{HalEigEqu}) for three unknown parameters $K$, $\chi$, and $\varepsilon$. 
When these equations have degenerate two solutions with $|e^{iK_n}|=e^{-\kappa_n}<1$, we can obtain the wave functions 
satisfying the boundary condition.
Since the equations to be solved are complex-valued nonlinear equations, it is not so easy to determine whether they allow 
two independent solutions or not.  

\begin{figure}[htb]
\begin{center}
\begin{tabular}{cc}
\includegraphics[width=.5\linewidth]{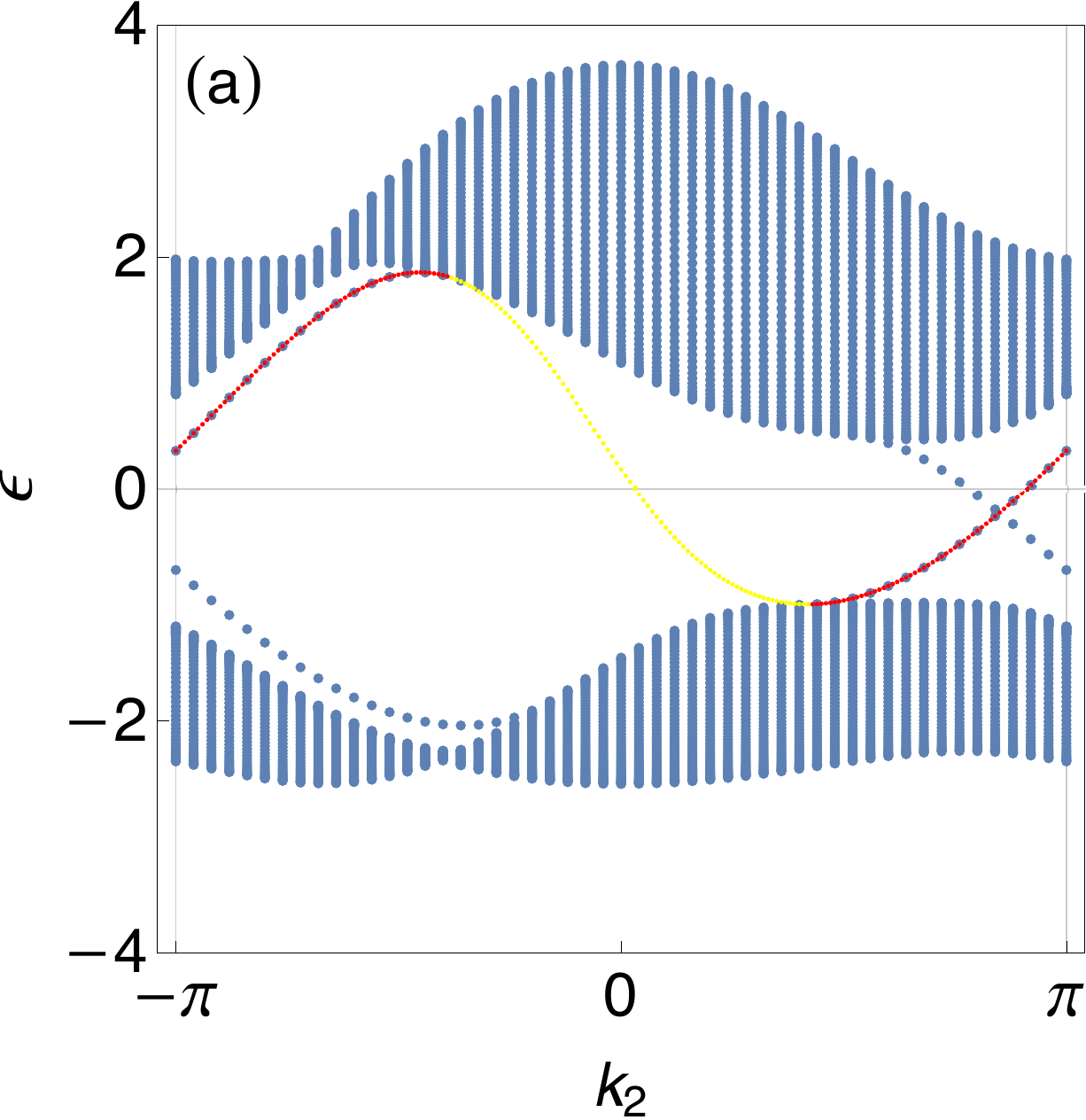}
&
\includegraphics[width=.5\linewidth]{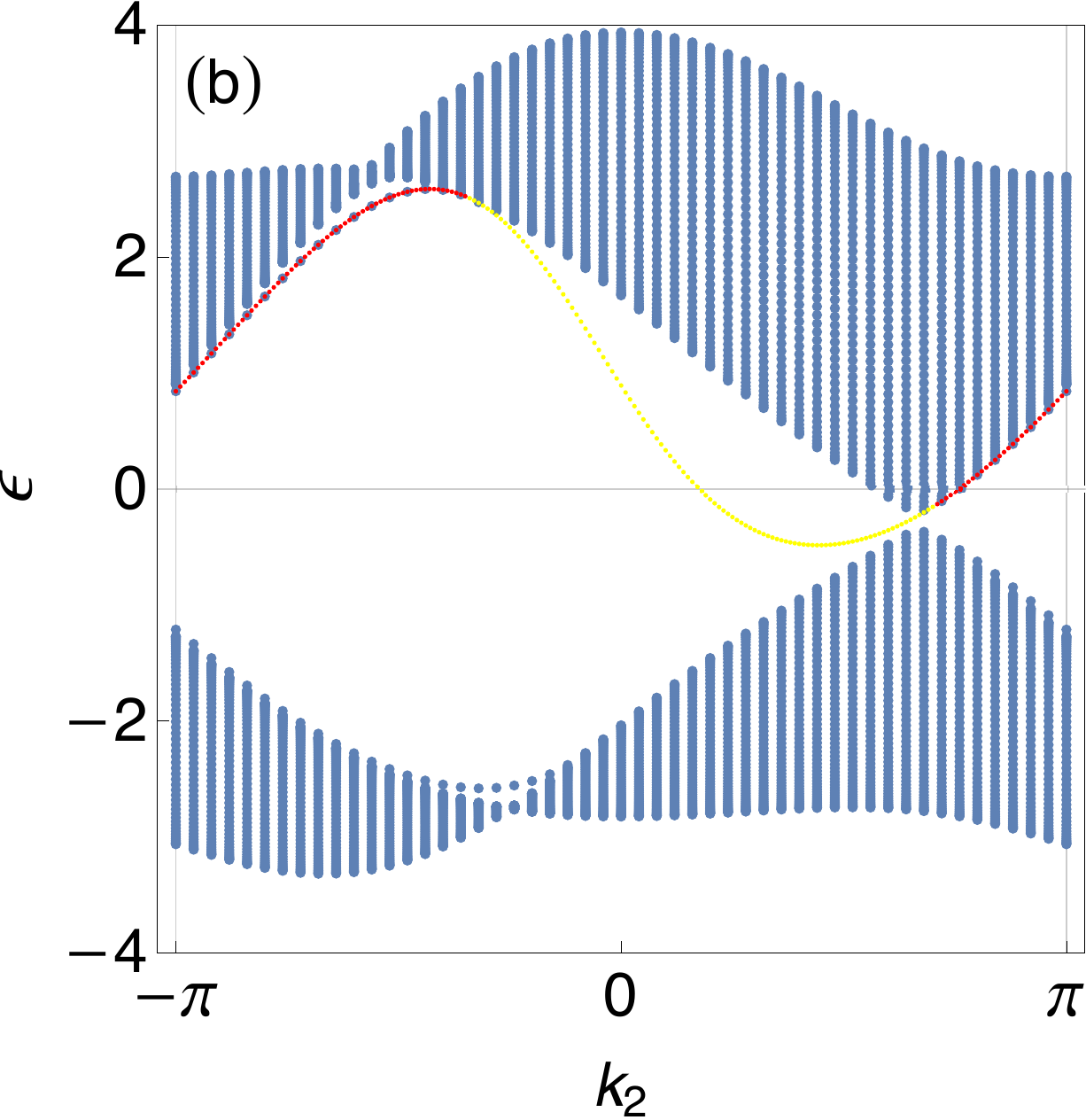}
\end{tabular}
\caption{
Red dots are energies of numerically obtained solutions of Eqs. (\ref{HalHerCon}) and (\ref{HalEigEqu}).
Parameters used are  $t=1$, $t'=0.3$, $\phi=2\pi/5$, and (a)
$m=3\sqrt{3}t'/2$ which belongs to the Chern insulating phase, whereas  (b) $m=3\sqrt{3}t'$ which
belongs to the trivial insulating phase.
The yellow dots show unphysical states with $e^{-\kappa_n}>1$ which do not satisfy the localization condition.
The background shows the total spectrum of the system of cylindrical geometry with zigzag edges.
}
\label{f:hal}
\end{center}
\end{figure}

In Fig. \ref{f:hal}, we show the spectra of the edge states by red dots, which are obtained by numerical calculations.
Figure \ref{f:hal} (a) and (b) are in the nontrivial and trivial phases, respectively.
Even in the trivial phase, there appear edge states localized at boundaries, which can actually be captured by 
the present method based on the Hermiticity  of the Hamiltonian (\ref{HalHerCon}).

\section{Higher-order topological insulators}\label{s:hoti}

This section is devoted to the application of our formulation to the HOTIs
\cin{\cite{Kunst:2018aa,Kunst:2019aa}}.
The HOTIs have been attracting much current interest, providing us a new concept of the bulk-edge correspondence. 
We study two typical models on the square lattice \cite{Benalcazar:2017aa,Benalcazar:2017ab,Liu:2017aa}
and on the breathing kagome lattice \cite{Ezawa:2018aa}.

\subsection{Model on the square lattice}\label{s:bbh}

\begin{figure}[htb]
\begin{center}
\begin{tabular}{c}
\includegraphics[width=.5\linewidth]{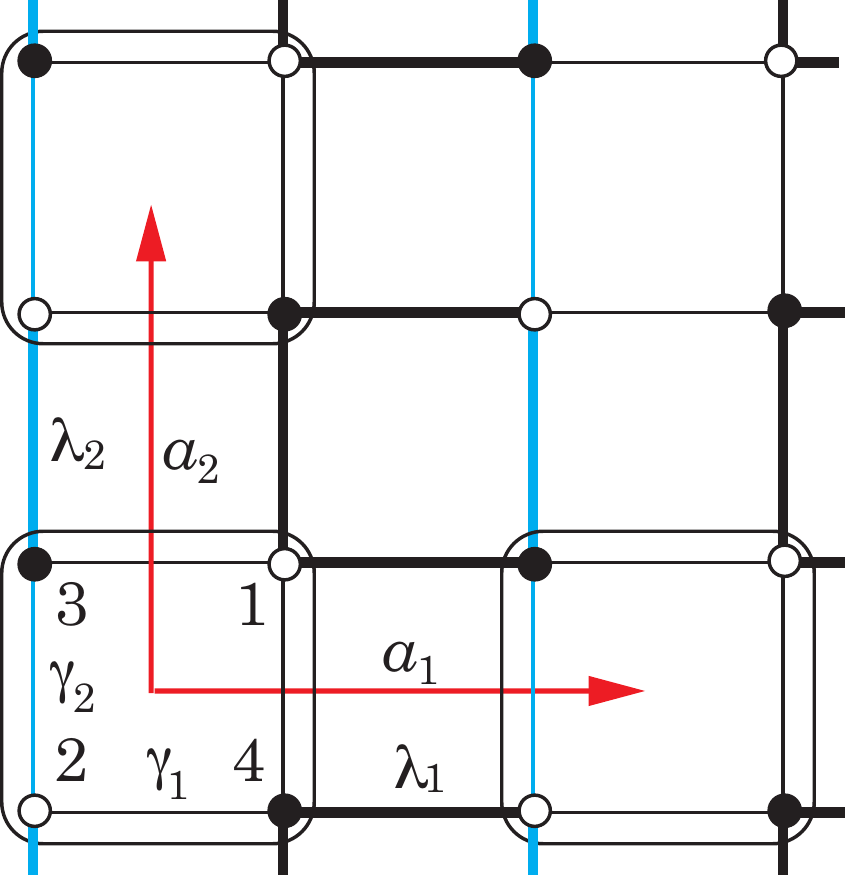}
\end{tabular}
\caption{
The thin and thick lines show the bond-alternation with strength $\gamma_j$ and $\lambda_j$ ($j=1,2$), respectively.
When we consider the BBH model, minus signs representing the $\pi$ flux are attached to the blue lines.
}
\label{f:bbh_lat}
\end{center}
\end{figure}

Let us start with the square lattice with bond-alternating hoppings towards the 1- and 2-directions,
as illustrated in Fig. \ref{f:bbh_lat}. 
On this lattice,  we introduce zero flux and $\pi$-flux, which defines two different models.
These models have nontrivial gapped edge states with nontrivial polarizations, and in particular,
in the so-called BBH model with $\pi$-flux, these edge states yield corner states within a bulk gap \cite{Benalcazar:2017aa,Benalcazar:2017ab,Liu:2017aa}.

From Fig. \ref{f:bbh_lat}, it follows that the Hamiltonian operator is defined by
\begin{alignat}1
\hat{\cal H}=
\left(
\begin{array}{cccc}
&&\hat\Delta_1&\hat\Delta_2\\
&&\pm\hat\Delta_2^*&\hat\Delta_1^*\\
\Delta_1^*&\pm\hat\Delta_2&&\\
\hat\Delta_2^*&\hat\Delta_1&&\\
\end{array}
\right), 
\label{BbhHamOpe}
\end{alignat}
where $\hat\Delta_j=\gamma_j+\lambda_j\delta_j$ and $\hat\Delta_j^*=\gamma_j+\lambda_j\delta_j^*$,
and $-$ signs
representing $\pi$-flux are for the BBH model.
From the coefficients of $\delta_j^*$, we derive the $\cal K$ matrices for the 1- and 2-directions,
\begin{alignat}1
{\cal K}_1=\lambda_1
\left(
\begin{array}{cccc}
&&0&0\\\vspace*{0mm}
&&0&1\\
1&0&&\\
0&0&&
\end{array}
\right),
\quad {\cal K}_2=\lambda_2
\left(
\begin{array}{cccc}
&&0&0\\
&&\pm1&0\\
0&0&&\\
1&0&&
\end{array}
\right).
\label{KBBH}
\end{alignat}
Since each $\cal K$ matrix has two nonzero matrix elements, the models
seem not to belong to class A in Sec. \ref{s:class}. However, due to the particularity of the models, we can apply our formulation to these models as class A, as shown below.

\subsubsection{Edge states}

To begin with, let us consider the system defined on the half-plane, $j_1\ge1$ and  
discuss the edge states along the left end.
The Fourier transformation toward the $2$-direction can be carried out by replacing $\delta_2\rightarrow e^{ik_2}$, 
Then, when we choose
\begin{alignat}1
\psi_{0n}=\left(\begin{array}{c}0\\ \chi_n \\0\end{array}\right),
\label{BBHWavEdg}
\end{alignat}
where $\chi_n=(\chi_{1,n},\chi_{2,n})^T$ is a two-component vector, the Hermiticity of the Hamiltonian and the boundary condition 
at $j_1=1$ are ensured by ${\cal K}_1\psi_{0n}=0$.
Generically, such a reference state yields two constraints on the eigenfunction of the ESH,
which allows no solutions. However, for the present models, two constraints reduce to only one.

To be more specific, based on the reference state (\ref{BBHWavEdg}), 
we assume a Bloch-type wave function $\psi_{j_1n}=\psi_{0n} e^{iK_{1,n}j_1}$, and then, 
the eigenvalue equation becomes,
\begin{widetext}
\begin{alignat}1
\left(
\begin{array}{c|cc|c}
&&\gamma_1+\lambda_1e^{iK_{1,n}}&\gamma_2+\lambda_2e^{ik_2}\\
\hline
&&\pm(\gamma_2+\lambda_2e^{-ik_2})&\gamma_1+\lambda_1e^{-iK_{1,n}}\\
\gamma_1+\lambda_1e^{-iK_{1,n}}&\pm(\gamma_2+\lambda_2e^{ik_2})&&\\
\hline
\gamma_2+\lambda_2e^{-ik_2}&\gamma_1+\lambda_1e^{iK_{,1n}}&&\\
\end{array}
\right)
\left(\begin{array}{c}0 \\\hline \multirow{2}{*}{$\chi_n$}\\ \\ \hline0\end{array}\right)
=\varepsilon_n
\left(\begin{array}{c}0 \\\hline  \multirow{2}{*}{$\chi_n$}\\ \\ \hline0\end{array}\right),
\label{EigEquBBH}
\end{alignat}
\end{widetext}
From this equation, we see that the middle $2\times2$ matrix serves as the ESH,
\begin{alignat}1
{\cal H}_{\rm e}=\pm
\left(
\begin{array}{cc}
&\gamma_2+\lambda_2e^{-ik_2}\\
\gamma_2+\lambda_2e^{-ik_2}&
\end{array}
\right),
\label{BbhEdgHam}
\end{alignat}
provided that the following two constraints, which are from the first and last components of Eq. (\ref{EigEquBBH}), are satisfied: 
\begin{alignat}1
&(\gamma_1+\lambda_1e^{iK_{1,n}})\chi_{2,n}=0,
\nonumber\\
&(\gamma_1+\lambda_1e^{iK_{1,n}})\chi_{1,n}=0.
\label{BbhCon0}
\end{alignat}
Generically, two conditions on the eigenfunctions $\chi_n$ are incompatible with the eigenvalue equation of the Hamiltonian 
(\ref{BbhEdgHam}), ${\cal H}_{\rm e}\chi_n=\varepsilon_n\chi_n$.
However, in the present case, the two conditions (\ref{BbhCon0})
reduce to one, which imposes moreover no constraint on the eigenfunction,
i.e., $\gamma_1+\lambda_1e^{iK_{1,n}}=0$. 
Thus, it turns out that when 
\begin{alignat}1
|e^{iK_{1,n}}|=e^{-\kappa_{1,n}}=\left|\frac{\gamma_1}{\lambda_1}\right|<1,
\label{BBHCon}
\end{alignat}
the edge states localized near $j_1=1$  exists, which is governed by the SSH Hamiltonian 
toward the 2-direction (\ref{BbhEdgHam}), regardless of whether the model includes $\pi$-flux or not.
The difference of the two models lies in the bulk spectrum. 
Since there are no constraints on the wave functions $\chi_n$, 
all of the eigenstates of the ESH (\ref{BbhEdgHam}) are edge states of the present models,
whose eigenvalues are $\varepsilon_\pm(k_2)=\pm|\gamma_2+\lambda_2e^{ik_2}|$.

\begin{figure}[htb]
\begin{center}
\begin{tabular}{cc}
\includegraphics[width=.5\linewidth]{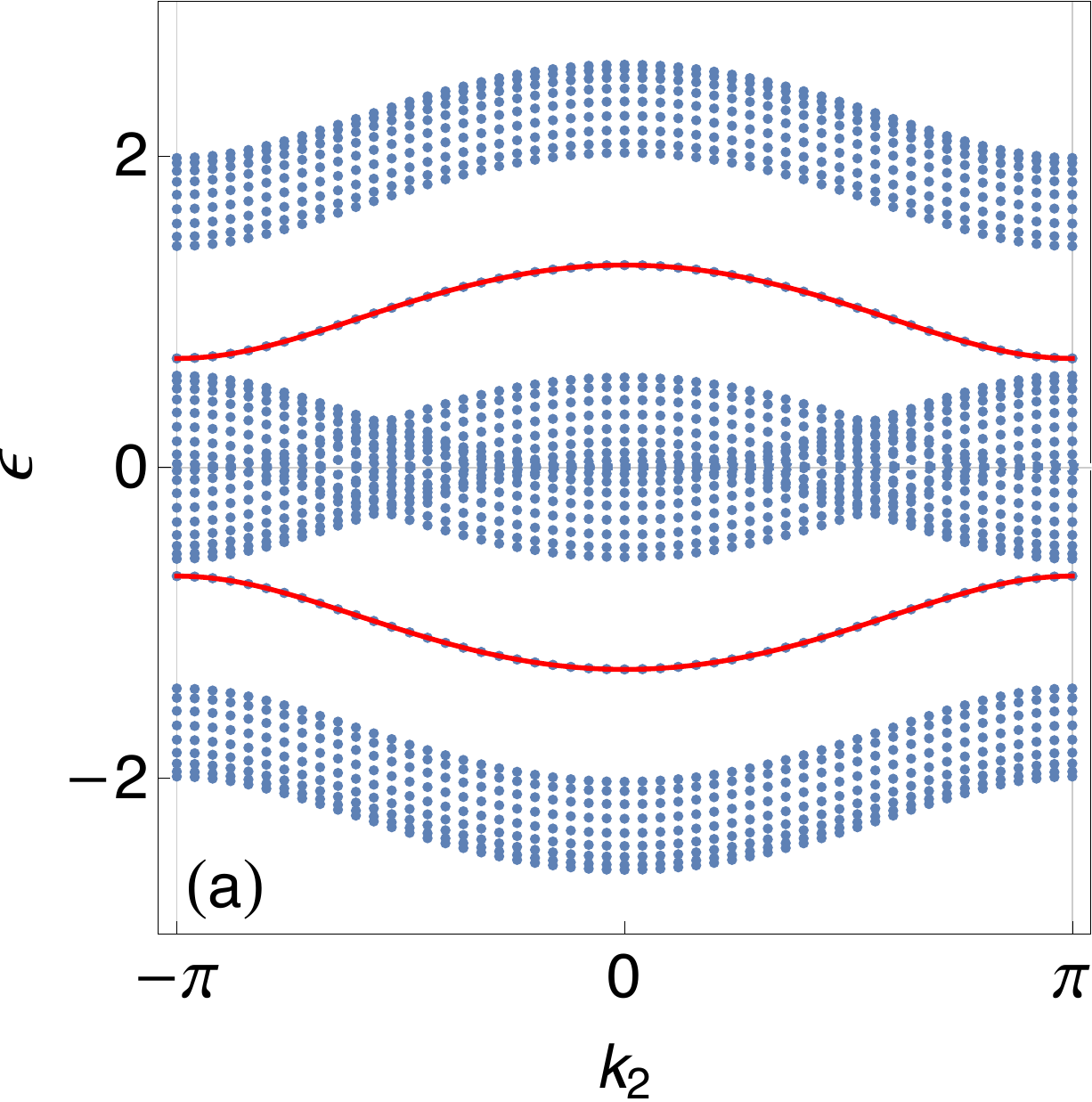}
&
\includegraphics[width=.5\linewidth]{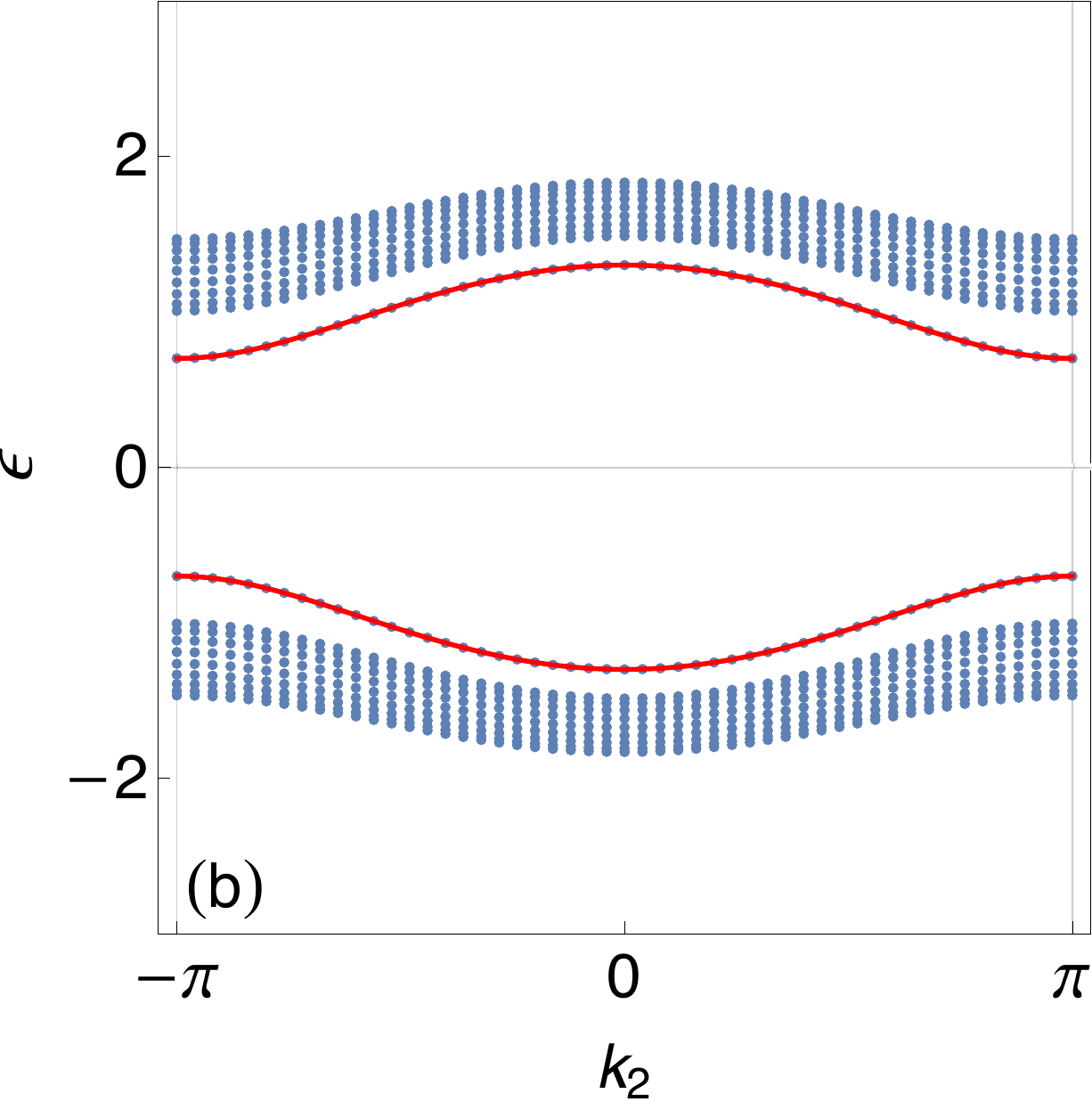}
\end{tabular}
\caption{
Spectra of the models on the square lattice for $\lambda_1=\lambda_2=1$ and $\gamma_1=\gamma_2=0.3$.
(a)  0-flux model, (b) $\pi$-flux model. 
The red curves are eigenvalues of the Hamiltonian (\ref{BbhEdgHam}), and hence show the 
edge states localized at the left end.
The background gray dots
denote the total spectrum obtained numerically for cylindrical system.
}
\label{f:bbh}
\end{center}
\end{figure}

In Fig. \ref{f:bbh}, we show by red curves the spectra $\varepsilon_\pm(k_2)$ of the edge states (\ref{BbhEdgHam}). 
In both figures (a) and (b), they are of course the same and the difference is just the bulk spectra shown by gray dots in the background.
The edge states for the cylindrical system, i.e., red curves in Fig. \ref{f:bbh},
 are doubly-degenerate: One is at the left end, and the other is at the right end.
In the 0-flux model, these edge states are partially embedded in the bulk bands for 
larger $\gamma_j$-parameters.

\subsubsection{Corner state}

In order to study the corner state, we consider the system on the quarter plane defined 
by $j_1\ge1$ and $j_2\ge1$. For the Hermiticity of the Hamiltonian, we have to impose two conditions
associated with ${\cal K}_1$ and ${\cal K}_2$ in Eq. (\ref{KBBH}) on the reference wave function $\psi_{0}$, 
${\cal K}_1\psi_0={\cal K}_2\psi_0=0$. 
Then, we have 
\begin{alignat}1
\psi_{0}=\left(\begin{array}{c}0\\ \chi \\0\\0\end{array}\right),
\end{alignat}
where $\chi$ is just one component vector.
Generically, these conditions are too strong to yield some meaningful solutions. 
However, the present models actually allow one solution, which is nothing but the corner state, as shown below.
Let $\psi_j=\psi_0e^{iK_1j_1+iK_2j_2}$ be the wave function of the corner state.
For such a Bloch-type wave function, 
the eigenvalue equation of the Hamiltonian (\ref{BbhHamOpe}) becomes
\begin{widetext}
\begin{alignat}1
\left(
\begin{array}{c|c|cc}
&&\gamma_1+\lambda_1e^{iK_1}&\gamma_2+\lambda_2e^{iK_2}\\
\hline
&0&\pm(\gamma_2+\lambda_2e^{-iK_2})&\gamma_1+\lambda_1e^{-iK_1}\\
\hline
\gamma_1+\lambda_1e^{-iK_1}&\pm(\gamma_2+\lambda_2e^{iK_2})&&\\
\gamma_2+\lambda_2e^{-iK_2}&\gamma_1+\lambda_1e^{iK_1}&&\\
\end{array}
\right)
\left(\begin{array}{c}0 \\ \hline \chi\\ \hline \multirow{2}{*}{0}\\ \\ \end{array}\right)
=\varepsilon
\left(\begin{array}{c}0 \\ \hline \chi\\\hline \multirow{2}{*}{0}\\ \\ \end{array}\right).
\end{alignat}
\end{widetext}
Thus, it turns out that the Hamiltonian for the corner state is the middle diagonal element 
denoted explicitly by $0$ in the above equation, provided that  
the third and fourth components are satisfied for $\chi\ne0$:
\begin{alignat}1
\gamma_j+\lambda_je^{iK_j}=0, \quad (j=1,2),
\end{alignat}
which reduces to 
\begin{alignat}1
e^{-\kappa_j}=\left|\frac{\gamma_j}{\lambda_j}\right|<1,\quad (j=1,2).
\label{Cor}
\end{alignat}
Namely,  a single corner state appears at $\varepsilon=0$ for the system with the parameters
(\ref{Cor}).
It is indeed an in-gap state for the BBH model, 
whereas it is embedded in the bulk band for the zero flux model.

\subsection{Model on the breathing kagome lattice}

\begin{figure}[htb]
\begin{center}
\begin{tabular}{c}
\includegraphics[width=.6\linewidth]{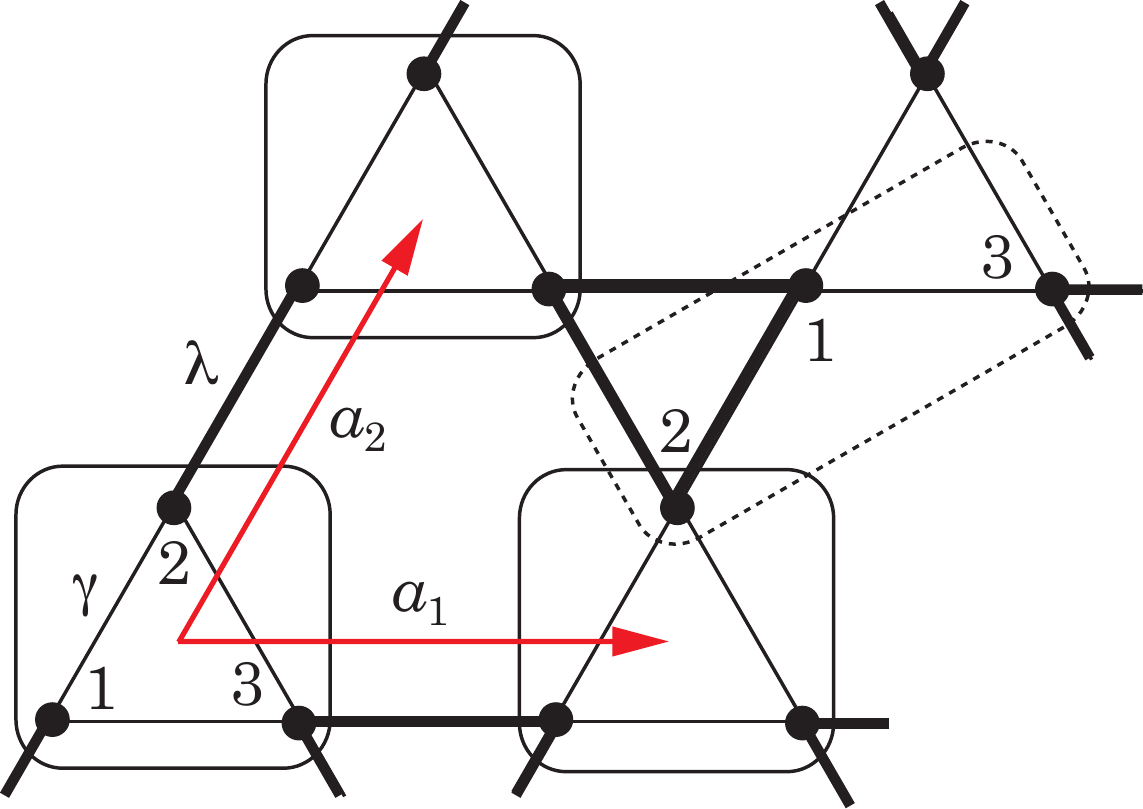}
\end{tabular}
\caption{
Unit cell on the breathing kagome lattice. The dashed-square is another unit cell used 
in Sec. \ref{s:kagome_obtuse}.
}
\label{f:kagome_lat}
\end{center}
\end{figure}

As another example of corner states, we consider the model reported in Ref. \cite{Ezawa:2018aa}.
The unit cell illustrated in Fig. \ref{f:kagome_lat} leads to 
the Hamiltonian operator given by
\begin{alignat}1
\hat{\cal H}=
-\left(
\begin{array}{ccc}
&\gamma+\lambda\delta_2^*&\gamma+\lambda\delta_1^*\\
\gamma+\lambda\delta_2&&\gamma+\lambda\delta_1^*\delta_2\\
\gamma+\lambda\delta_1&\gamma+\lambda\delta_1\delta_2^*&
\end{array}
\right), 
\label{KagHamOpe}
\end{alignat}
from which the $\cal K$ matrices for 1- and 2-directions read
\begin{alignat}1
\newfont{\bg}{cmr10 scaled\magstep2}
\newcommand{\bigzero}{\smash{\hbox{\bg 0}}}
\hat{\cal K}_1=\lambda
\left(
\begin{array}{ccc}
&&1\\
~~&~~&\delta_2\\
&&
\end{array}
\right),\quad
\hat{\cal K}_2=\lambda
\left(
\begin{array}{ccc}
&1&\\
~~&&~~\\
&\delta_1&\\
\end{array}
\right).
\label{KagK}
\end{alignat}
We have written them as operator forms, but when we apply them to edge or corner states, 
$\delta_2$ in $\hat{\cal K}_1$ is replaced by $e^{ik_2}$, and so on.
As already mentioned, two or more nonzero components of $\cal K$ matrices reduce models into class B or C generically.
A major feature of Eq. (\ref{KagK}) is that 
although each $\cal K$ matrix includes two nonzero elements, they are aligned vertically.
This enables us to treat this model as the one in class A, as shown below.

\subsubsection{Edge states}\label{s:kagome_edge}

Let us consider the system defined on the half-plane $j_1\ge1$, and study the 
edge state localized around the left end $j_1=1$. 
Then, we can replace the shift operator $\delta_2$ by $e^{ik_2}$.
It follows from Eq. (\ref{KagK}) that the ${\cal K}_1$ matrix allows the following reference state 
which satisfies the Hermiticity Eq. (\ref{HerCon}) or (\ref{HerConBlo})
\begin{alignat}1
\psi_{0n}=\left(\begin{array}{c}\multirow{2}{*}{$\chi_n$}\\  \\0\end{array}\right),
\label{KagEdgRef}
\end{alignat}
where $\chi_n$ is a two-component vector. 
Note that this state satisfies ${\cal K}_1\psi_{0n}=0$, implying that the model with the present 
edge, i.e., a straight line toward the 1-direction,  belongs to class A.
Assume a Bloch-type wave function of the edge state as $\psi_{j_1n}=\psi_{0n} e^{iK_{1,n}j_1}$ using Eq. (\ref{KagEdgRef}).
Then, the eigenvalue equation becomes
\begin{widetext}
\begin{alignat}1
-\left(
\begin{array}{cc|c}
&\gamma+\lambda e^{-ik_2}&\gamma+\lambda e^{-iK_{1,n}}\\
\gamma+\lambda e^{ik_2}&&\gamma+\lambda e^{-iK_{1,n}}e^{ik_2}\\
\hline
\gamma+\lambda e^{iK_{1,n}}&\gamma+\lambda e^{iK_{1,n}}e^{-ik_2}&
\end{array}
\right)
\left(\begin{array}{c}\multirow{2}{*}{$\chi_n$}\\  \\\hline0\end{array}\right)
=\varepsilon_n
\left(\begin{array}{c}\multirow{2}{*}{$\chi_n$}\\  \\\hline0\end{array}\right).
\label{KagHamEig}
\end{alignat}
\end{widetext}
Thus, the ESH is the upper $2\times2$ matrix defined by
\begin{alignat}1
{\cal H}_{\rm e}=
-\left(
\begin{array}{cc}
&\gamma+\lambda e^{-ik_2}\\
\gamma +\lambda e^{ik_2}&
\end{array}
\right),
\label{KagEdgHam}
\end{alignat}
which gives the eigenvalues $\varepsilon_\pm=\pm|\gamma+\lambda e^{ik_2}|$.
This is the same SSH Hamiltonian as the ESH of the model on the square lattice in Eq. (\ref{BbhEdgHam}).
However, the condition on the wave function given by the last component in (\ref{KagHamEig})
yields nontrivial constraint, 
$(\gamma+\lambda e^{iK_{1,\pm}})\chi_{1,\pm}+(\gamma+\lambda e^{iK_{1,\pm}}e^{-ik_2})\chi_{2,\pm}=0$, 
from which it follows
\begin{alignat}1
|e^{iK_{1,\pm}}|=e^{-\kappa_{1,\pm}}=
\left|\frac{\gamma}{\lambda}\right|\left|
\frac{\varepsilon_\pm-(\gamma +\lambda e^{ik_2})}{\varepsilon_\pm-(\gamma e^{-ik_2}+\lambda)}
\right|<1 .
\label{KagCon}
\end{alignat}
\begin{figure}[htb]
\begin{center}
\begin{tabular}{cc}
\includegraphics[width=.5\linewidth]{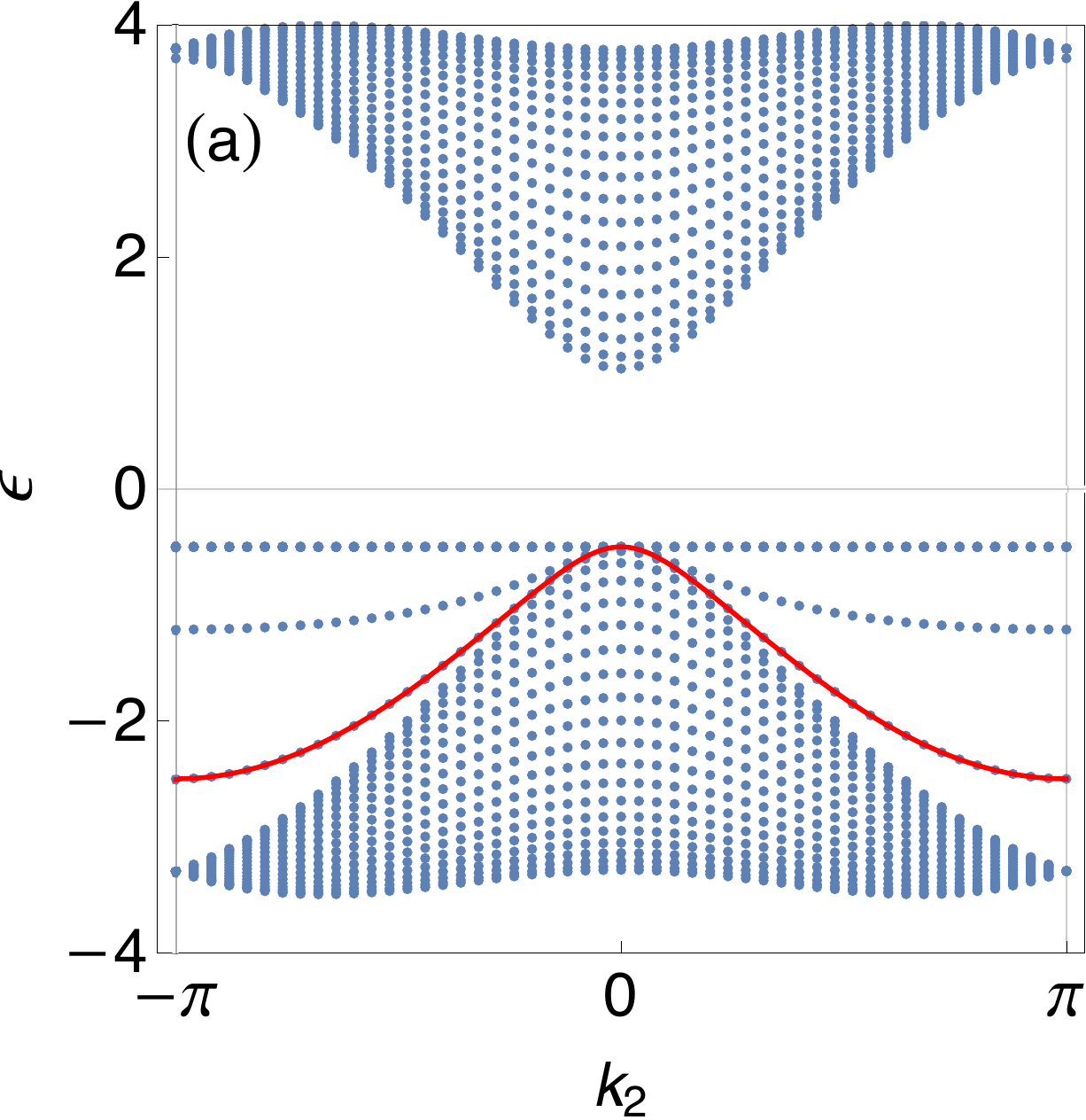}
&
\includegraphics[width=.5\linewidth]{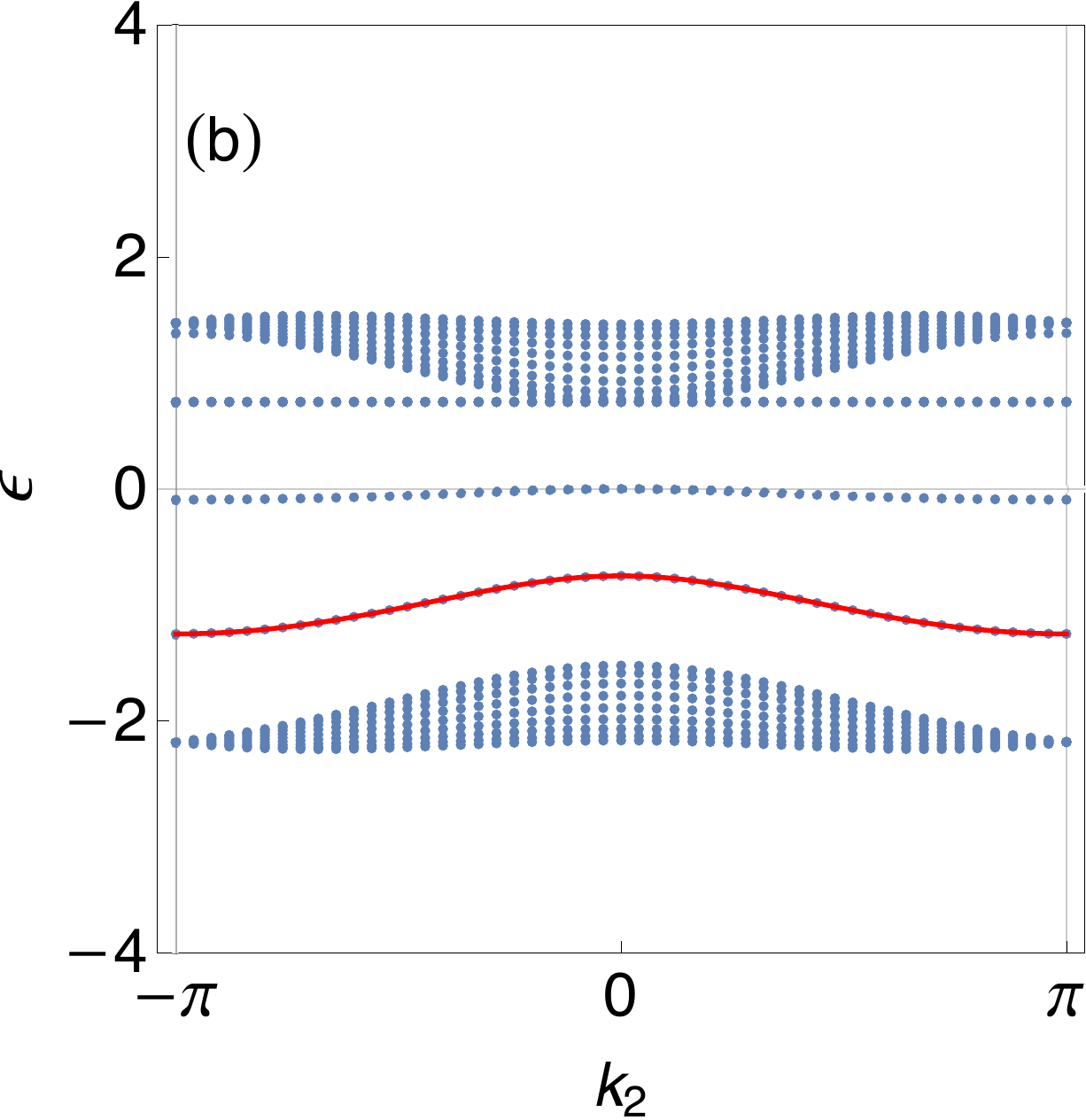}
\\
\includegraphics[width=.5\linewidth]{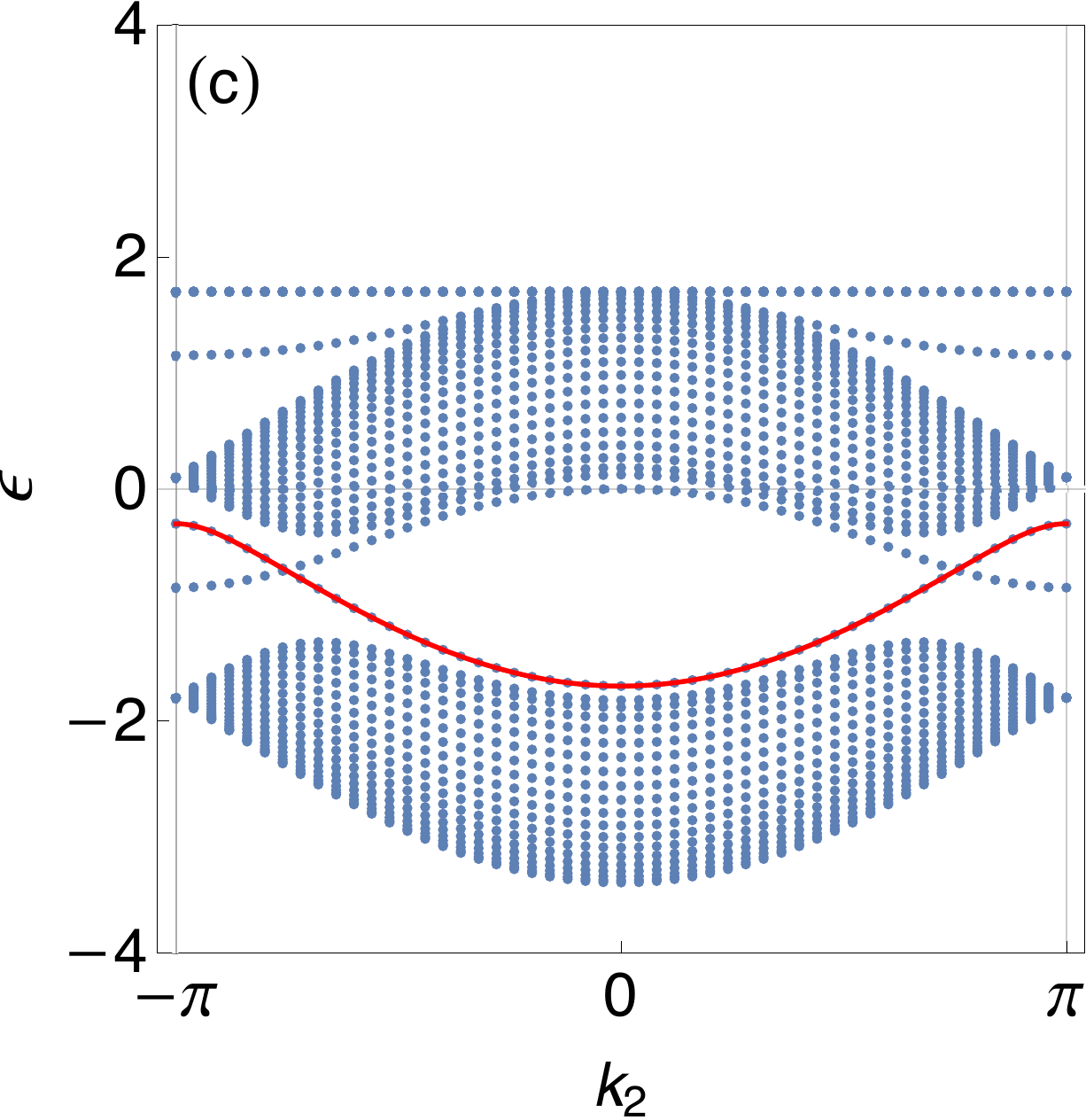}
&
\includegraphics[width=.5\linewidth]{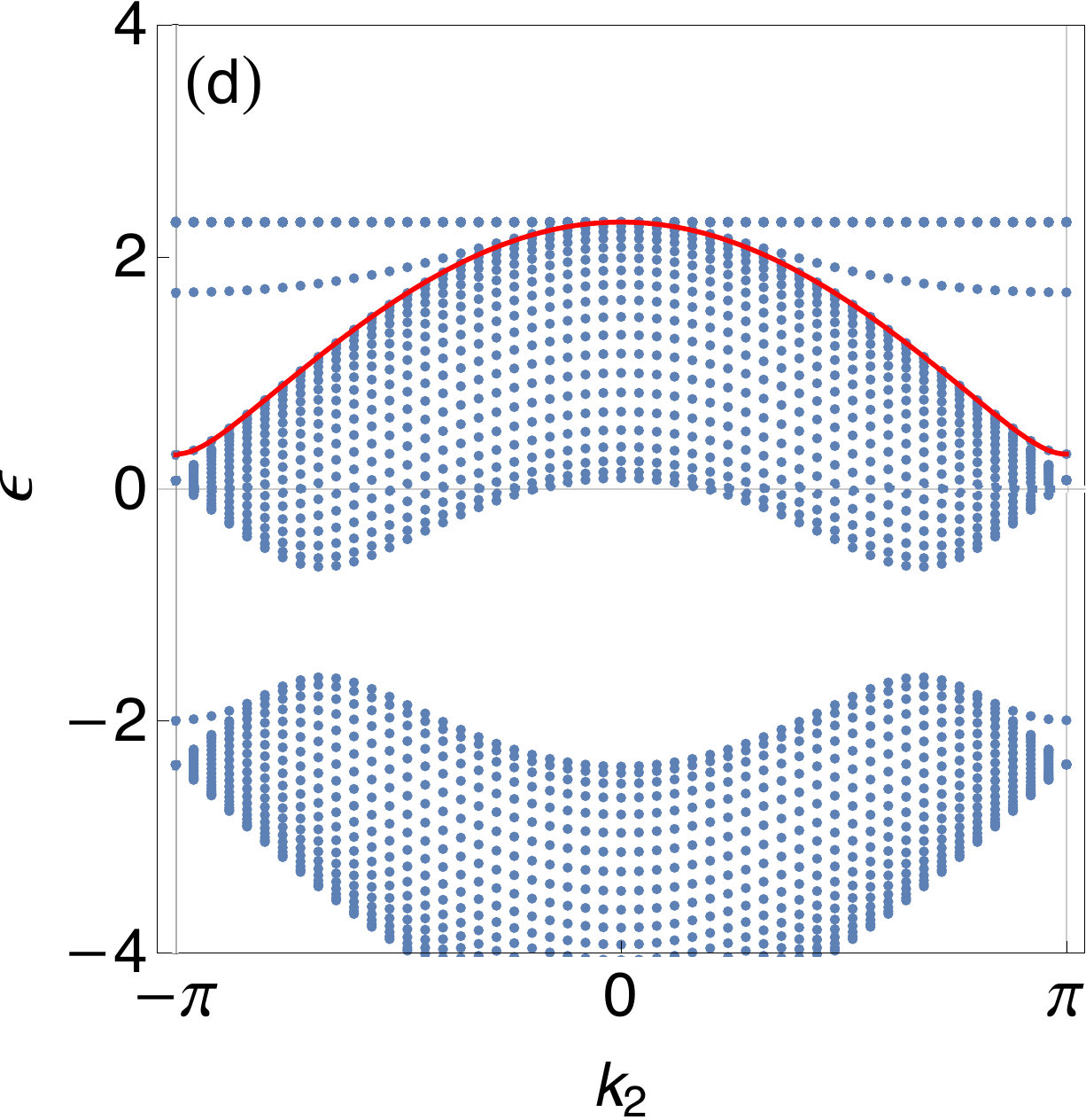}
\end{tabular}
\caption{
Spectra of the model on the breathing kagome lattice.
(a)  $\gamma/\lambda=-1.5$,  (b) $-0.25$, (c) $0.7$, and (d)$1.3$.
The red curves denote the eigenvalues of the ESH (\ref{KagEdgHam}) satisfying the 
localization condition (\ref{KagCon}).
}
\label{f:kagome}
\end{center}
\end{figure}
In Fig. \ref{f:kagome}, we show the  eigenvalues of Eq. (\ref{KagEdgHam})
with the constraint (\ref{KagCon}) by red curves. 
The edge state spectra of this model are the same as those of the models on the square lattice in the previous
Sec. \ref{s:bbh}.
However,  due to the condition (\ref{KagCon}) different from (\ref{BBHCon}) on the square lattice, 
the model on the kagome lattice has only one edge state at \cin{the left} end \cin{on semi-infinite plane
$j_1\ge1$}.
In Figs. \ref{f:kagome} (a)-(c),  the lower branch of the eigenstates of the ESH (\ref{KagEdgHam}) 
forms the physical band localized  at the left end,
whereas in (d), the upper branch becomes the band of the edge states. 
The transition occurs at $\gamma/\lambda=1$.

The background spectra of the cylindrical system in Fig. \ref{f:kagome} show the edge states at the right end as well.
For example, in Fig. \ref{f:kagome} (a), there appear a band in between the edge state 
at the left end and the flat band. This is the edge state at the right end.
Since we consider the unit cell illustrated in Fig. \ref{f:kagome_lat}, the right end has 
a zigzag-like shape. Unfortunately, the problem to determine the right edge states of such a shape 
is of class C type in Sec. \ref{s:class},
and hence, it may be too difficult for the present approach.

\subsubsection{Acute angle corner state}\label{s:kagome_corner}
Suppose that  the system is defined on $j_1\ge1$ and $j_2\ge1$. 
This corresponds to the corner with angle $\pi/3$ on the kagome lattice.
In this case, the reference state $\psi_0$ can be chosen as
\begin{alignat}1
\psi_0=
\left(\begin{array}{c}\chi\\0  \\0 \end{array}\right),
\label{KagCorRef}
\end{alignat}
where $\chi$ is a one component vector.
This reference state satisfies both ${\cal K}_1\psi_0={\cal K}_2\psi_0=0$.
Assuming a Bloch-type state $\psi_{j_1,j_2}=\psi_0e^{i(K_1j_1+K_2j_2)}$ on the reference state 
(\ref{KagCorRef}), the eigenvalue equation becomes
\begin{widetext}
\begin{alignat}1
\left(
\begin{array}{c|cc}
0&\gamma+\lambda e^{-iK_2}&\gamma+\lambda e^{-iK_1}\\
\hline
\gamma+\lambda e^{iK_2}&&\gamma+\lambda e^{-iK_1}e^{iK_2}\\
\gamma+\lambda e^{iK_1}&\gamma+\lambda e^{iK_1}e^{-iK_2}&
\end{array}
\right)
\left(\begin{array}{c}\chi\\\hline0  \\0 \end{array}\right)
=\varepsilon
\left(\begin{array}{c}\chi\\\hline0  \\0 \end{array}\right).
\label{KagHamEigCor}
\end{alignat}
\end{widetext}
This equation tells that the upper diagonal component in the Hamiltonian matrix denoted explicitly by 0
is the corner state Hamiltonian.
Therefore, a single corner state localized around the corner $j=(1,1)$ exists at zero energy, provided that
the following conditions are satisfied,
\begin{alignat}1
|e^{iK_1}|=|e^{iK_2}|=\left|\frac{\gamma}{\lambda}\right|<1,
\end{alignat} 
which are due to the second and third components in (\ref{KagHamEigCor}).
While it appears within a bulk gap for $-1<\gamma/\lambda<1/2$, it is embedded in the bulk states for
$1/2<\gamma/\lambda<1$ \cite{Ezawa:2018aa}.

\subsubsection{Obtuse angle corner state}\label{s:kagome_obtuse}

In order to discuss whether the model allows a corner state at the corner with obtuse angle $2\pi/3$,
we  replace the unit cell by the dashed box in Fig. \ref{f:kagome}. 
The Hamiltonian operator is then
\begin{alignat}1
\hat{\cal H}=
-\left(
\begin{array}{ccc}
&\lambda+\gamma\delta_2&\gamma+\lambda\delta_1^*\\
\lambda+\gamma\delta_2^*&&\lambda\delta_1^*+\gamma\delta_2^*\\
\gamma+\lambda\delta_1&\lambda\delta_1+\gamma\delta_2&
\end{array}
\right), 
\label{KagHamOpe2}
\end{alignat}
from which the $\cal K$ matrices read
\begin{alignat}1
\newfont{\bg}{cmr10 scaled\magstep2}
\newcommand{\bigzero}{\smash{\hbox{\bg 0}}}
\hat{\cal K}_1=\lambda
\left(
\begin{array}{ccc}
&&1\\
~~&~~&1\\
&&
\end{array}
\right),\quad
\hat{\cal K}_2=\gamma
\left(
\begin{array}{ccc}
&&\\
1&~~&1\\
&&
\end{array}
\right).
\label{KagK2}
\end{alignat}

Edge states: In exactly the same way as in the previous Sec. \ref{s:kagome_edge},
a single edge state exists, governed by the same Hamiltonian ${\cal H}_{\rm e}$ in Eq. (\ref{KagEdgHam})
but with $\gamma\leftrightarrow\lambda$ and $k_2\rightarrow-k_2$,
and by the same localization condition (\ref{KagCon}), which of course gives the same energies,
as it should be.

Corner state: Let us consider the system defined on $j_1\ge1$ and $j_2\le-1$.
The reference state should satisfy ${\cal K}_1\psi_0={\cal K}_2^\dagger\psi_0=0$, from which 
it follows that the same state as in Sec. \ref{s:kagome_corner} (\ref{KagCorRef}) can be the reference state.
However, due to the different localization condition,
\begin{alignat}1
|e^{iK_1}|=\left|\frac{\gamma}{\lambda}\right|<1,\quad |e^{-iK_2}|=\left|\frac{\lambda}{\gamma}\right|<1,
\end{alignat} 
the corner with the obtuse angle has no corner  state for any parameters.

\section{Summary and discussion}\label{s:con}

We studied the edge state of various 2D tight-binding systems with particular emphasis on the
Hermiticity of the first-quantized Hamiltonian operators.
We showed that even for tight-binding models defined on latices,  Hamiltonians described by 
the shift  operators, i.e., discrete versions of the differential operators are quite useful.
Then, it turned out that  with boundaries, Hamiltonians on lattices are also not necessarily Hermitian, 
as is often the case with the Dirac Hamiltonians in the continuum space.
We pointed out that the Hermiticity of Hamiltonians has close relationship with boundary conditions.
In particular, for models with NN hoppings only, both conditions can be satisfied simultaneously by specific wave functions.
It then follows that 
we can develop a Bloch-type theory for edge states based on such wave functions.
%
This enables us to extract {\it Hamiltonians describing edge states at one end} from the 
total states including the bulk contributions. 
Namely, we can study edge states appearing at one end for semi-infinite system, or
in other words, we can treat edge states at the left and the right ends separately.

On the other hand, for models including NNN hoppings, we found 
it still difficult to apply our formulation.
This may be due to the fact that while the boundaries of NN models are realized by simple lattice 
terminations, for those including NNN hoppings, how to realized the boundary conditions
is not so obvious. 
Nevertheless, we showed, using Haldane model, 
that even if the boundary conditions are not so clear, 
the Hermiticity of Hamiltonians serves as a guiding principle to determine edge states of
tight-binding models defined on lattices.

Our formulation would have many potential applications.
For example, the ESHs derived by our method enable us to investigate 
the bulk-edge correspondence for various kind of tight-binding models, 
as carried out by Pletyukhov {\it et. al. } for a variant of the Hofstadter model \cite{Pletyukhov:2020aa}.
In this paper, we focused our attention to technical details of 
deriving ESHs for well-known models. 
On the other hand, 
in the discussion of the bulk-edge correspondence, it is important to reveal how ESHs are 
embedded in the bulk Hamiltonians. In the subsequent paper, we would like to clarify this point \cite{inpreparation}.

As demonstrated by Kunst {\it et. al.} \cite{Kunst:2018aa,Kunst:2019aa,Kunst:2019ab}, 
it may also be interesting to construct various models which yield ESHs by exploring the first-quantized 
Hamiltonian operators $\hat{\cal H}$ and corresponding $\cal K$ matrices. 
Even if models do not look NN models, some specific forms of $\cal K$ matrices allow ESHs, as shown  in Sec. \ref{s:hoti}. 
Based on $\hat{\cal H}$, one can construct a lot of tight-binding models to give ESHs.


\acknowledgements

This work was supported in part by Grants-in-Aid for Scientific Research Numbers 17K05563 and 17H06138
from the Japan Society for the Promotion of Science.


\end{document}